\documentclass[preprint,12pt]{elsarticle}

\usepackage{caption}
\usepackage{amsmath}
\usepackage{amssymb}
\usepackage{amsthm}
\usepackage{mathrsfs}

\usepackage{subcaption}

\usepackage{booktabs} 
\usepackage{multirow} 

\usepackage{url}
\usepackage{hyperref}
\usepackage{xcolor}
\usepackage{cleveref} 

\usepackage{lineno}

\usepackage{color}
\usepackage{epstopdf}
\usepackage{float}
\usepackage[top=2.5cm, bottom=2.5cm, left=2cm, right=2cm]{geometry}
\renewcommand{\vec}[1]{\boldsymbol{#1}}

\begin{document}


\begin{frontmatter}



\title{Adaptive wave-particle decomposition in UGKWP method for high-speed flow simulations}

\author[a]{Yufeng Wei}
\author[b]{Junzhe Cao}
\author[c]{Xing Ji}
\author[a,d,e]{Kun Xu\corref{cor1}}

\cortext[cor1] {Corresponding author.}
\ead{makxu@ust.hk}

\address[a]{Department of Mathematics, Hong Kong University of Science and Technology, Hong Kong, China}
\address[b]{School of Aeronautics, Northwestern Polytechnical University, Xi'an, Shaanxi 710072, China}
\address[c]{Shaanxi Key Laboratory of Environment and Control for Flight Vehicle, Xi'an Jiaotong University, Xi'an, China}
\address[d]{Department of Mechanical and Aerospace Engineering, Hong Kong University of Science and Technology, Hong Kong, China}
\address[e]{HKUST Shenzhen Research Institute, Shenzhen, 518057, China}

\begin{abstract}
With wave-particle decomposition, a unified gas-kinetic wave-particle (UGKWP) method has been developed for the multiscale flow simulations. With the variation of cell's Knudsen number, the UGKWP method captures the transport process in all flow regimes without kinetic solver's constraint on the numerical mesh size and time step being less than the particle mean free path and collision time. In the current UGKWP method, the cell's Knudsen number, which is defined as the ratio of particle collision time to numerical time step, is used to distribute the components in the wave-particle decomposition. However, the adaptation of particle in UGKWP is mainly for the capturing of the non-equilibrium transport, and the cell's Knudsen number alone is not enough to identify the non-equilibrium state. For example, in the equilibrium flow regime with a Maxwellian distribution function, even at a large cell's Knudsen number, the flow evolution can be still modelled by the Navier-Stokes solver. More specifically, in the near space environment both the hypersonic flow around a space vehicle and the plume flow from a satellite nozzle will encounter a far field rarefied equilibrium flow in a large computational domain. In the background dilute equilibrium region, the large particle collision time and a uniform small numerical time step can result in a large local cell's Knudsen number and make UGKWP track a huge number of particles for the far field background flow. But, in this region the analytical wave representation can be legitimately used in UGKWP to capture the nearly equilibrium flow evolution. Therefore, to further improve the efficiency of UGKWP for multiscale flow simulations, an adaptive UGKWP (AUGKWP) method will be developed with the introduction of an additional local flow variable gradient-dependent Knudsen number. As a result, the wave-particle decomposition in UGKWP will be determined by both cell's and gradient's Knudsen numbers, and the use of particle in UGKWP is solely to capture the non-equilibrium flow transport. The current AUGKWP becomes much more efficient than the previous one with the cell's Knudsen number only in the determination of wave-particle composition. Many numerical tests, including Sod shock tube, normal shock structure, hypersonic flow around cylinder, flow around reentry capsule, and an unsteady nozzle plume flow, have been conducted to validate the accuracy and efficiency of the AUGKWP method. Compared with the original UGKWP method, the AUGKWP method achieves the same accuracy, but has advantages in memory reduction and computational efficiency in the simulation for the flow with the co-existing of multiple regimes.
\end{abstract}


\end{frontmatter}



\section{Introduction}\label{sec:intro}

Around a high-speed flying vehicle in near flight, the highly compressed gas at the leading edge and the strong expansion wave in the trailing edge can cover all flow regimes from continuum to rarefied with several order differences in the magnitude of particle mean free path \cite{bird1994molecular}. In the control system of moving satellite, the expansion flow in a nozzle from high pressure gas inside the nozzle to the background vacuum can take a rapid and unsteady transition from continuum to the free-molecule flow with a variation of local Knudsen number on the order of ten magnitude. Multiscale flow involves a large variation of Knudsen number and significant changes of the degrees of freedom in the description of flow physics.  In aerospace applications, an accurate and efficient multiscale method with the capability of simulating both continuum and rarefied flow is of great importance \cite{xu2021unified}.

The Boltzmann equation is the fundamental governing equation for rarefied gas dynamics.
Theoretically, it can capture multiscale flow physics in all Knudsen regimes, with the enforcement to the modeling scales of the Boltzmann equation, such as the particle mean free path and mean collision time.
For non-equilibrium flow, there are mainly two kinds of numerical methods to solve the Boltzmann equation, i.e., the stochastic particle method and the deterministic discrete velocity method.
The stochastic methods employ discrete particles to simulate the statistical behavior of molecular gas dynamics \cite{bird1994molecular,fan2001statistical,shen2006rarefied,sun2002direct,baker2005variance,homolle2007low,degond2011moment,pareschi2000asymptotic,ren2014asymptotic,dimarco2011exponential}. This kind of Lagrangian-type scheme achieves high computational efficiency and robustness in rarefied flow simulation, especially for hypersonic flow.
However, it suffers from statistical noise in the low-speed flow simulation due to its intrinsic stochastic nature.
Meanwhile, in the near continuum flow regime, the particle method becomes very expensive due to the requirement of small cell size and time step in the computational space and the tracking of a large amount of particles with intensive collisions.
The deterministic approaches use discrete particle velocity space to solve the kinetic equation and naturally obtain accurate solutions without statistical noise \cite{chu1965kinetic,JCHuang1995,Mieussens2000,tcheremissine2005direct,Kolobov2007,LiZhiHui2009,ugks2010,wu2015fast,aristov2012direct,li2004study,li2019gas,guo2013discrete,chen2017unified,chen2015comparative}.
At the same time, the deterministic method can achieve high efficiency using numerical acceleration techniques, such as implicit algorithms \cite{yang1995rarefied,Mieussens2000,zhu2016implicit,zhu2017implicit,zhu2018implicit,jiang2019implicit},
memory reduction techniques \cite{chen2017unified}, adaptive refinement method \cite{chen2012unified}, and fast evaluation of the Boltzmann collision term \cite{mouhot2006fast,wu2013deterministic}.
Asymptotic preserving (AP) schemes \cite{filbet2010class,dimarco2013asymptotic} and unified preserving (UP) schemes \cite{guo2023unified} can be developed to release the stiffness of the collision term at the small Knudsen number case.
However, for most AP schemes only the Euler solution in the hydrodynamic limit is recovered, where the NS limit can be obtained in the AP schemes.
Additionally, for hypersonic and rarefied flow, the deterministic methods have to use a gigantic amount of grid points in the particle velocity space to cover a large variation of particle velocity and resolve the non-equilibrium distribution.
The three-dimensional hypersonic flow calculation can be hardly conducted due the huge memory consumption and computational cost.
Moreover, for both stochastic and deterministic methods,
with the operator-splitting treatment of particle free transport and instant collision,
a numerical dissipation proportional to the time step is usually unavoidable. Therefore, the mesh size and the time step in these schemes have to be less than the mean free path and the particle mean collision time, respectively, to avoid the physical dissipation being overwhelmingly taken over by the numerical one, especially in the continuum flow regime, such as the laminar boundary layer computation at high Reynolds number.
In order to remove the constraints on the mesh size and time step
in the continuum flow regime, the unified gas-kinetic scheme (UGKS) and discrete UGKS (DUGKS) with the coupled particle transport and collision in the flux evaluation have been constructed successfully \cite{ugks2010,jiang2019implicit,guo2021progress}.
At the same time, the multiscale particle methods have been constructed as well \cite{fei2020unified,fei2021efficient}.

Combining the advantages of the deterministic and the stochastic methods,
a unified gas-kinetic wave-particle (UGKWP) method \cite{liu2020unified, zhu2019unified} has been proposed under the UGKS framework \cite{xu-book,xu2021unified}, as well as simplified versions \cite{PhysRevE.102.013304,DUGKWP-yang}.
The method has included molecular rotation and vibration \cite{xu2021rot,wei2022}, and extended to other multiscale transports,
such as radiation, plasma, and multiphase flow \cite{li2020unified,liu2020plasma,yang2022unified}.
Taking advantage of the evolution solution of kinetic model equation \cite{BGK1954} model in the scheme construction,
the UGKWP method can capture the flow physics in all flow regimes and release the restrictions on the mesh size and time step
which are previously imposed on the kinetic solvers.
In rarefied flow region with a large cell's Knudsen number, the UGKWP method becomes a particle method for the capturing of peculiar gas distribution function. In the continuum region with a small cell's Knudsen number,
the UGKWP method gets back to a hydrodynamic flow solver, where the gas-kinetic scheme (GKS) for the Navier-Stokes solution \cite{xu2001gas}
is fully recovered.
 Different from domain decomposition methods with buffer zones, the UGKWP method employs wave-particle decomposition in each cell with a unified treatment in the whole computational domain. The essential criteria used in the UGKWP method to identify different flow regime according to
 the cell’s Knudsen number ${\rm Kn}_c = \tau / {\Delta t}$, where $\tau$ is the local particle collision time and $\Delta t$ is the numerical time step.
 Naturally the cell's Knudsen number controls the weights in the wave-particle decomposition.
 However, besides identifying the real non-equilibrium flow region, the large cell's Knudsen number also picks up a dilute equilibrium state as the rarefied regime and uses particles to simulate the background equilibrium flow evolution.
 For example, the far field of hypersonic flow around a space vehicle is usually in an equilibrium state with a Maxwellian distribution, and
the cell's Knudsen number in the background equilibrium state is large.
 The large cell's number will increase the weight of particles while in reality the analytical wave can be faithfully used in this region.
 In this paper, besides the cell's Knudsen number, a new parameter for identifying the local non-equilibrium state will be introduced
and added to the previous criteria for wave-particle decomposition.
As a result, an adaptive unified gas-kinetic wave-particle method (AUGKWP) will be constructed with a clear identification of non-equilibrium region. The new method will avoid using particles to the background dilute equilibrium region and further improve the computational efficiency
of UGKWP.

The paper is organized as follows. Since the UGKWP is an enhanced version of the solely particle-based unified gas-kinetic particle (UGKP) method by employing both wave and particle compositions, the UGKP method will be introduced first in Section \ref{sec:kp}.
The UGKWP method will be discussed in Section \ref{sec:wp}. Then the adaptive UGKWP by mainly concentrating
particle distribution to the non-equilibrium flow region will be presented in Section \ref{sec:etaKn}.
Numerical validation of the current method will be carried out in Section \ref{sec:test} and a conclusion will be drawn in Section \ref{sec:conclusion}.

\section{Unified gas-kinetic particle method}\label{sec:kp}

\subsection{General framework}

The unified gas-kinetic particle (UGKP) method is a particle implementation of the UGKS under the finite volume framework, where the discrete particles are employed to describe the non-equilibrium gas distribution function, and the evolution of particles recovers the multiscale nature in different flow regimes. The kinetic equation
with BGK relaxation model is
\begin{equation}\label{eq:BGK-vib-2}
\frac{{\partial {f}}}{{\partial t}} + {\vec{u}} \cdot \frac{{\partial {f}}}{{\partial {\vec{r}}}} = \frac{ g - f}{\tau },
\end{equation}
where the equilibrium state $g$ is the Maxwellian distribution function
\begin{equation*}
g= \rho \left(\frac{\lambda}{\pi}\right)^{\frac{d}{2}} e^{-\lambda {\vec{c}}^2},
\end{equation*}
where $d$ is the degrees of freedom, and $\lambda$ is related to the temperature $T$ by $\lambda = m_0/2k_BT$. Here, $m_0$ and $k_B$ are the molecular mass and Boltzmann constant, respectively. $\vec{c} = \vec{u} - \vec{U}$ denotes the peculiar velocity.
Along the characteristic line, the integral solution of the kinetic model equation gives
\begin{equation}\label{eq:integral-solution1}
f(\vec{r},t)
=
\frac{1}{\tau}\int_{0}^t e^{-(t-t')/\tau}
g(\vec{r}',t')
{\rm{d}} t'
+ e^{-t/\tau}f_0(\vec{r}-\vec{u}t),
\end{equation}
where $ f_0(\vec{r}) $ is the initial distribution function at the beginning of each step $t_n$, and
$g(\vec{r}, t)$ is the equilibrium state distributed in space and time around $\vec{r}$ and $t$. The integral solution describes
an evolution process from non-equilibrium to equilibrium state through particle collision.

In the UGKS, with the expansion of initial distribution function and equilibrium state
\begin{equation}\label{eq:2nd-expansion}
\begin{aligned}
f_0(\vec{r}) &= f_0 + \vec{r} \cdot \frac{\partial f}{\partial \vec{r}}, \\
g(\vec{r}, t) &= g_0 + \vec{r} \cdot \frac{\partial g}{\partial \vec{r}} + \frac{\partial g}{\partial t} t,
\end{aligned}	
\end{equation}
the second-order accurate flux for macroscopic flow variables across cell interface $ij$ can be constructed from the integral solution
\begin{equation} \label{eq:flux}
\begin{aligned}
\vec{F}_{ij} &= \frac{1}{\Delta t} \int_0^{\Delta t} \int \vec{u} \cdot \vec{n}_{ij} f_{ij}(t) {\vec{\psi}} {\rm d}\vec{\Xi} {\rm d}t \\
&= \int \vec{u} \cdot \vec{n}_{ij} \left[C_1 g_0
+ C_2 \vec{u} \cdot \frac{\partial g}{\partial \vec{r}} + C_3 \frac{\partial g}{\partial t} \right]{\vec{\psi}} {\rm d}\vec{\Xi} +
  \int \vec{u} \cdot \vec{n}_{ij} \left[C_4 f_0 + C_5 \vec{u} \cdot \frac{\partial f}{\partial \vec{r}} \right]{\vec{\psi}} {\rm d}\vec{\Xi} \\
&= {\vec{F}}_{ij}^{eq} + {\vec{F}}_{ij}^{fr},
\end{aligned}
\end{equation}
where ${\vec{n}_{ij}}$ is the normal vector of the cell interface, and
\begin{equation*}
\vec{\psi} = \left( 1, \vec{u}, \frac{1}{2} {\vec{u}}^2 \right)^{T}.
\end{equation*}
$\vec{F}^{fr}_{ij}$ and $\vec{F}^{eq}_{ij}$ are the macroscopic fluxes from the free transport and collision processes, respectively. The integrated time coefficients are
\begin{equation*}
\begin{aligned}
C_1 &= 1 - \frac{\tau}{\Delta t} \left( 1 - e^{-\Delta t / \tau} \right) , \\
C_2 &= -\tau + \frac{2\tau^2}{\Delta t} - e^{-\Delta t / \tau} \left( \frac{2\tau^2}{\Delta t} + \tau\right) ,\\
C_3 &=  \frac12 \Delta t - \tau + \frac{\tau^2}{\Delta t} \left( 1 - e^{-\Delta t / \tau} \right) , \\
C_4 &= \frac{\tau}{\Delta t} \left(1 - e^{-\Delta t / \tau}\right), \\
C_5 &= \tau  e^{-\Delta t / \tau} - \frac{\tau^2}{\Delta t}(1 -  e^{-\Delta t / \tau}).
\end{aligned}
\end{equation*}

The UGKS updates both the gas distribution function and macroscopic flow variables under a finite volume framework.
In the UGKP method, the particle will be used to follow the evolution of gas distribution function directly and keep the finite volume version for the updates of macroscopic flow variables.
On the microscopic scale, the particle evolution follows the evolution solution in Eq.\eqref{eq:integral-solution1}, where the particle free transport and collision will be taken into account. On the macroscopic scale, the fluxes across the cell interface for the updates of macroscopic flow variables inside each control volume are evaluated by Eq.\eqref{eq:flux}.

Denote a simulation particle as $P_k(m_k,\vec{r}_k,\vec{u}_k)$, which represents a package of real gas molecules at location $\vec{r}_k$ with particle mass $m_k$ and microscopic velocity $\vec{u}_k$. According to the evolution solution, the cumulative distribution function of particle's collision is
\begin{equation*}\label{tc-distribution}
\mathcal{G}(t)=1-\exp(-t/\tau),
\end{equation*}
then the free transport time of a particle within one time step $\Delta t$ will be
\begin{equation}\label{freetime}
t_{f} = \min(-\tau \ln \eta, \Delta t),
\end{equation}
where $\eta$ is a random number uniformly distributed in $(0,1)$. In a numerical time step from $t^{n}$ to $t^{n+1}$, according to the free transport time $t_f$, the simulation particles can be categorized into collisionless particles ($t_f = \Delta t$) and collisional particles ($t_f < \Delta t$).

In the free transport process, i.e., $t < t_f$, no collisions will happen, and the particles move freely and carry the initial information. The trajectory of particle $P_k$ could be fully tracked by
\begin{equation}\label{stream}
\vec{x}_{k} = \vec{x}_{k}^{n} + \vec{u}_k t_{f,k}.
\end{equation}
During the free transport process, the effective net flux across interfaces of cell $i$ can be evaluated by
\begin{equation}\label{particleevo}
\vec{W}_{i}^{fr}=\frac{1}{\Delta t}\left( \sum_{\vec{x}_k \in \Omega_i}\vec{\phi}_k - \sum_{\vec{x}_k^n \in \Omega_i} \vec{\phi}_{k}\right),
\end{equation}
where $\vec{\phi}_{k} = (m_k, m_k \vec{u}_k, \frac{1}{2} m_k \vec{u}_k^2)^T$.
The free transport flux $\vec{F}_{ij}^{fr}$ in Eq.\eqref{eq:flux} has been recovered by the particles' movement.

In the free transport process, the particle during the time interval $(0, t_f)$ is fully tracked.
The collisionless particles with $t_f = \Delta t$ are kept at the end of the time step.
The collisional particles with $t_f < \Delta t$ would encounter collision at $t_f$ and they are only tracked up to this moment.
Then, all collisional particles are removed, but their accumulating mass, momentum, and energy inside each cell can be still updated through the evolution of macroscopic variables. These collisional particles can be re-sampled from the updated macroscopic variables at the beginning of next time step from equilibrium state if needed.

The equilibrium flux $\vec{F}_{ij}^{eq}$ in Eq.\eqref{eq:flux} contains three terms, i.e., $g$, $\partial_{\vec{r}} g$ and $\partial_t g$, which are only related to the equilibrium states and can be fully determined by the macroscopic flow variables. Once the Maxwellian distribution and its derivatives around the cell interface are determined, the equilibrium flux $\vec{F}^{eq}_{ij}$ can be obtained by
\begin{equation}\label{eq:Feq}
\vec{F}^{eq}_{ij} =
\int \vec{u} \cdot \vec{n}_{ij} \left[C_1 g_0
+ C_2 \vec{u} \cdot \frac{\partial g_t}{\partial \vec{r}}
+ C_3 \frac{\partial g_t}{\partial t} \right]{\vec{\psi}} {\rm d}\vec{\Xi}.
\end{equation}
The macroscopic variables for the determination of equilibrium state $g_0$ at cell interface $ij$ are coming from the
colliding particles from both sides of the cell interface
\begin{equation*}
\vec{W}_{ij}=\int  \left[ g_{t}^l H[\bar{u}_{ij}] + g_{t}^r (1 - H[\bar{u}_{ij}])\right] \vec{\psi} {\rm d} \vec{\Xi},
\end{equation*}
where $\bar{u}_{ij}=\vec{u}\cdot\vec{n}_{ij}$ and $H[x]$ is the Heaviside function.
The gradient of the equilibrium state is obtained from the gradient of macroscopic flow variables ${\partial \vec{W}_{ij}}/{\partial {\vec{r}}}$.
In this study, the spatial reconstruction of macroscopic flow variables is carried out by the least-square method with Venkatakrishnan limiter \cite{venkatakrishnan1995convergence}. As to the temporal gradient, the compatibility condition on Eq.\eqref{eq:BGK-vib-2}
\begin{equation*}
	\int {(g - f) \vec{\psi} {\rm d}\Xi} = 0
\end{equation*}
is employed to give
\begin{equation*}
\frac{\partial\vec{W}_{ij}}{\partial t}
= - \int \vec{u} \cdot \frac{\partial g}{\partial \vec{r}} \vec{\psi} {\rm d}\vec{\Xi}.
\end{equation*}
Correspondingly, the temporal gradient of equilibrium state $\partial_t g$ can be evaluated from the above ${\partial\vec{W}_{ij}}/{\partial t}$. With $g_0$, $\partial_{\vec{r}} g$ and $\partial_t g$, the equilibrium flux $\vec{F}_{ij}^{eq}$ can be fully determined.

\subsection{Updates of macroscopic variables and discrete particles}\label{subsec:update}

Under the finite volume framework, according to the conservation law, the updates of macroscopic variables can be written as
\begin{equation}\label{eq:update}
{\vec {W}}_i^{n + 1}
=
{\vec {W}}_i^n
- \frac{\Delta t}{\Omega_i}
\sum\limits_{j \in N(i)} {{\vec{F}}^{eq}_{ij}{\cal A}_{ij}}
+ \frac{\Delta t}{\Omega_i} \vec{W}^{fr}_i,
\end{equation}
where ${\vec{W}}^{fr}_{i}$ is the net free streaming flow of cell $i$ calculated by particle tracking in the free transport process in Eq.~\eqref{particleevo}, the equilibrium flux ${\vec{F}}^{eq}_{ij}$ is evaluated from macroscopic flow variables and their gradients in Eq.\eqref{eq:Feq}.

Substituting Eq.~\eqref{eq:2nd-expansion} into the integral solution Eq.~\eqref{eq:integral-solution1} of kinetic model equation, the time evolution of distribution function along the characteristic line is
\begin{equation*}
	f(\vec{r},t) = (1-e^{-t/\tau}) g(\vec{r}^\prime, t^\prime) + e^{-t/\tau} f_0(\vec{r} - \vec{u} t),
\end{equation*}
where
\begin{equation*}
	\vec{r}^\prime = \vec{u} \left(\frac{t e^{-t/\tau}}{1-e^{-t/\tau}} -\tau\right), \quad t^\prime = \left( \frac{t e^{-t/\tau}}{1-e^{-t/\tau}}-\tau \right) + t.
\end{equation*}
It indicates that the collisional particles will follow the near-equilibrium state $g(\vec{r}^\prime,t^\prime)$ after collision within the time step $t_f < \Delta t$. With the updated macroscopic flow variables, these untracked collisional particles within the time  $t \in (t_f, \Delta t)$ can be re-sampled from the hydro-particle macroscopic  quantities
\begin{equation}\label{eq:hydro-kp}
	  \vec{W}_i^{h, n+1}
	= \vec{W}_i^{n+1} - \vec{W}_i^{p, n+1}
	= \vec{W}_i^{n+1} - \frac{1}{\Omega_i} \sum_{x_k^{n+1} \in\Omega_i} \vec{\phi}_k,
\end{equation}
where $\vec{W}_i^{p, n+1}$ is from the the collisionless particles remaining in cell $i$.
With the macroscopic quantities and the form of equilibrium state $g$, the corresponding particles can be generated.
Details of sampling from a given distribution function are provided in \cite{zhu2019unified}.

The free transport and collision processes for both microscopic discrete particles and macroscopic flow variables have been described above.
Here, we give a summary of the procedures of the UGKP method. Following the illustration in \cite{zhu2019unified}, the algorithm of UGKP method for diatomic gases with molecular translation, rotation and vibration can be
summarized as follows:

\begin{figure}[H]
	\centering
	\begin{subfigure}[b]{0.24\textwidth}
		\includegraphics[width=\textwidth]{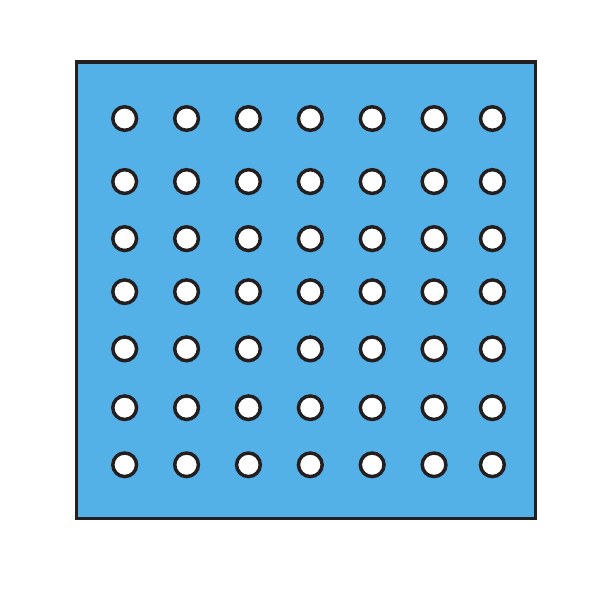}
		\caption{}
		\label{ugkp1}
	\end{subfigure}
	\begin{subfigure}[b]{0.24\textwidth}
		\includegraphics[width=\textwidth]{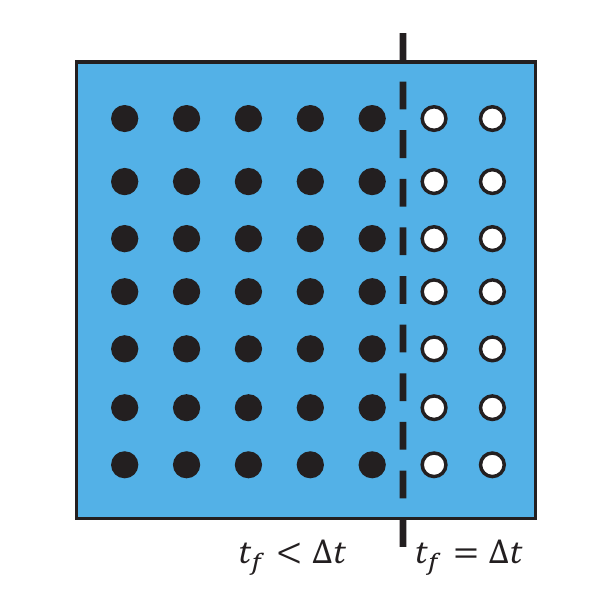}
		\caption{}
		\label{ugkp2}
	\end{subfigure}
	\begin{subfigure}[b]{0.24\textwidth}
		\includegraphics[width=\textwidth]{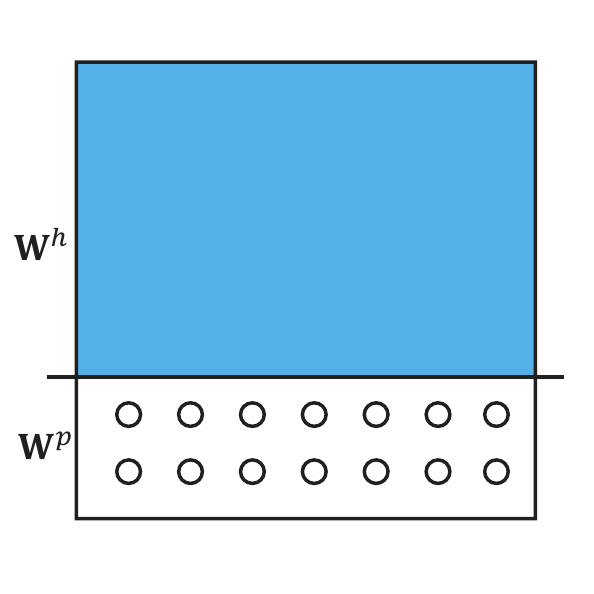}
		\caption{}
		\label{ugkp3}
	\end{subfigure}
	\begin{subfigure}[b]{0.24\textwidth}
		\includegraphics[width=\textwidth]{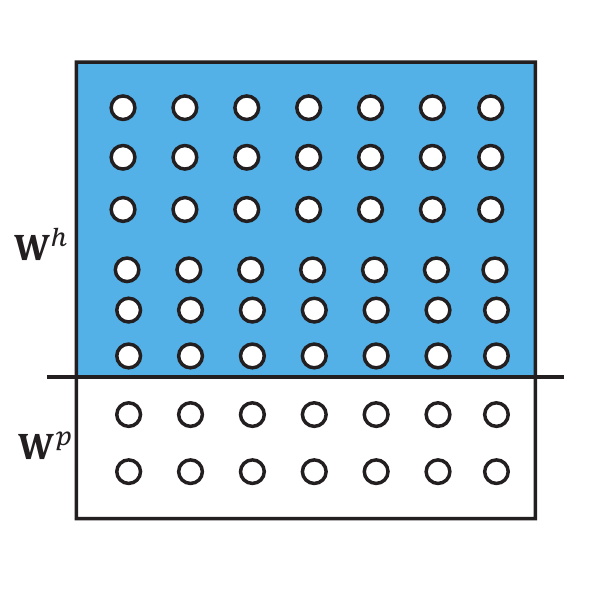}
		\caption{}
		\label{ugkp4}
	\end{subfigure}
	\caption{Diagram to illustrate the composition of the particles during time evolution in the UGKP method. (a) Initial field, (b) classification of the collisionless particles (white circle) and collisional particles (solid circle) according to the free transport time $t_f$, (c) update solution at the macroscopic level, and (d) update solution at the microscopic level.}
	\label{fig:ugkp}
\end{figure}

\begin{description}
	\item[Step 1]
	For the initialization, sample particles from the given initial conditions as shown in Fig.~\ref{fig:ugkp}(a).
	\item[Step 2]
		Generate the free transport time $t_f$ for each particle by Eq.~\eqref{freetime}, and classify the particles into collisionless particles (white circles in Fig.~\ref{fig:ugkp}(b)) and collisional ones (solid circles in Fig.~\ref{fig:ugkp}(b)).
		Stream the particles for free transport time by Eq.~\eqref{stream}, and evaluate the net free streaming flow $\vec{W}^{fr}_i$ by Eq.~\eqref{particleevo}.
	\item[Step 3]
	    Reconstruct macroscopic flow variables and compute the equilibrium flux $\vec{F}^{eq}_{ij}$ by Eq.~\eqref{eq:Feq}.		
	\item[Step 4]
		Update the macroscopic flow variables $\vec{W}_i$ by Eqs~\eqref{eq:update}.
		Obtain the updated hydro-particle macroscopic quantities of collisional particles $\vec{W}^h_i$ by extracting the macro-quantities of collisionless particles
		$\vec{W}^p_i$ from the total flow variables $\vec{W}_i$ in Eq.~\eqref{eq:hydro-kp} as shown in Fig.~\ref{fig:ugkp}(c).		
	\item[Step 5]
		Delete the collisional particles at $t_f$ and re-sample these particles from the updated hydro-particle macroscopic  variables $\vec{W}^h_i$ as shown in Fig.~\ref{fig:ugkp}(d), which becomes the initial state in Fig.~\ref{fig:ugkp}(a) at the beginning of next time step.
	\item[Step 6] Go to Step 2. Continue time step evolution until the finishing time.
\end{description}

\section{Unified gas-kinetic wave-particle method}\label{sec:wp}

In the UGKP method, based on the updated hydro-particle macroscopic variables $\vec{W}^h_i$  of collisional particles, these particles
will be re-sampled from equilibrium state at the beginning of next time step. However, some of these re-sampled particles will get collision in the next time step and get eliminated again.
Therefore, in the unified gas-kinetic wave-particle (UGKWP) method, only free transport particles in the next time step will be re-sampled from $\vec{W}^h_i$. In the continuum regime at very small Knudsen number, it is possible that no free particles will get re-sampled.

\begin{figure}[H]
	\centering
	\begin{subfigure}[b]{0.24\textwidth}
		\includegraphics[width=\textwidth]{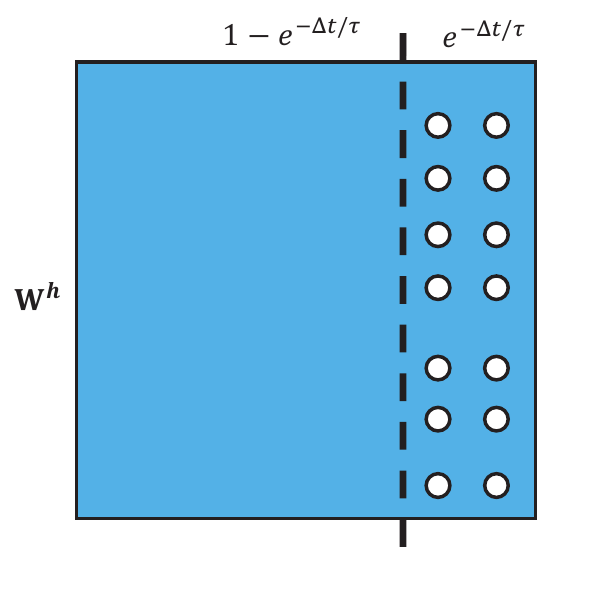}
		\caption{}
		\label{ugkwp1}
	\end{subfigure}
	\begin{subfigure}[b]{0.24\textwidth}
		\includegraphics[width=\textwidth]{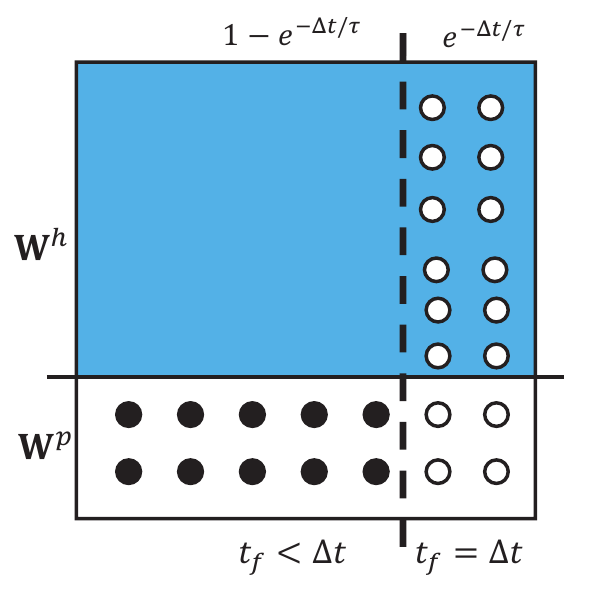}
		\caption{}
		\label{ugkwp2}
	\end{subfigure}
	\begin{subfigure}[b]{0.24\textwidth}
		\includegraphics[width=\textwidth]{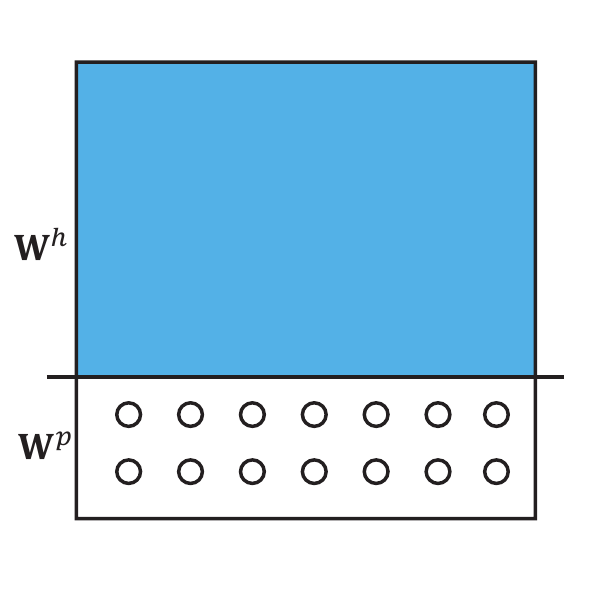}
		\caption{}
		\label{ugkwp3}
	\end{subfigure}
	\begin{subfigure}[b]{0.24\textwidth}
		\includegraphics[width=\textwidth]{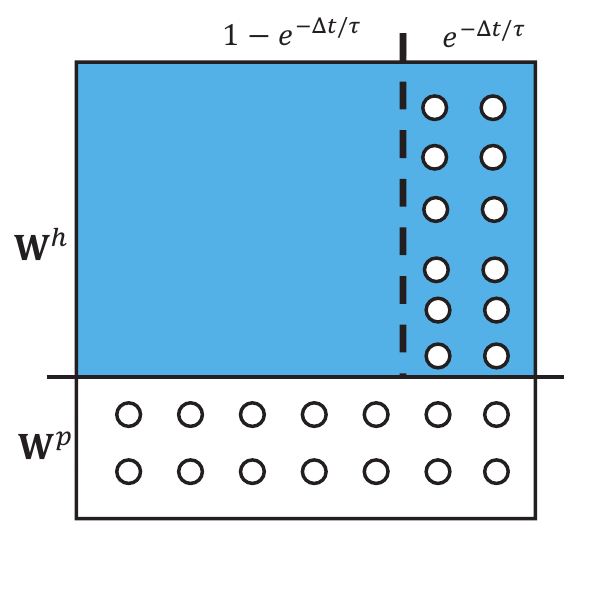}
		\caption{}
		\label{ugkwp4}
	\end{subfigure}
	\caption{Diagram to illustrate the composition of the particles during time evolution in the UGKWP method. (a) Initial field, (b) classification of the collisionless and collisional particles for $\vec{W}^p_i$, (c) update on the macroscopic level, and (d) update on the microscopic level.}
	\label{fig:ugkwp}
\end{figure}

The collisionless particles with $t_f = \Delta t$ will be sampled from  $\vec{W}^h_i$.
According to the integral solution, the collisionless particles will take a fraction of $\vec{W}^h_i$ by the amount
\begin{equation}\label{origWP}
{\vec{W}}^{hp}_i = e^{\frac{-\Delta t}{\tau}}{\vec{W}}^{h}_{i}.
\end{equation}
As shown in Fig.~\ref{fig:ugkwp}, there is no need to sample these collisional particles from the hydrodynamic portion with macroscopic variables $(\vec{W}^h_i -{\vec{W}}^{hp}_i)$.
The free transport flux from these un-sampled collisional particles can be evaluated analytically
\begin{equation}\label{Ffrh}
\vec{F}^{fr,h}_{ij}=\int
\vec{u}\cdot\vec{n}_{ij}
\left[
C_4^\prime g_0
+ C_5^\prime \vec{u} \cdot \frac{\partial g_t}{\partial \vec {r}}\right]
\vec{\psi} {\rm{d}} \vec{\Xi} ,
\end{equation}
where
\begin{equation*}
\begin{aligned}
C_4^\prime &= \frac{\tau}{\Delta t} \left(1 - e^{-\Delta t / \tau}\right)
            -  e^{-\Delta t / \tau}, \\
C_5^\prime &= \tau  e^{-\Delta t / \tau}
- \frac{\tau^2}{\Delta t}(1 -  e^{-\Delta t / \tau})
+ \frac12\Delta t e^{-\Delta t / \tau}.
\end{aligned}
\end{equation*}
Then, the update of macroscopic flow variables in the UGKWP method becomes
\begin{equation}\label{eq:update-Wp}
	{\vec{W}}_i^{n + 1}
	= {\vec{W}}_i^n
		- \frac{\Delta t}{\Omega_i}
		  \sum\limits_{j \in N(i)}  {\vec{F}}^{eq}_{ij}{\cal A}_{ij}
		- \frac{\Delta t}{\Omega_i}
		  \sum\limits_{j \in N(i) } {\vec{F}}^{fr,h}_{ij} {\cal A}_{ij}
		+ \frac{\Delta t}{\Omega _i} \vec{W}^{fr,p}_{i}
		+ \vec{S}_i.
\end{equation}

The algorithm of the UGKWP method for diatomic gases can be summarized as follows.
\begin{description}
	\item[Step 1]
		For the initialization, sample collisionless particles from $\vec{W}_i^{hp}$ with $t_f = \Delta t$ as shown in Fig.~\ref{fig:ugkwp}(a).
		For the first step, $\vec{W}^h_i = \vec{W}_i^{n = 0}$.		
	\item[Step 2]
		Generate the free transport time $t_f$ by Eq.~\eqref{freetime} for the remaining particles from previous step evolution with total amount $\vec{W}_i^p$, and classify the particles into collisionless particles (white circles in Fig.~\ref{fig:ugkwp}(b)) and collisional ones (solid circles in Fig.~\ref{fig:ugkwp}(b)).
		Stream the particles for free transport time by Eq.~\eqref{stream}, and evaluate the net free streaming flow $\vec{W}^{fr}_i$ by Eq.~\eqref{particleevo}.
	\item[Step 3]
		Reconstruct macroscopic flow variables and compute the free transport flux of collisional particles $\vec{F}_{ij}^{fr,h}$ by Eq.~\eqref{Ffrh} and the equilibrium flux $\vec{F}^{eq}_{ij}$ by Eq.~\eqref{eq:Feq}.			
	\item[Step 4]
		Update the macroscopic flow variables $\vec{W}_i$ by Eqs~\eqref{eq:update-Wp}.
		Obtain the updated macroscopic quantities for collisional particles $\vec{W}^h_i$ by extracting the macro-quantities of collisionless particles
		$\vec{W}^p_i$ from the total flow variables $\vec{W}_i$ in Eq.~\eqref{eq:hydro-kp} as shown in Fig.~\ref{fig:ugkwp}(c).	
	\item[Step 5]
	Delete the collisional particles at $t_f$ ($t_f < \Delta t $).
Re-sample the collisionless particles from $\vec{W}^{hp}_i$ with $t_f = \Delta t$ at the beginning of next time step, as shown in Fig.~\ref{fig:ugkwp}(d).
	\item[Step 6]
	Go to Step 2, continue time evolution until the output time.
\end{description}

The UGKP method uses particles to represent the gas distribution function. However, the UGKWP method adopts a hybrid formulation of wave and particles to recover the gas distribution function. Within the time step, the evolution of wave part for the collisional particles ($t_f < \Delta t$ ) can be described analytically by the time accurate solution of macroscopic flow variables without sampling these particles explicitly.
The evolution of the remaining particles will track the non-equilibrium effect through particle free transport.
In the rarefied flow regime, the UGKWP method is dominated by particle evolution, which results in a particle method.
While in the continuum regime, the UGKWP method is mainly about the evolution of macroscopic variables,
and the scheme becomes a hydrodynamic NS solver, such as the so-called gas-kinetic scheme (GKS) \cite{xu2001}.
Therefore, the UGKWP method achieves much better computational efficiency and lower memory consumption than the purely particle method UGKP.

However, the weights of wave and particles in the current UGKWP are controlled by the cell's Knudsen number $\tau/\Delta t$.
The total mass fraction of particles is proportional to $e^{-\Delta t/\tau} \vec{W}$.
There are still weakness in the above formulation.
For example, even in the continuum flow regime, if a small time step $\Delta t$ is used, the particles will emerge automatically in the flow evolution.
At the same time, for a dilute background equilibrium distribution in near space environment with a large particle collision time $\tau$,
the UGKWP will use the particles to capture the background equilibrium flow evolution. Therefore, besides the cell's Knudsen number,
in order to use particle to really capture the evolution of non-equilibrium state, other parameter has to be designed as well in the determination of the distributions between wave and particle in UGKWP.

\section{Adaptive unified gas-kinetic wave-particle method}\label{sec:etaKn}
For the UGKWP method, the evaluation of flow regimes depends on the cell's Knudsen number. However, this criteria cannot capture the real non-equilibrium regime, especially for the rarefied undisturbed equilibrium flow and the flow simulation with a very small numerical time step for high resolution. Even in the equilibrium regime, the large particle collision time and the small time step can give a large the cell's Knudsen number for particle generation. Theoretically, the analytical wave can be used in those equilibrium regime.
Besides the cell's Knudsen number, the adaptive unified gas-kinetic wave-particle (AUGKWP) method will be developed by introducing another parameter to identify the real non-equilibrium region for the generation of particles.
This parameter is a gradient-length related local Knudsen number ${\rm{Kn}}_{Gll}$.
In other words, the analytical wave will take effect as well when ${\rm Kn}_{Gll}$ is small.
Therefore, the portion from the macroscopic flow variables to sample particles becomes
\begin{equation*}
	{\vec{W}}^{hp}_i = e^{\frac{-\Delta t}{\tau}}\eta({\rm{Kn}}_{Gll}){\vec{W}}^{h}_{i},
\end{equation*}
where
\begin{equation*}
	\eta({\rm Kn}_{Gll})
	= \frac{1}{2} \left[ \tanh \left( \frac{{\rm Kn}_{Gll}/{\rm Kn}_{ref} - 1 }
										   {{\rm Kn}_{ref}}
							   \right)
			             + 1 \right],
\end{equation*}
and the gradient-length local Knudsen number is defined by
\begin{equation*}
	{\rm Kn}_{Gll} = \frac{l_{mfp}}{\rho / \left| \nabla \rho\right|},
\end{equation*}
where $l_{mfp}$ is the local mean free path. The reference Knudsen number ${\rm Kn}_{ref}$ is included as a critical value to evaluate the equilibrium and non-equilibrium regimes and thus directly control the wave-particle decomposition in each control volume. Then, the analytical free transport flux from un-sampled particles is amended as
\begin{equation*}
	\vec{F}^{fr,h}_{ij}=\int
	\vec{u}\cdot\vec{n}_{ij}
	\left[
	C_4^{{\rm Kn}_{Gll}} g_0
	+ C_5^{{\rm Kn}_{Gll}} \vec{u} \cdot \frac{\partial g_t}{\partial \vec {r}}\right]
	\vec{\psi} {\rm{d}} \vec{\Xi} ,
	\end{equation*}
where
\begin{equation*}
	\begin{aligned}
	C_4^{{\rm Kn}_{Gll}} &= \frac{\tau}{\Delta t} \left(1 - e^{-\Delta t / \tau}\right)
				-   e^{-\Delta t / \tau} \eta({\rm Kn}_{Gll}), \\
	C_5^{{\rm Kn}_{Gll}} &= \tau  e^{-\Delta t / \tau}
	- \frac{\tau^2}{\Delta t}(1 -  e^{-\Delta t / \tau})
	+ \frac12\Delta t e^{-\Delta t / \tau} \eta({\rm Kn}_{Gll}).
	\end{aligned}
\end{equation*}

\begin{figure}[H]
	\centering
	\includegraphics[width=0.45\textwidth]{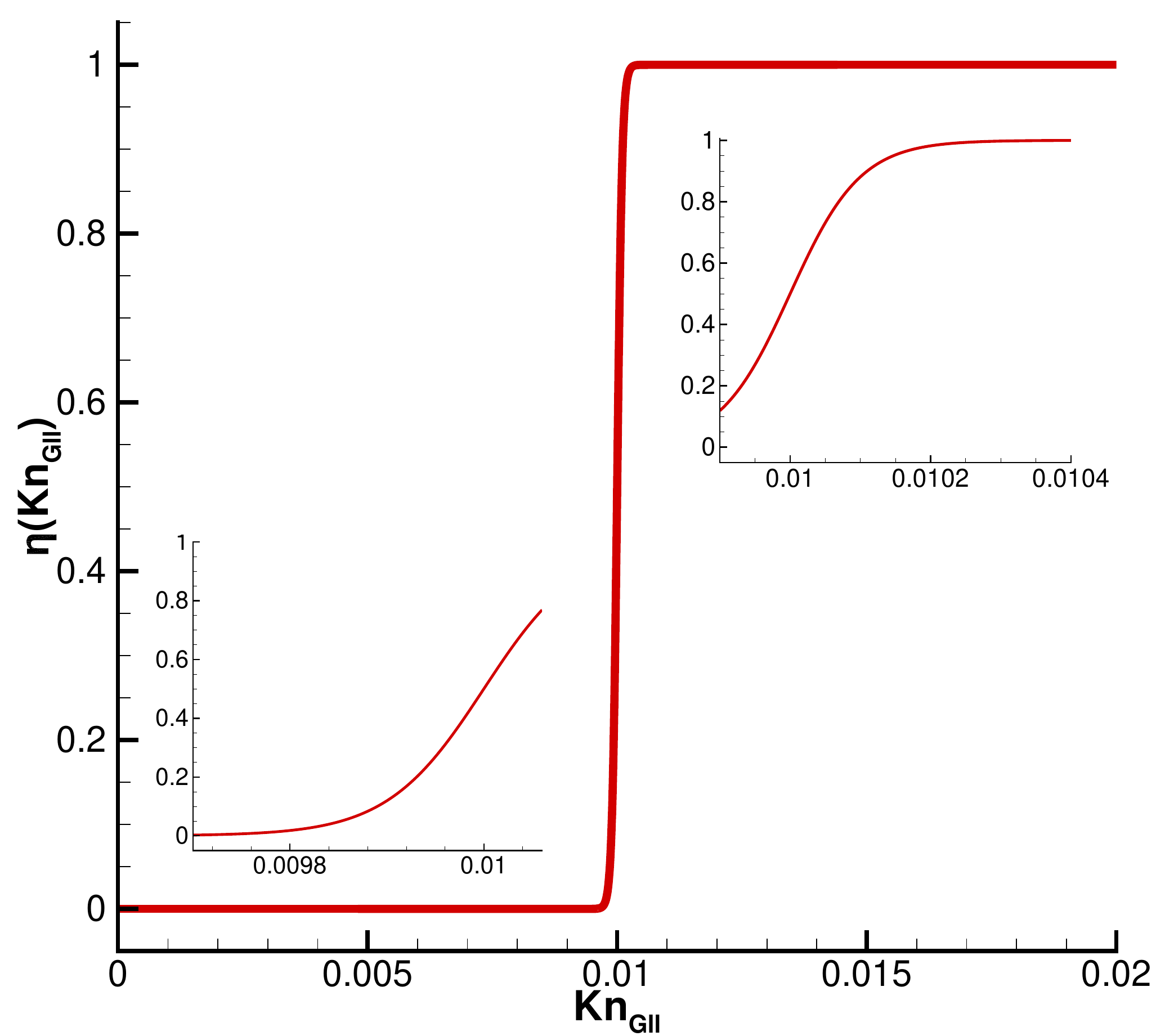}
	\caption{The function $\eta({\rm Kn}_{Gll})$ at a reference Knudsen number ${\rm Kn}_{ref} = 0.01$.}
	\label{fig:augkwp-etaKn}
\end{figure}

In the AUGKWP method, a hyperbolic tangent function shown in Fig.~\ref{fig:augkwp-etaKn} is used in the determination of the
wave-particle decomposition due to its boundness, smoothness, and convexity. Its boundness ensures a nature transition without further restriction, and the smoothness avoids the oscillation in the transition regime. Its convexity satisfies the expectation for sampling particle with slow rate of change in the continuum regime and a large rate otherwise. Moreover, the reference Knudsen number ${\rm Kn}_{ref}$ can be straightforwardly added in this function as a value to distinguish the flow regime, and can be conveniently adjusted according to the flow condition.

\begin{figure}[H]
	\centering
	\includegraphics[width=0.55\textwidth]{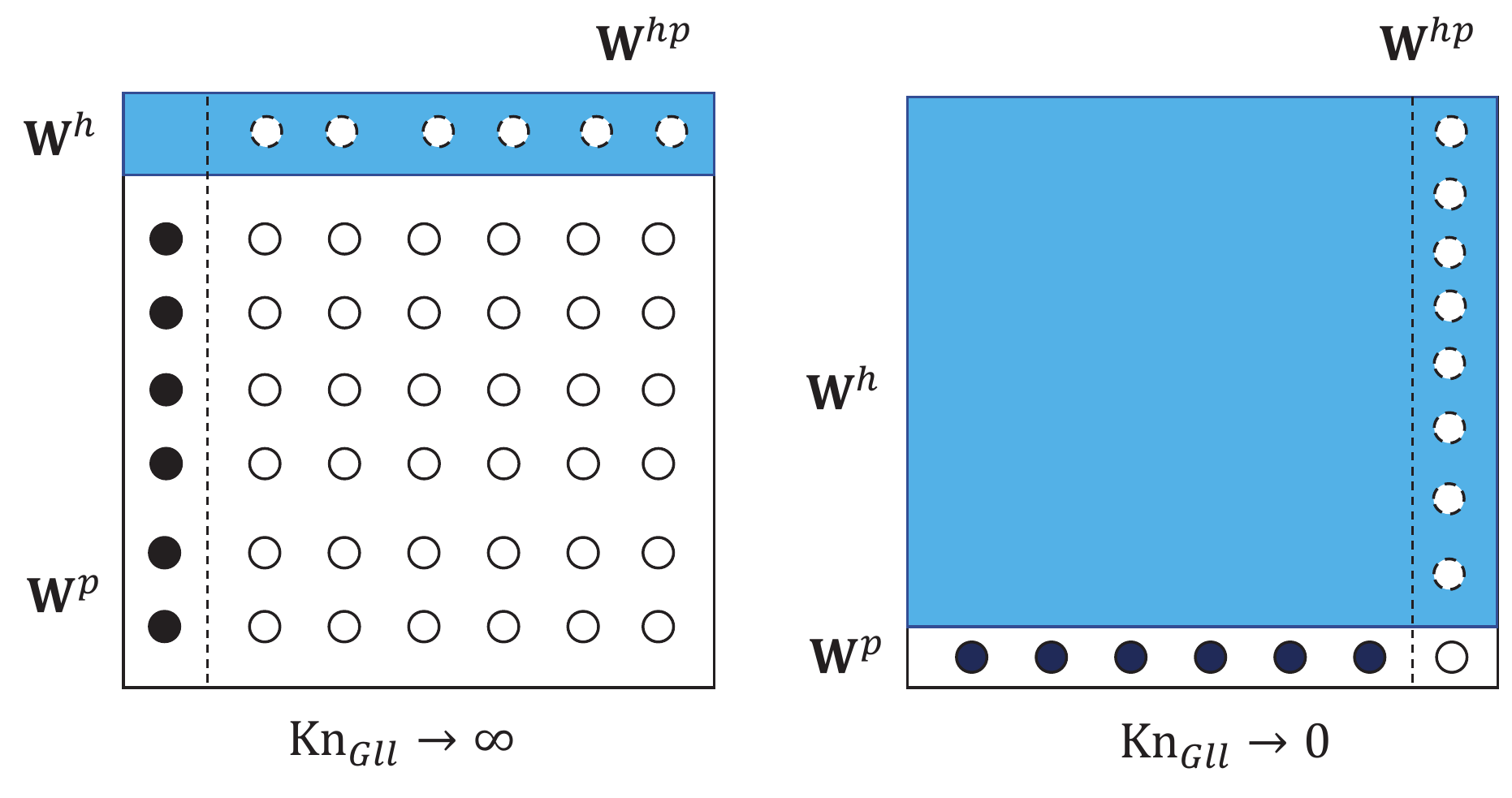}
	\caption{Illustration for the wave-particle decomposition of the AUGKWP method in  the limit of free molecular and continuum flow regimes.}
	\label{fig:augkwp}
\end{figure}

Figure~\ref{fig:augkwp} illustrates the wave-particle decomposition in the AUGKWP method in the limits of free molecular and continuum flow. Different from domain decomposition methods, the AUGKWP method employs an adaptive wave-particle formulation in each control volume with a unified treatment when the reference Knudsen number is fixed. In the AUGKWP method, the adjustment of reference Knudsen number influences the weights of wave and particles in each cell, instead of identifying different flow regimes according to this parameter in the conventional domain-decomposition approaches. In other words, the AUGKWP method has no buffer zone to distinguish and connect fluid and kinetic solvers.
Additionally, since the gradient-length local Knudsen number is not influenced by the numerical resolution in a computational domain, it only identifies the local non-equilibrium state.
The analytical wave will be mainly used in the equilibrium and near-equilibrium regimes whatever the cell resolution $(\Delta x, \Delta t)$ is adapted in AUGKWP, and the computational efficiency will be improved significantly.

\section{Numerical Validation}\label{sec:test}
In this section, the AUGKWP method is tested in many cases. Since most of the cases are external flow, the determination of the initial condition of free stream at different Knudsen number will be provided here first. For a specific gas, the density in the free stream corresponding to a given Knudsen number is
\begin{equation*}
	\rho = \frac{4\alpha(5-2\omega)(7-2\omega)}{5(\alpha+1)(\alpha+2)}\sqrt{\frac{m}{2\pi k_B T}}\frac{\mu}{L_{ref}{\rm Kn}},
\end{equation*}
where $m$ is the molecular mass and $L_{ref}$ is the reference length to define the Knudsen number. The dynamic viscosity is calculated from the translational temperature by the power law
\begin{equation*}\label{eq:pow-law}
	\mu  = \mu_{ref} \left( \frac{T}{T_{ref}} \right)^{{\omega}},
\end{equation*}
where $\mu_{ref}$ is the reference dynamic viscosity at the temperature $T_{ref}$.

In the tests, diatomic gas of nitrogen gas is employed with molecular mass $m=4.65\times 10^{-26}$ kg, $\alpha=1.0$, $\omega=0.74$, and the reference dynamic viscosity $\mu_{ref} = 1.65\times 10^{-5}$ ${\rm Nsm^{-2}}$ at the temperature $T_{ref} = 273$ K.
In the computations, the freestream or upstream values are used to non-dimensionalize the flow variables, i.e.,
\begin{equation*}
	\begin{aligned}
	\rho_0 &= \rho_\infty, \quad U_0 = \sqrt{2 k_B T_{\infty} / m}, \\
	T_0 &= T_\infty, \quad \text{or} \quad p_0 = p_\infty.
	\end{aligned}
\end{equation*}

\subsection{Sod Shock Tube Test}
The Sod shock tube problem is computed at different Knudsen numbers to verify the acceleration effect and capability for simulating the continuum and rarefied flows by the AUGKWP method. The non-dimensional initial condition is
\begin{equation*}
	(\rho, U, V, W, p)= \begin{cases}(1,0,0,0,1), & 0<x<0.5, \\ (0.125,0,0,0.1), & 0.5<x<1.\end{cases}
\end{equation*}
The spatial discretization is carried out by a one-dimensional structured mesh with $200$ uniform cells. The inlet and outlet of the tube are treated as far field. The Courant--Friedrichs--Lewy (CFL) number is taken as 0.5. The critical value for wave-particle decomposition is chosen as ${\rm Kn}_{ref} = 0.01$. The output time of the simulation is $t = 0.12$.

\begin{figure}[H]
	\centering
	\subfloat[]{\includegraphics[width=0.33\textwidth]{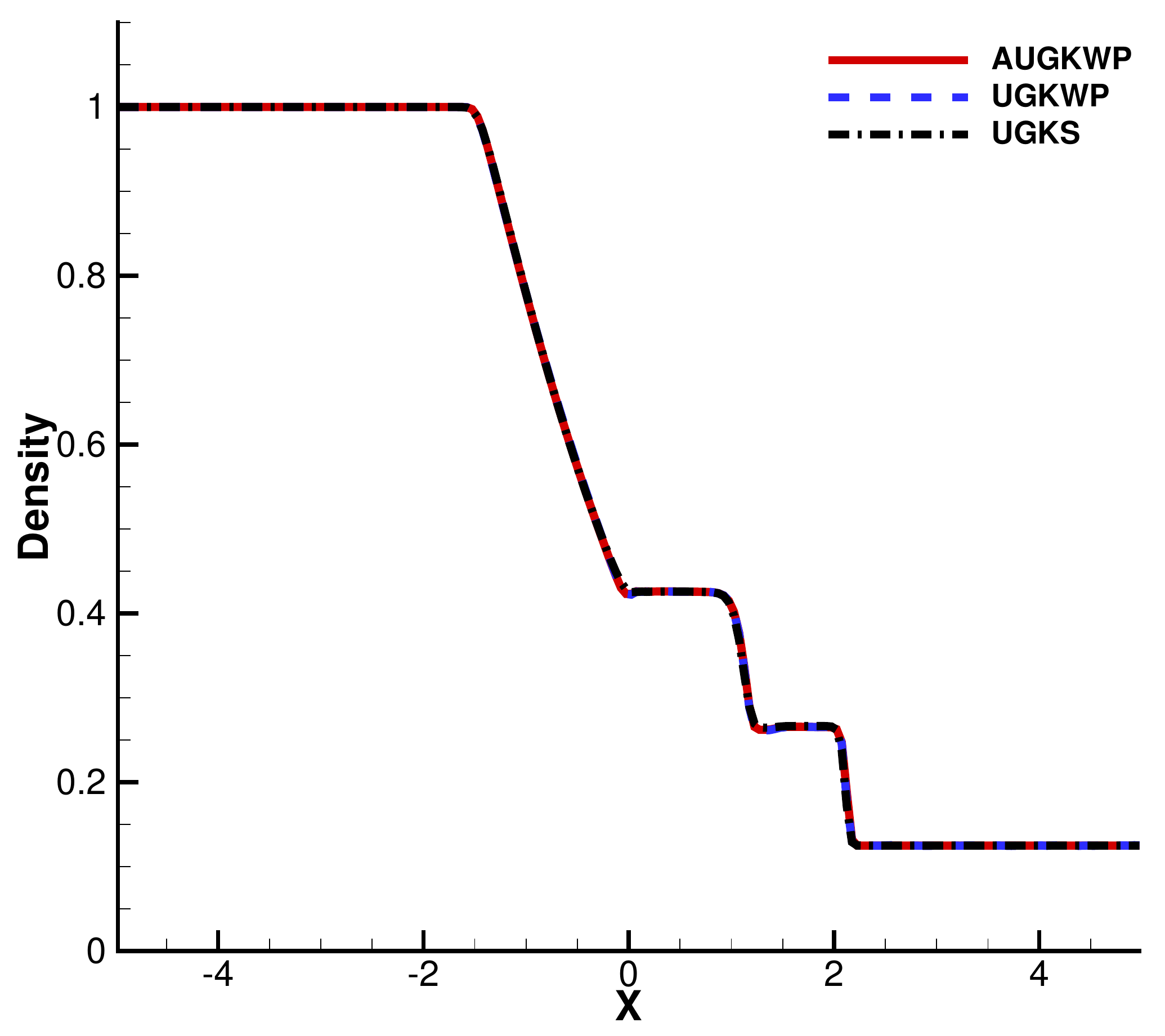}}
	\subfloat[]{\includegraphics[width=0.33\textwidth]{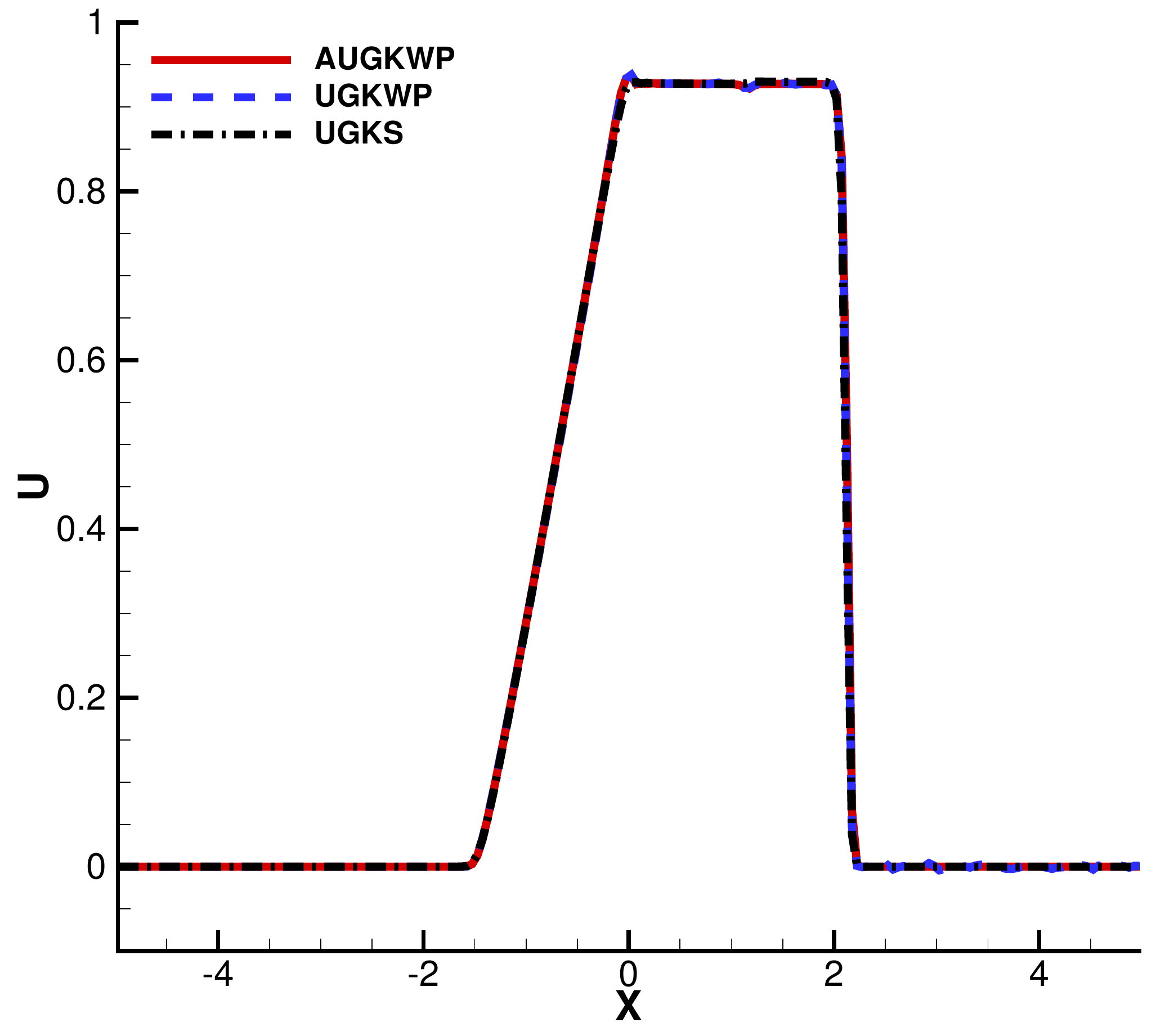}}
	\subfloat[]{\includegraphics[width=0.33\textwidth]{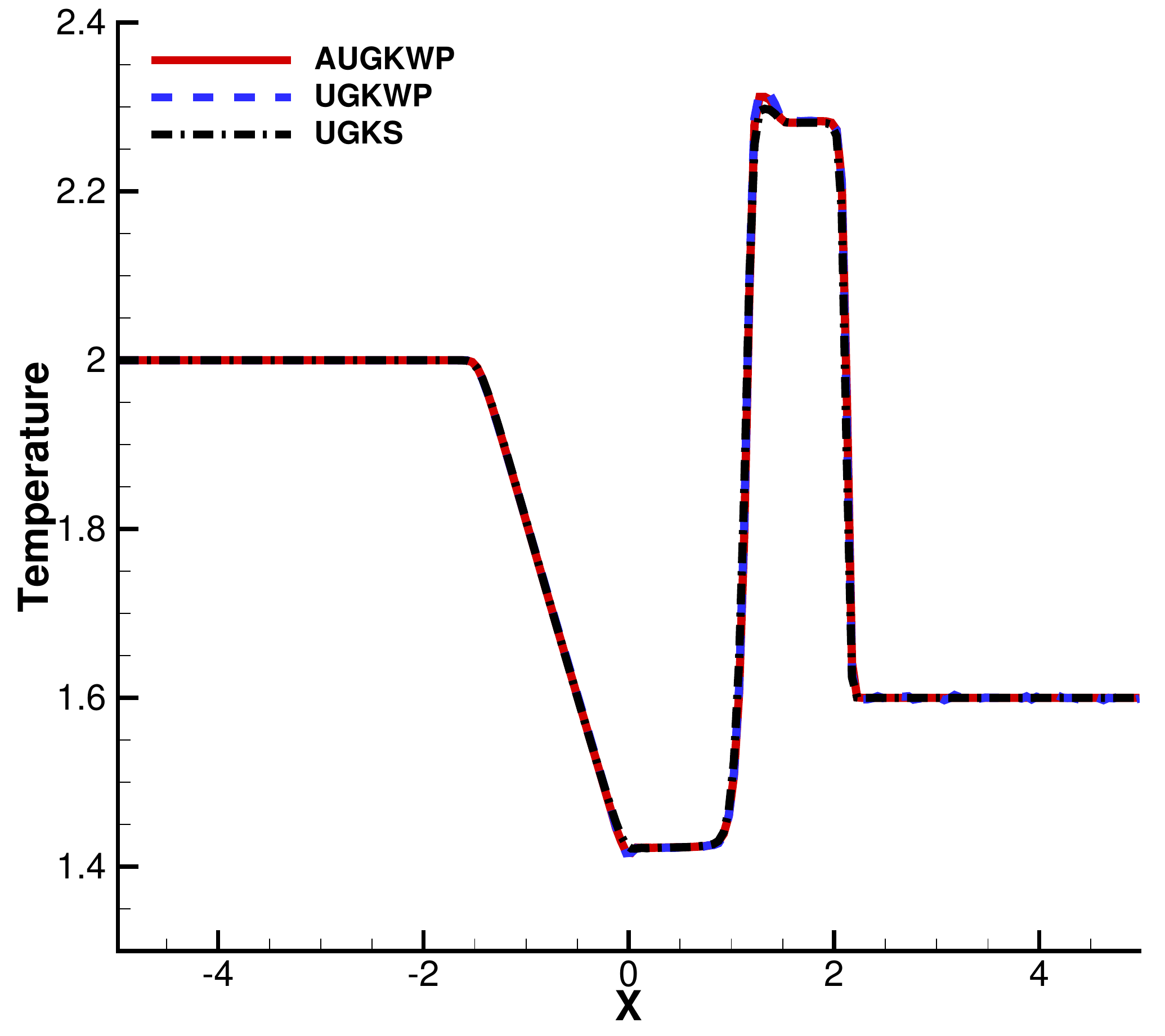}}
	\caption{Sod tube at ${\rm Kn} = 10^{-5}$. (a) Density, (b) velocity, and (c) temperature.}
	\label{fig:sod-kn1e-5}
\end{figure}

\begin{figure}[H]
	\centering
	\subfloat[]{\includegraphics[width=0.33\textwidth]{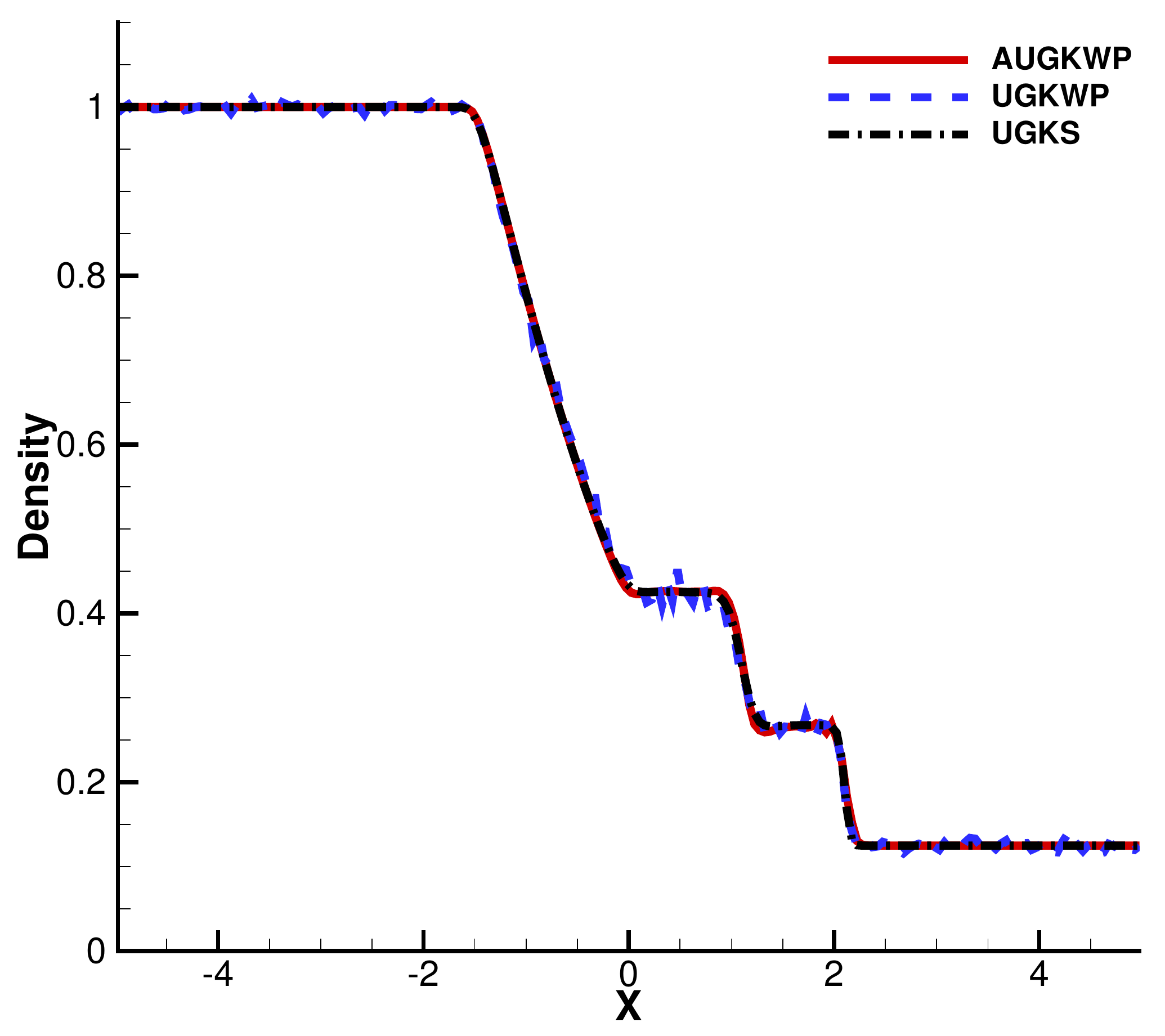}}
	\subfloat[]{\includegraphics[width=0.33\textwidth]{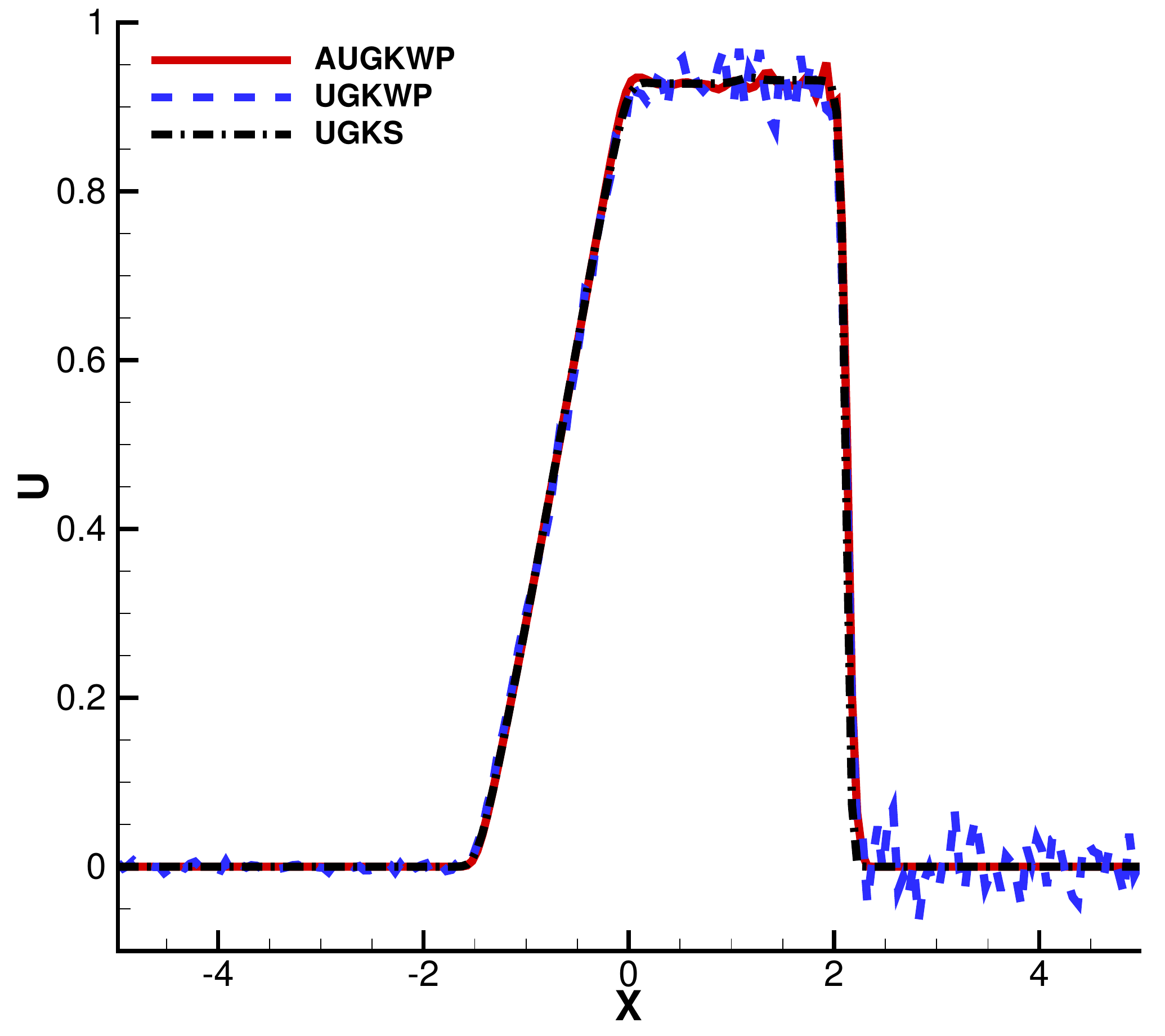}}
	\subfloat[]{\includegraphics[width=0.33\textwidth]{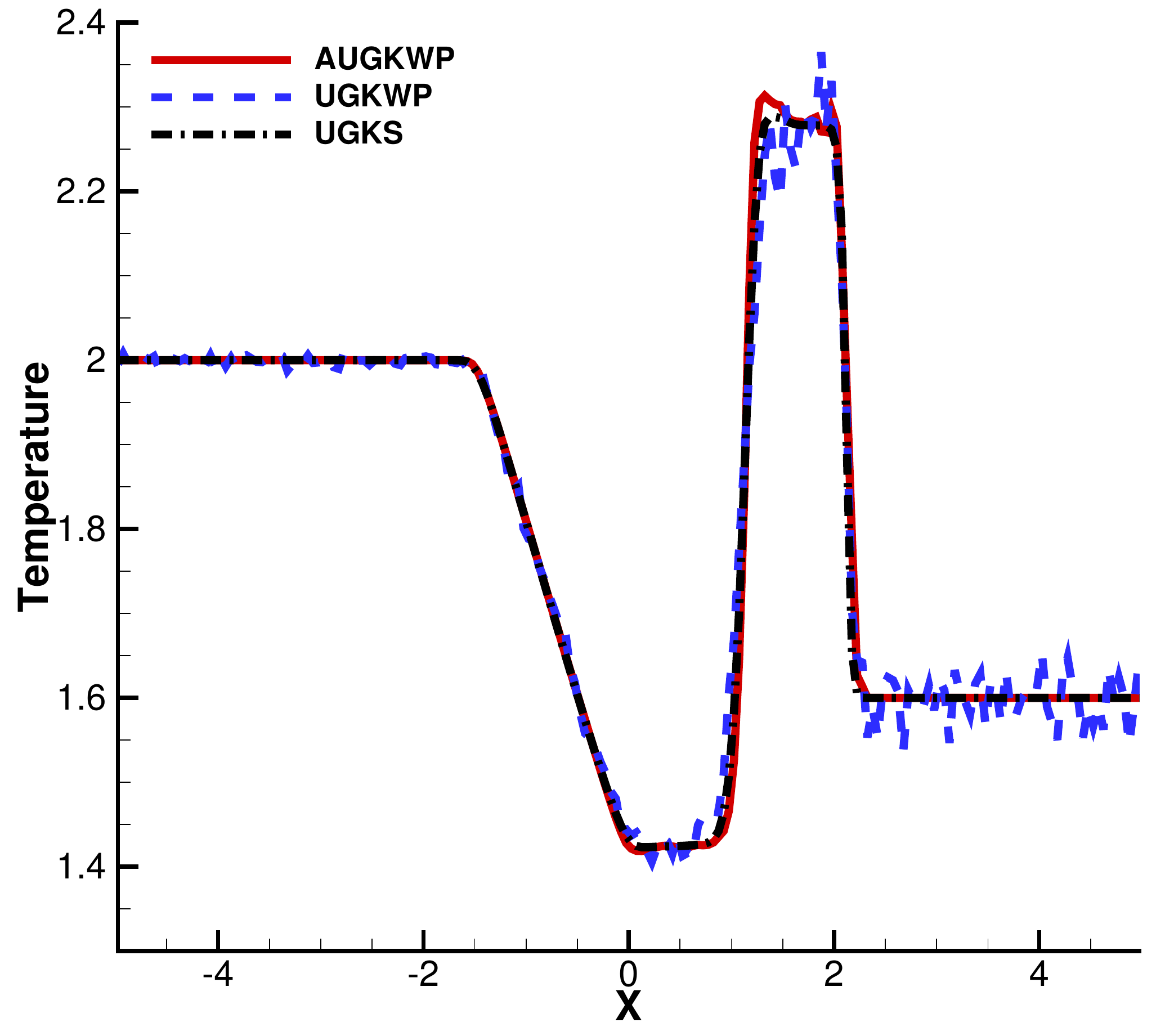}}
	\caption{Sod tube at ${\rm Kn} = 10^{-4}$. (a) Density, (b) velocity, and (c) temperature.}
	\label{fig:sod-kn1e-4}
\end{figure}

\begin{figure}[H]
	\centering
	\subfloat[]{\includegraphics[width=0.33\textwidth]{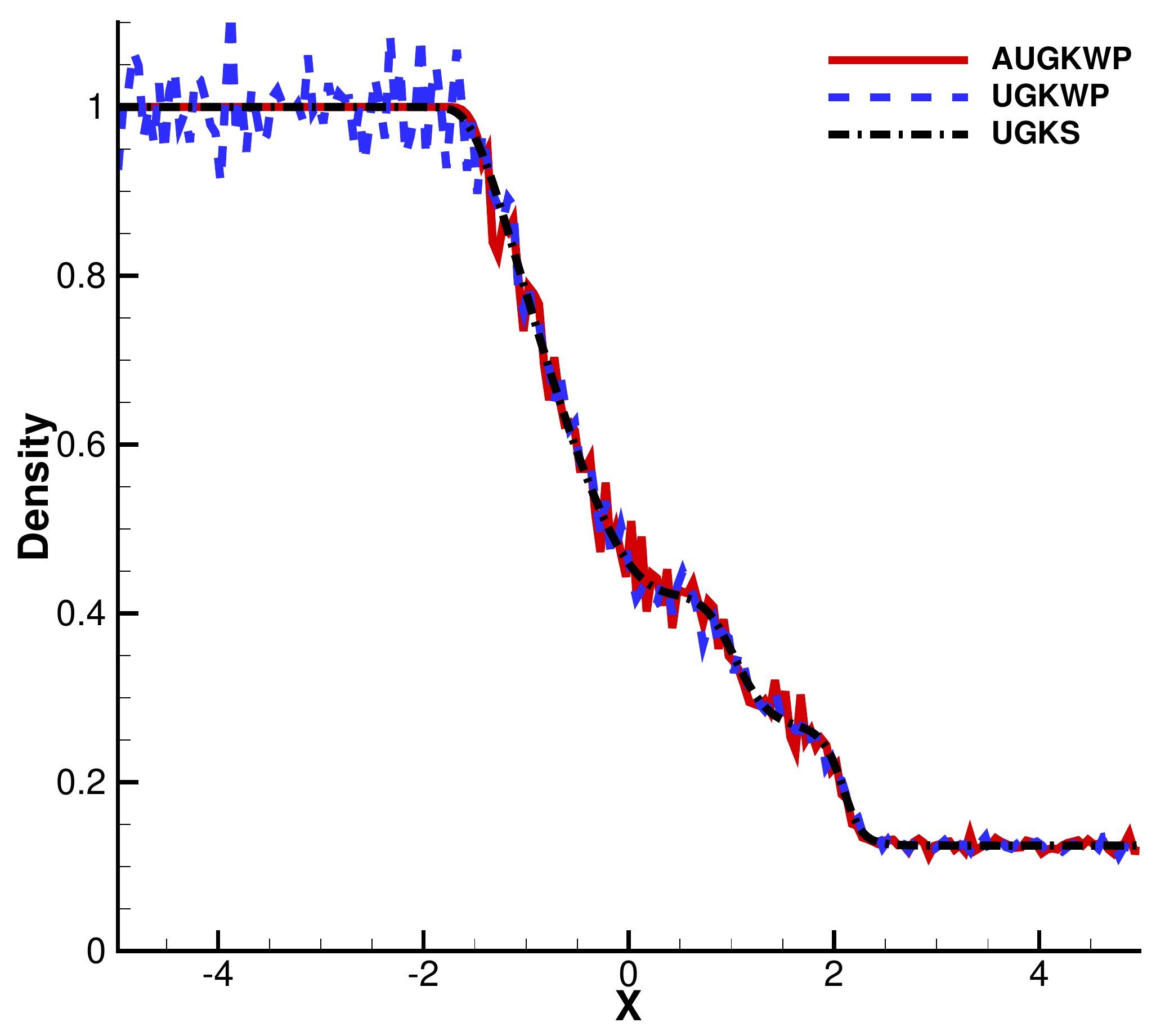}}
	\subfloat[]{\includegraphics[width=0.33\textwidth]{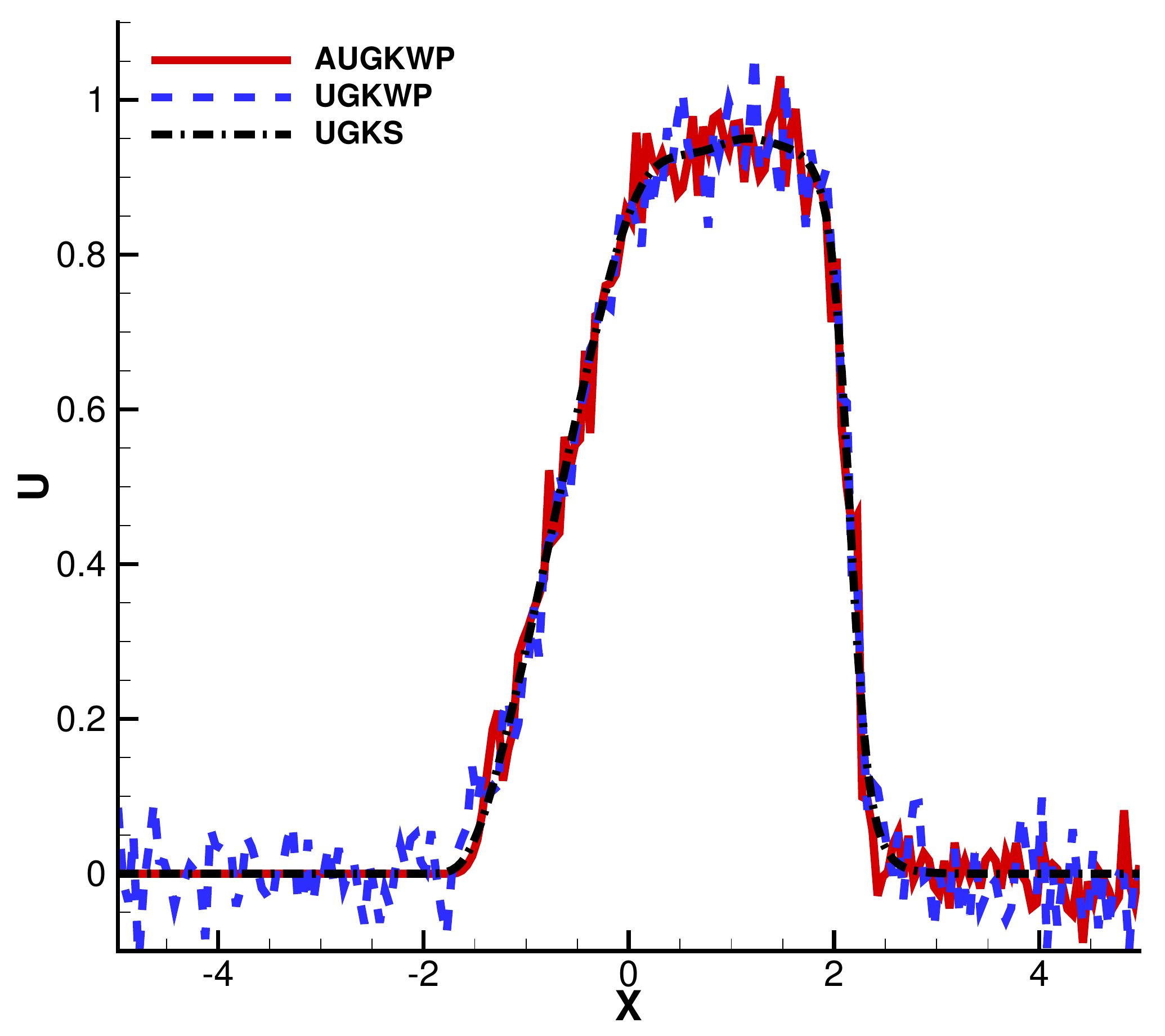}}
	\subfloat[]{\includegraphics[width=0.33\textwidth]{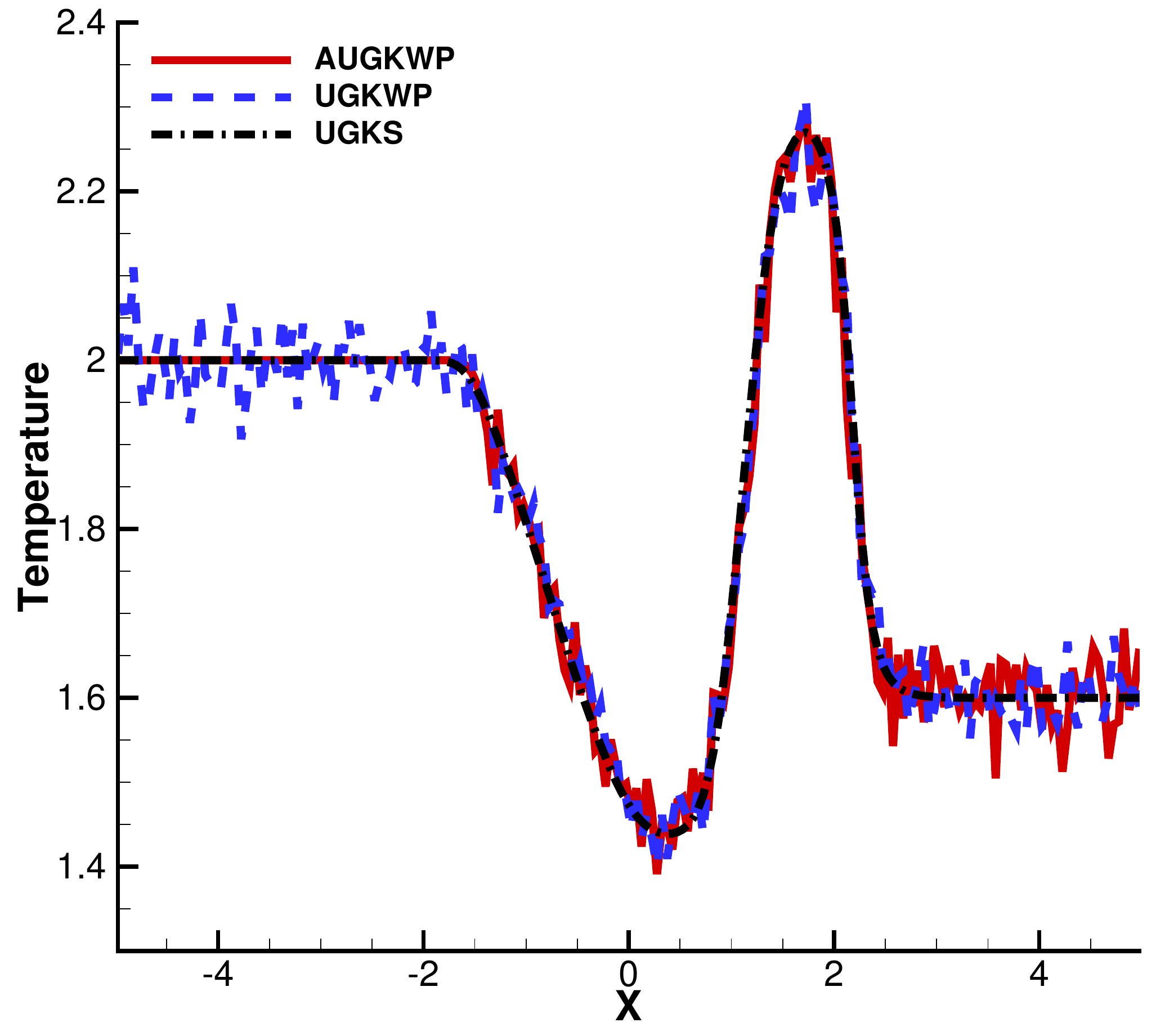}}
	\caption{Sod tube at ${\rm Kn} = 10^{-3}$. (a) Density, (b) velocity, and (c) temperature.}
	\label{fig:sod-kn1e-3}
\end{figure}

\begin{figure}[H]
	\centering
	\subfloat[]{\includegraphics[width=0.33\textwidth]{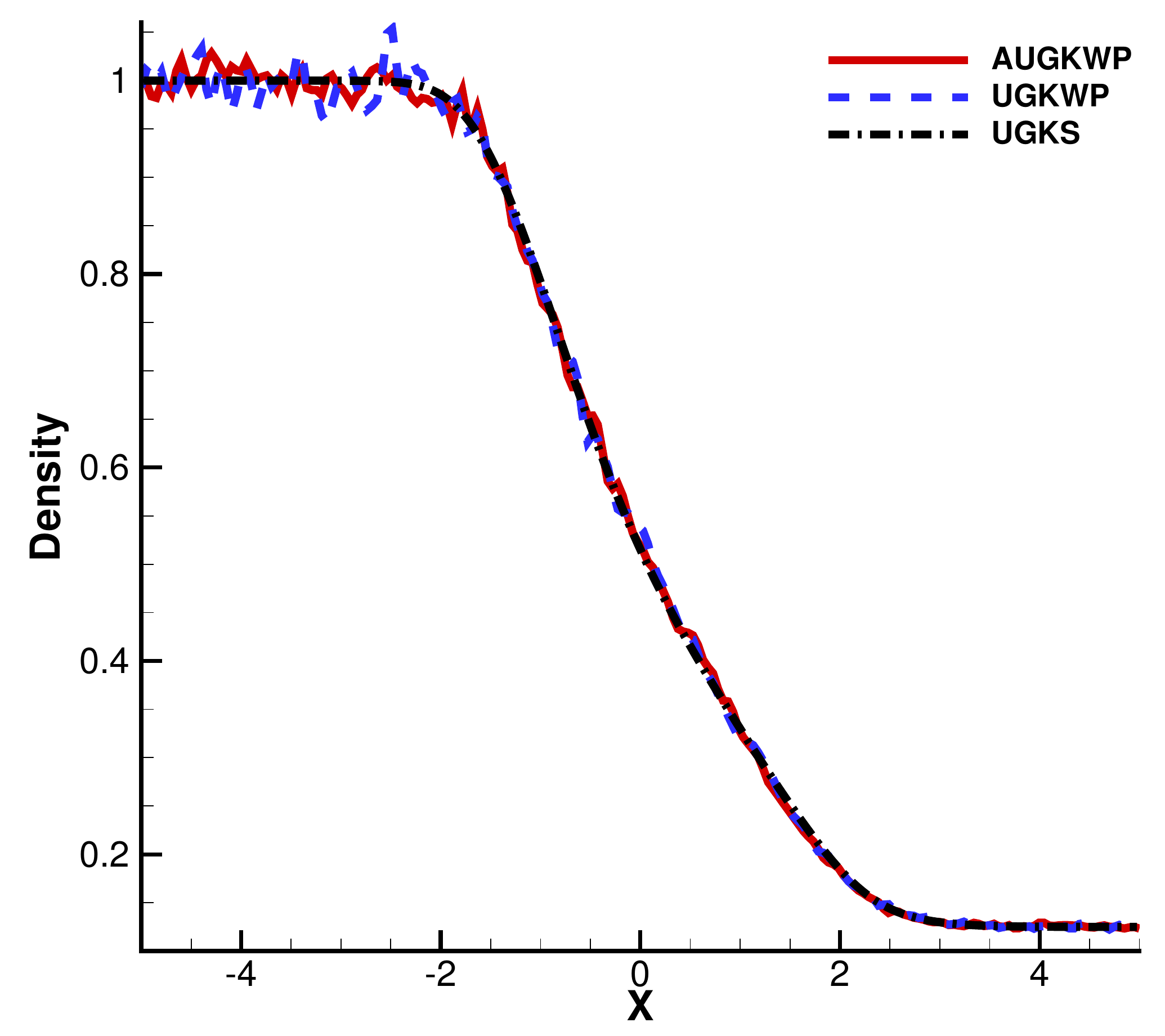}}
	\subfloat[]{\includegraphics[width=0.33\textwidth]{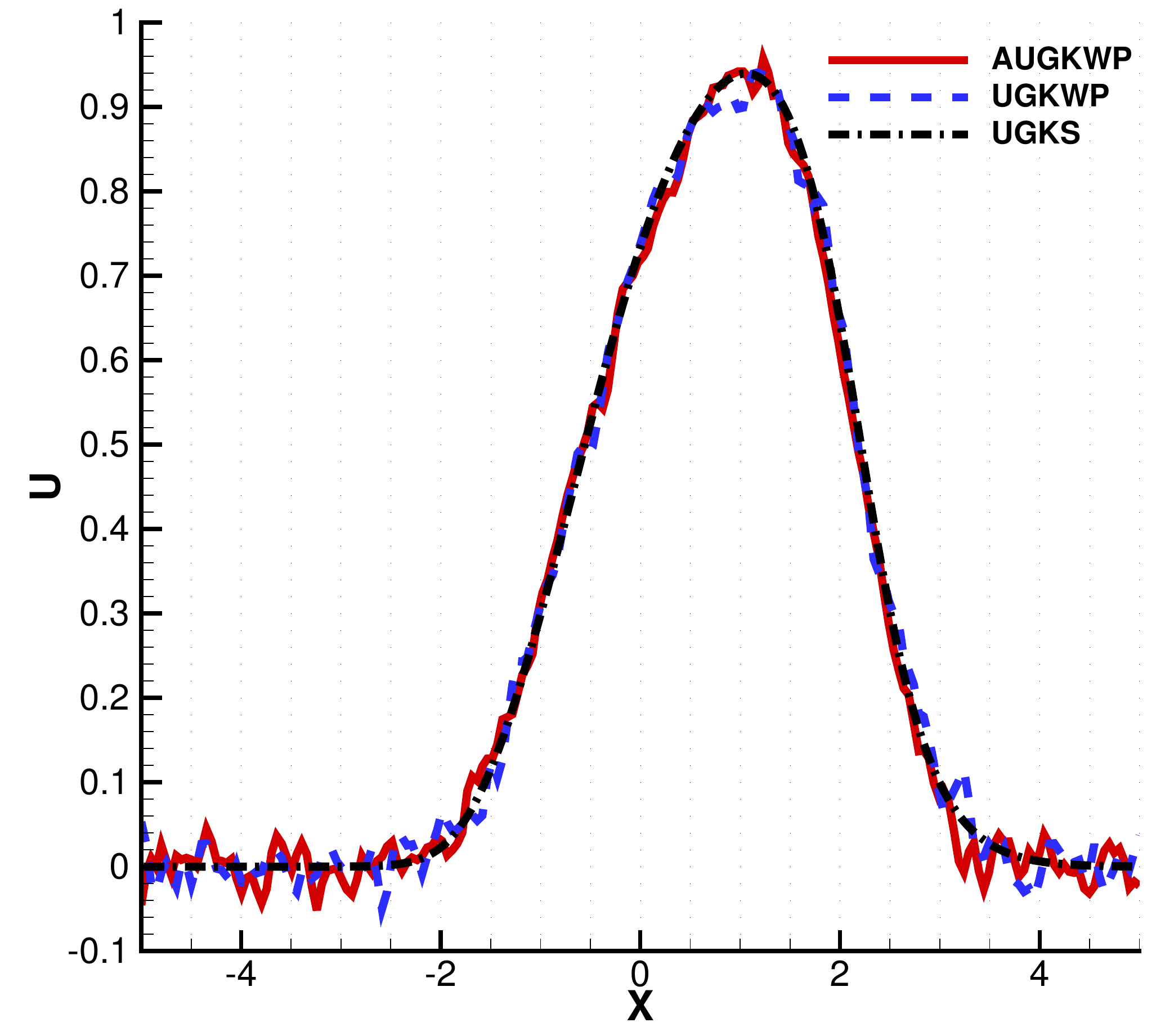}}
	\subfloat[]{\includegraphics[width=0.33\textwidth]{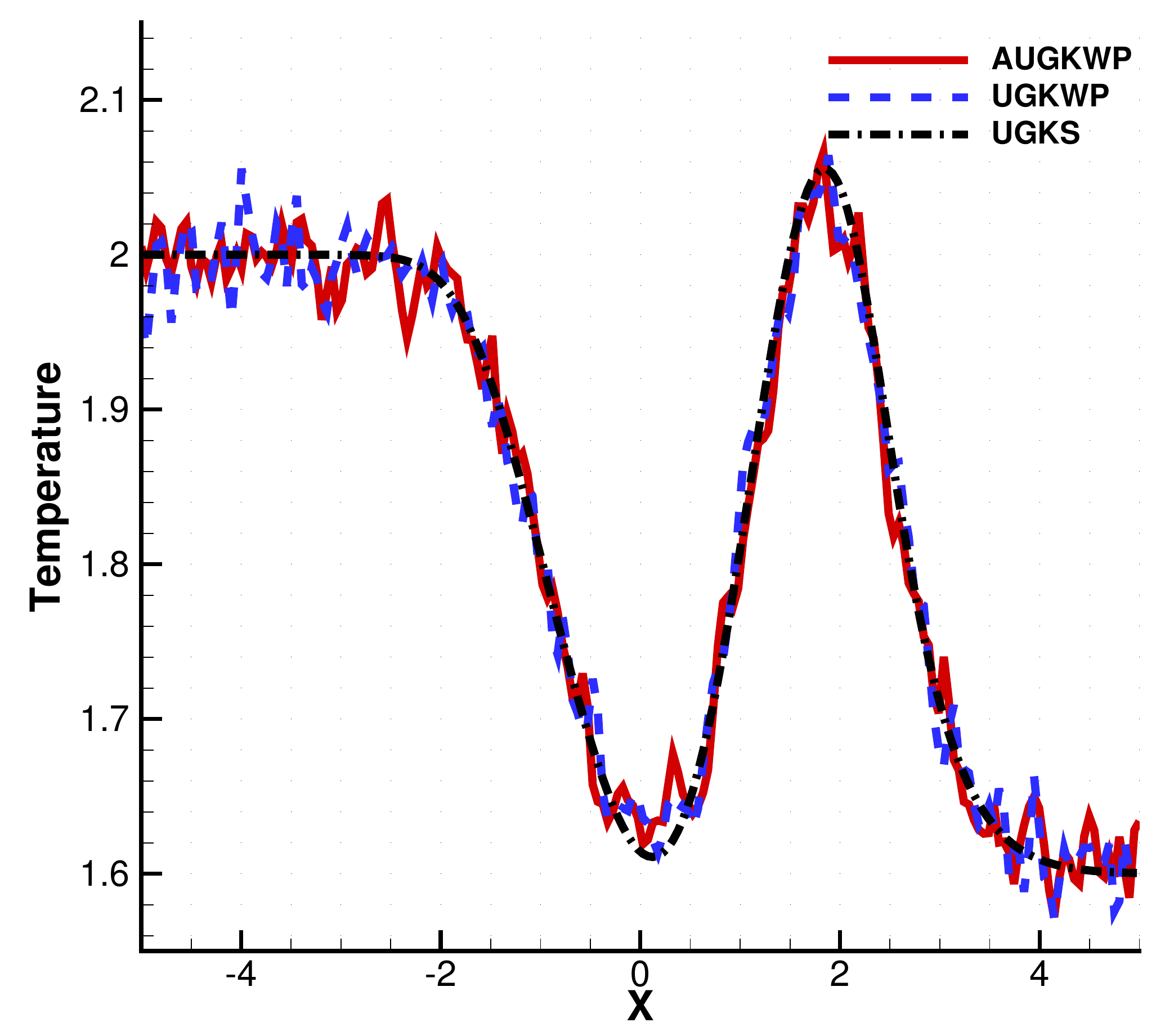}}
	\caption{Sod tube at ${\rm Kn} = 10^{-2}$. (a) Density, (b) velocity, and (c) temperature.}
	\label{fig:sod-kn1e-2}
\end{figure}

\begin{figure}[H]
	\centering
	\subfloat[]{\includegraphics[width=0.33\textwidth]{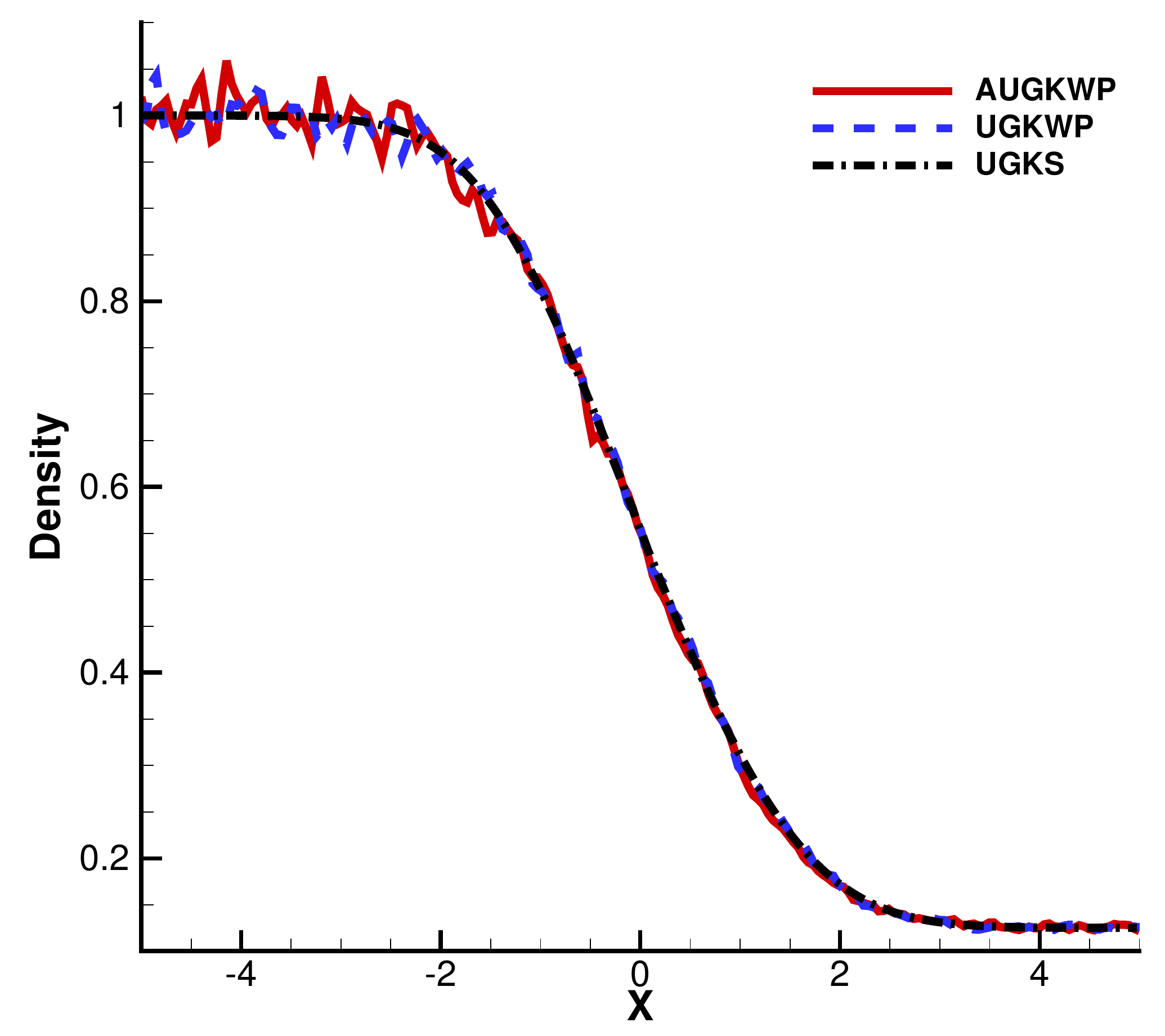}}
	\subfloat[]{\includegraphics[width=0.33\textwidth]{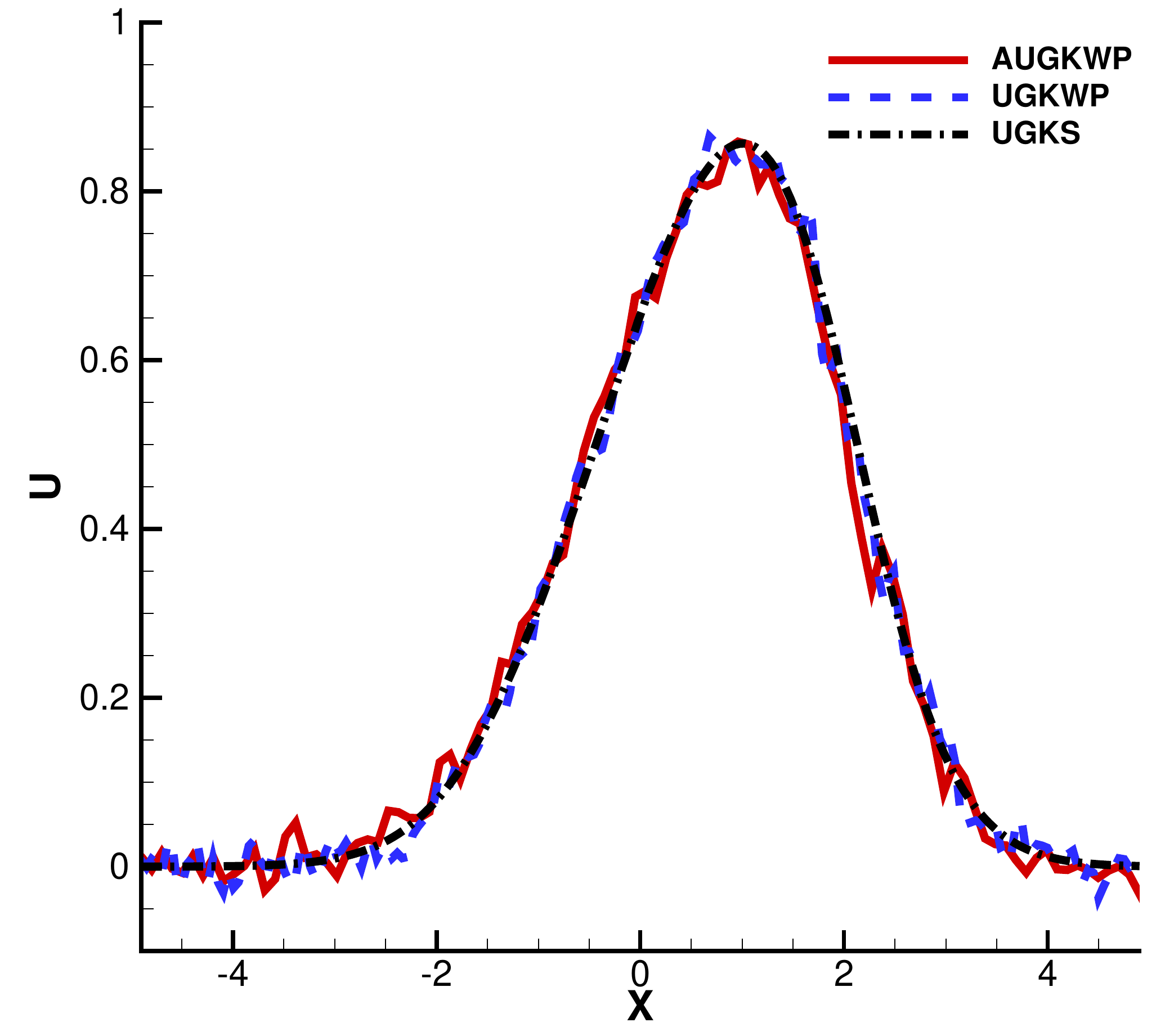}}
	\subfloat[]{\includegraphics[width=0.33\textwidth]{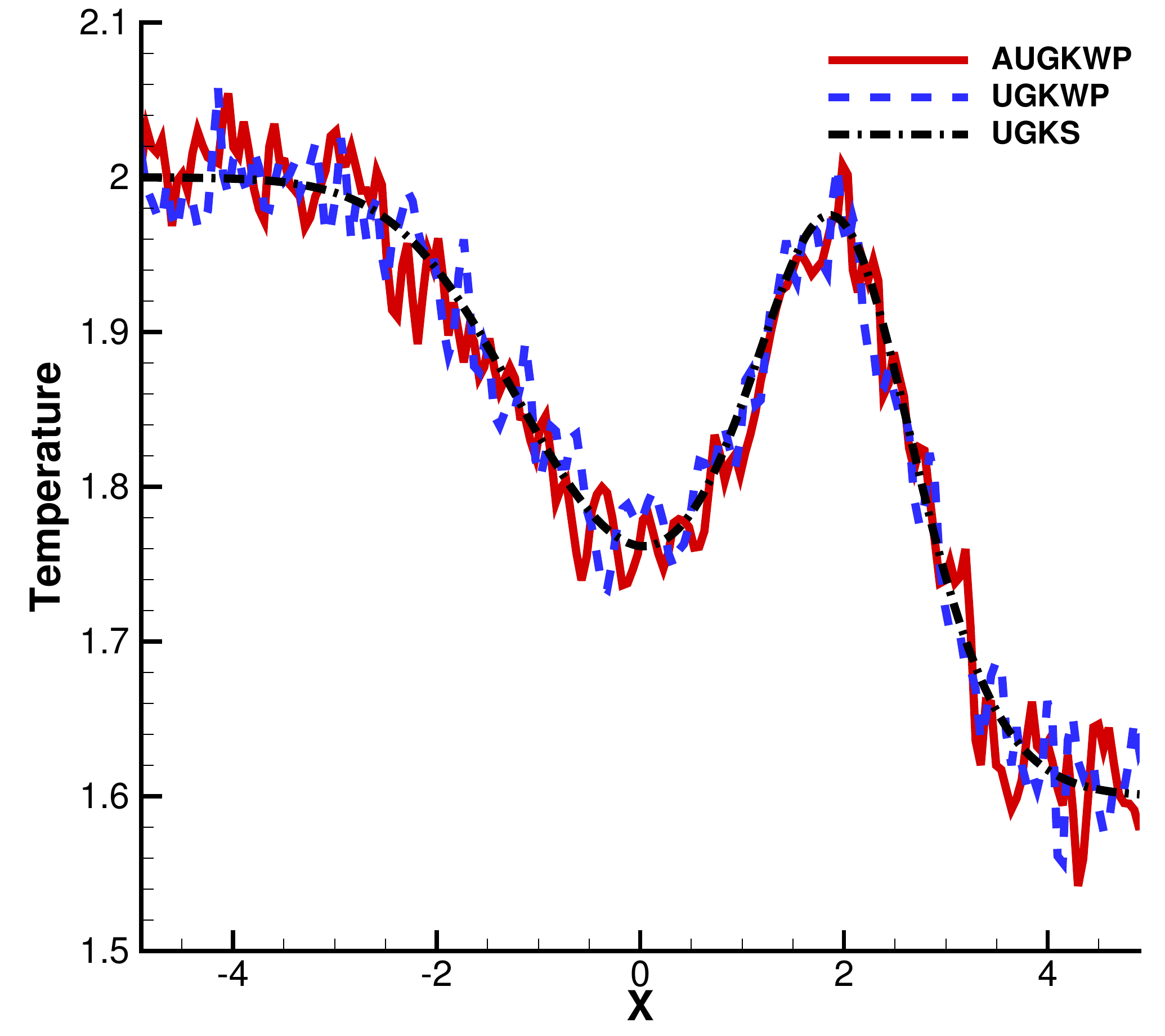}}
	\caption{Sod tube at ${\rm Kn} = 10^{-1}$. (a) Density, (b) velocity, and (c) temperature.}
	\label{fig:sod-kn1e-1}
\end{figure}

\begin{figure}[H]
	\centering
	\subfloat[]{\includegraphics[width=0.33\textwidth]{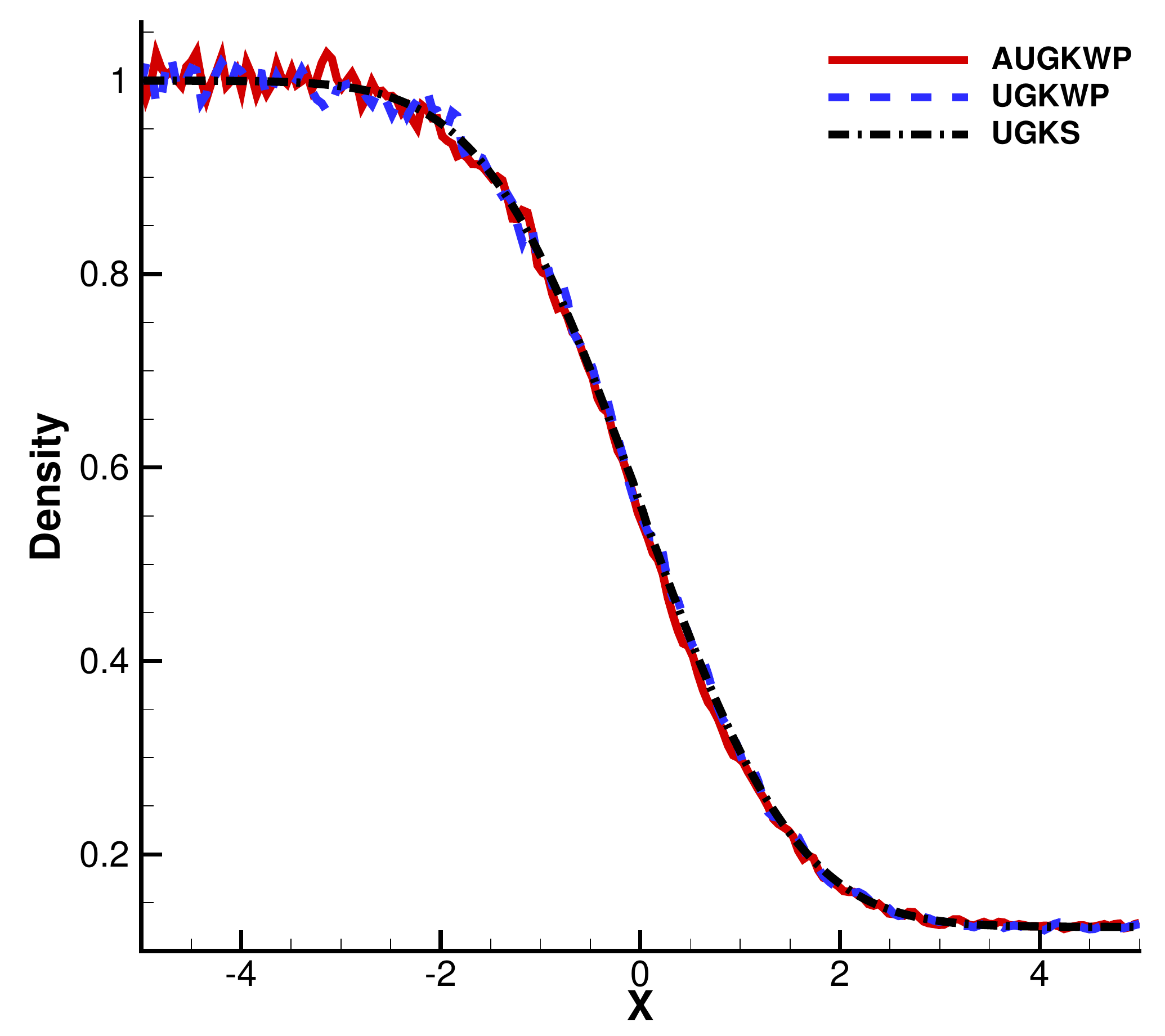}}
	\subfloat[]{\includegraphics[width=0.33\textwidth]{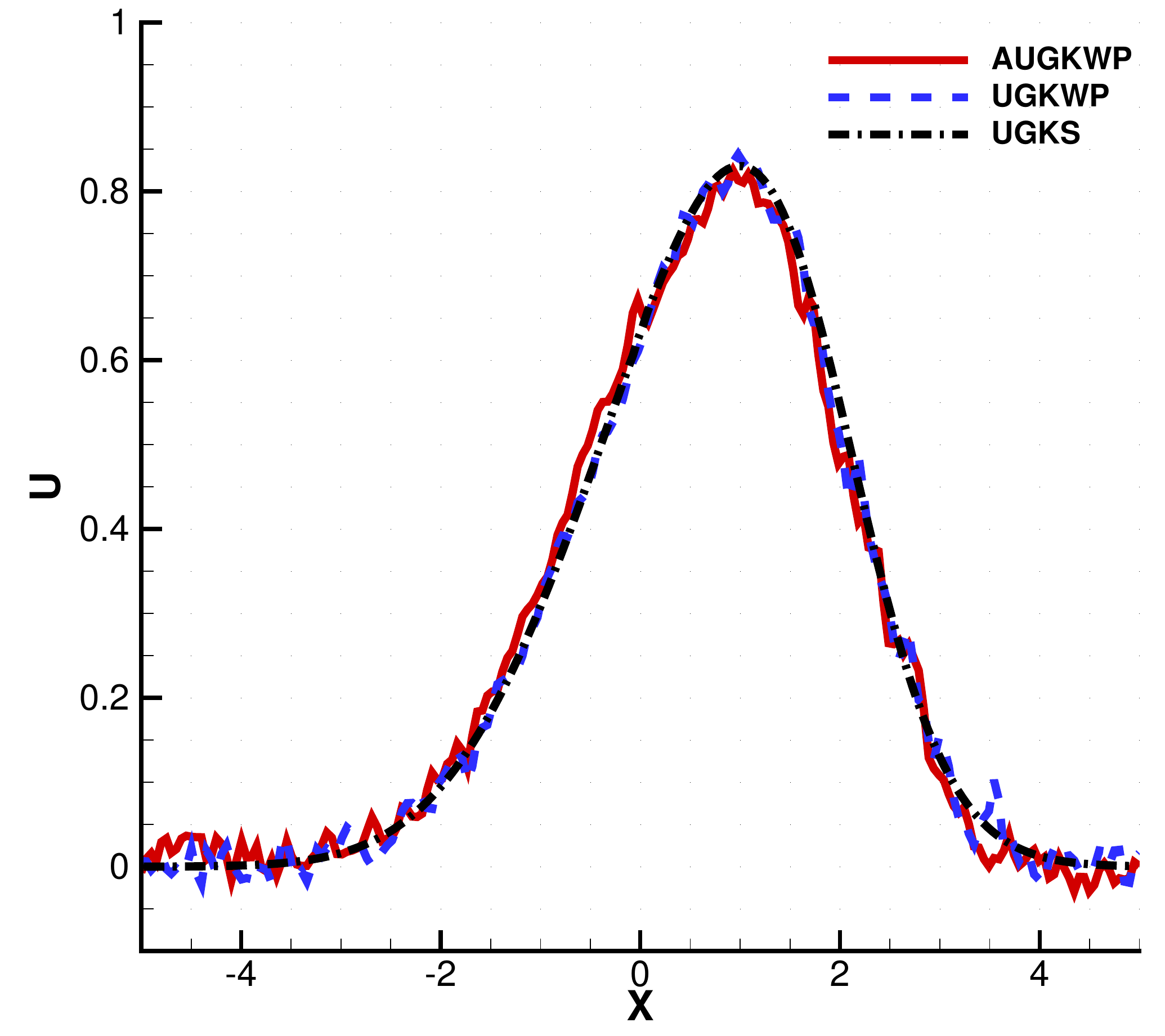}}
	\subfloat[]{\includegraphics[width=0.33\textwidth]{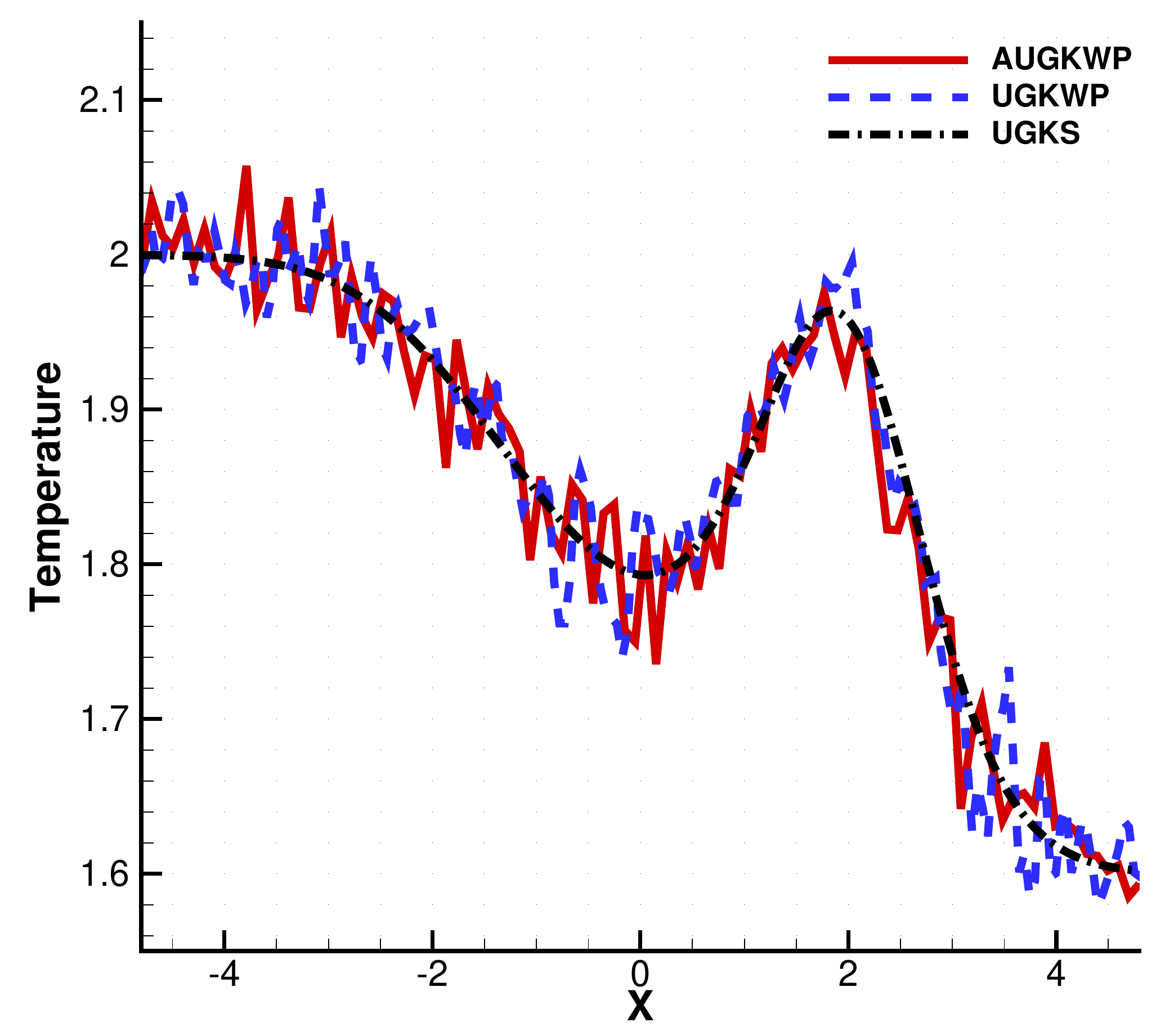}}
	\caption{Sod tube at ${\rm Kn} = 1$. (a) Density, (b) velocity, and (c) temperature.}
	\label{fig:sod-kn1}
\end{figure}

\begin{figure}[H]
	\centering
	\subfloat[]{\includegraphics[width=0.33\textwidth]{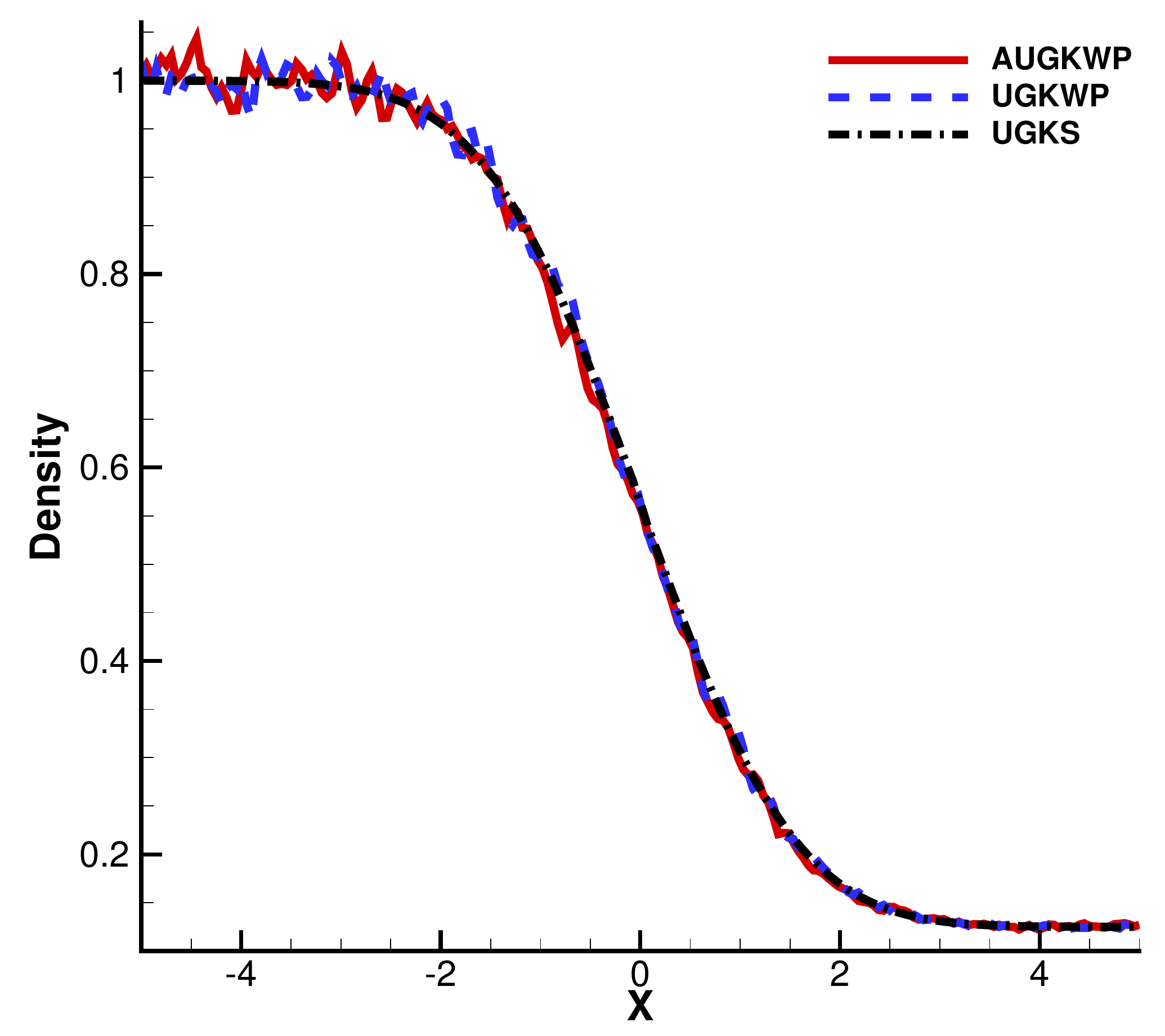}}
	\subfloat[]{\includegraphics[width=0.33\textwidth]{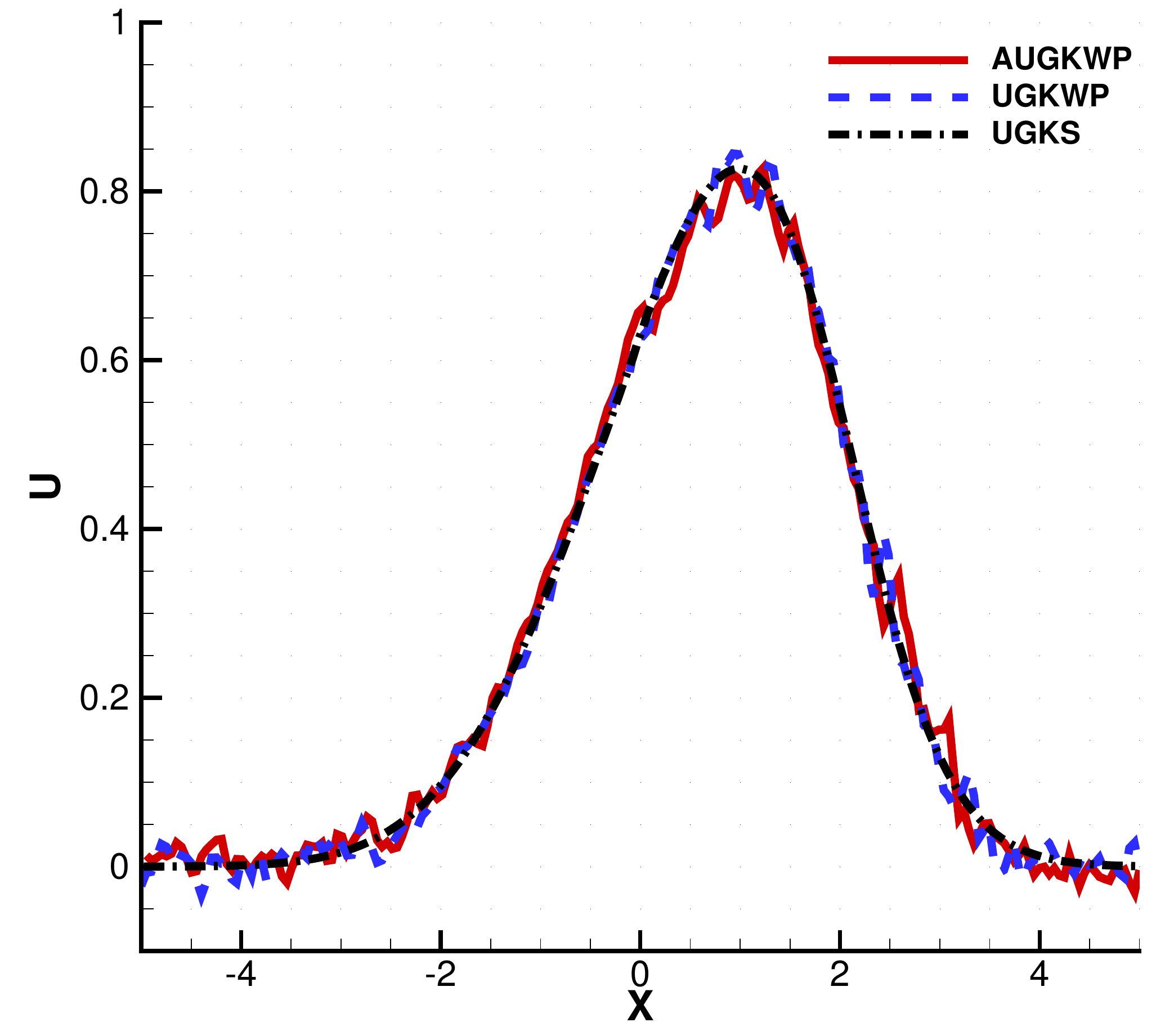}}
	\subfloat[]{\includegraphics[width=0.33\textwidth]{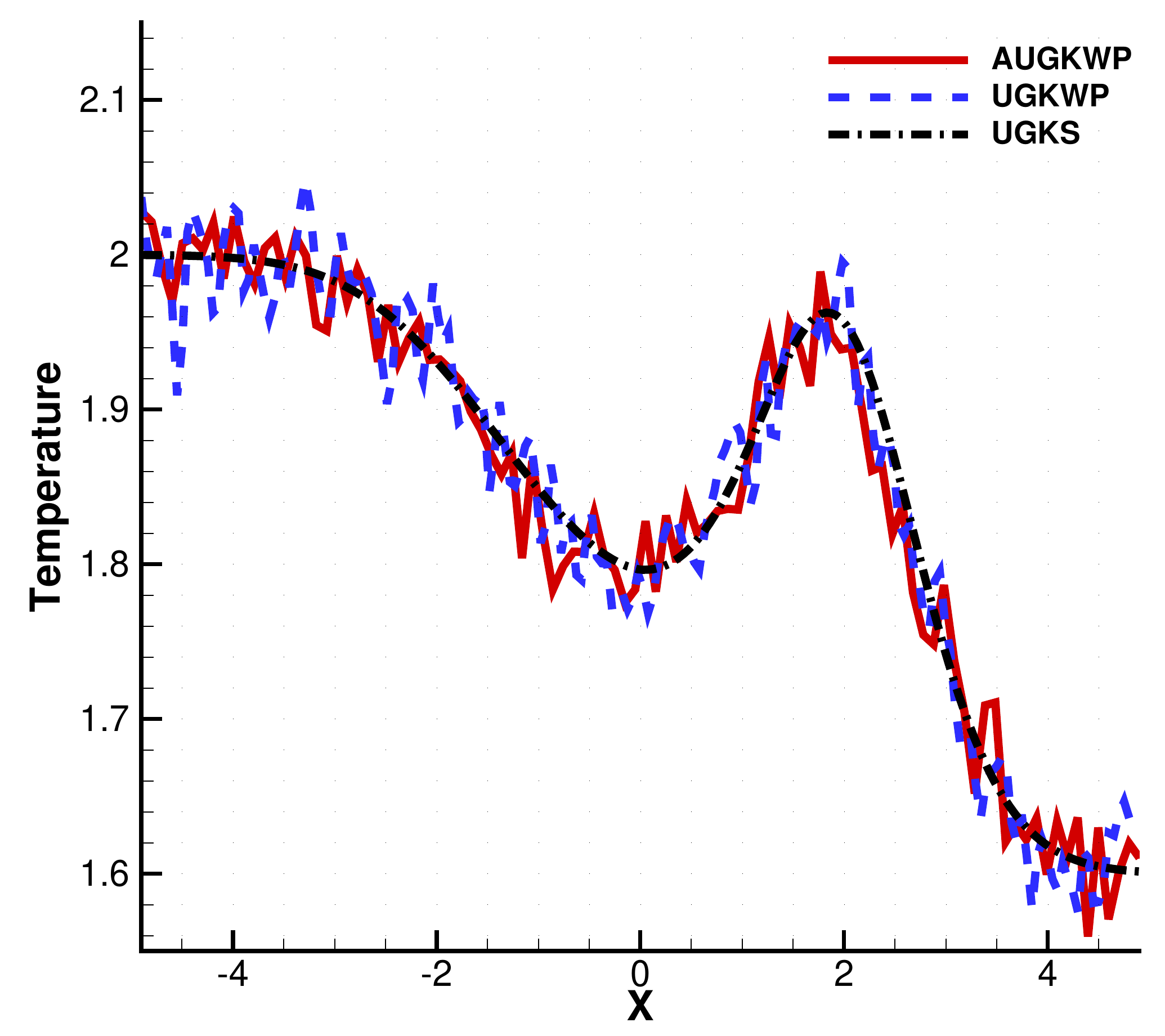}}
	\caption{Sod tube at ${\rm Kn} = 10$. (a) Density, (b) velocity, and (c) temperature.}
	\label{fig:sod-kn10}
\end{figure}

The density, velocity and temperature obtained by the original UGKWP method, the AUGKWP method, and UGKS at different Knudsen numbers from $10^{-5}$ to $10$ are plotted in Fig.~\ref{fig:sod-kn1e-5}--\ref{fig:sod-kn10}. The preset reference number of particles in both original UGKWP and AUGKWP methods are $400$ per cell. Here, the purpose of setting this number of particles in computation is to show the noise introduced in the AUGKWP method is acceptable compared with the original UGKWP method for unsteady flow when particles are not sufficiently enough. The results show the AUGKWP method can maintain the same solutions as the original UGKWP method and match with the UGKS solutions in all Knudsen regimes.


\begin{figure}[H]
	\centering
	\includegraphics[width=0.5\textwidth]{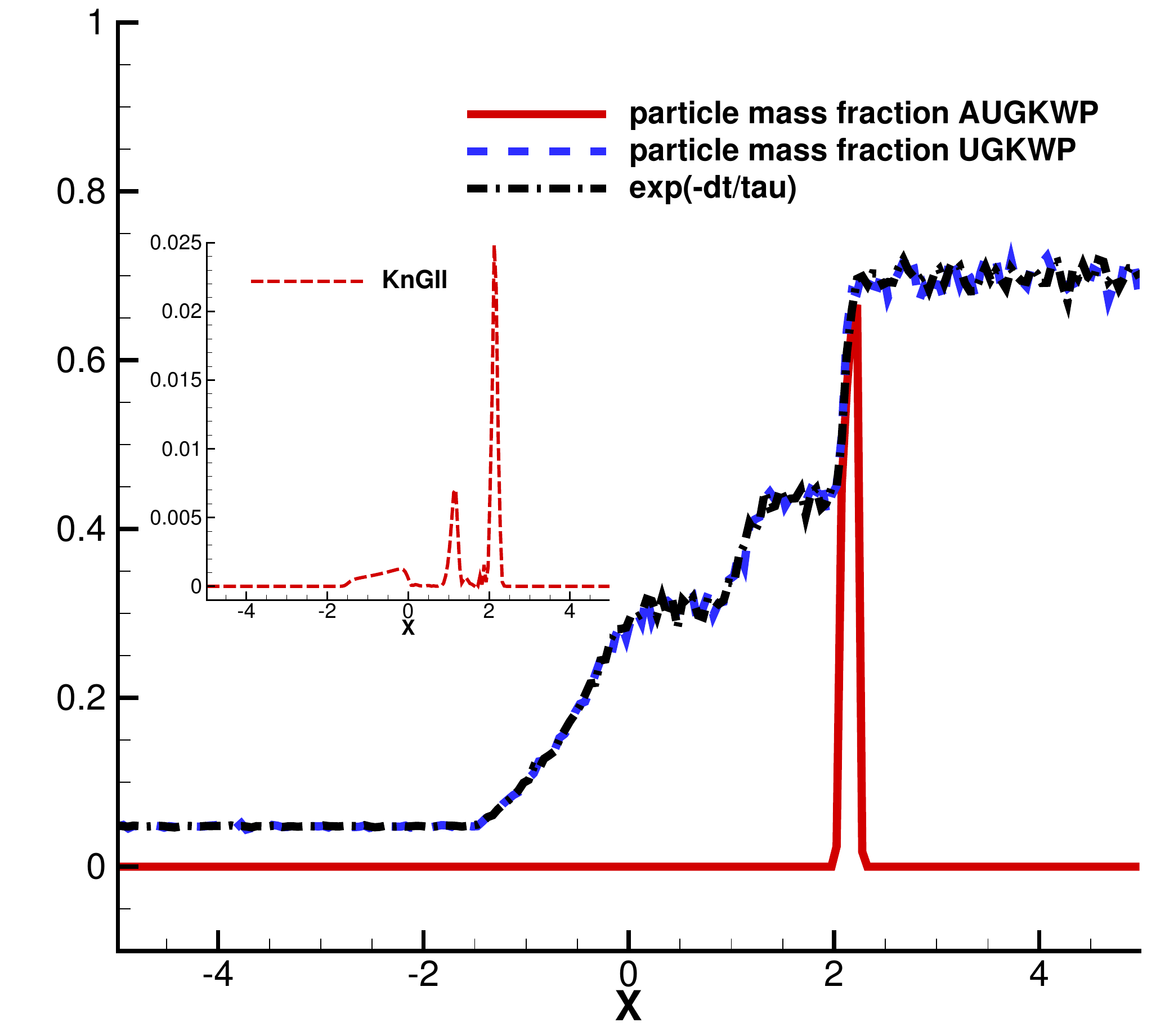}
	\caption{Particle mass fraction of the original UGKWP method and the AUGKWP method, the exponential function of cell's Knudsen number $e^{-\Delta t/\tau}$, and the gradient-length local Knudsen number ${\rm Kn}_{Gll}$ for Sod tube at ${\rm Kn}= 10^{-4}$.}
	\label{fig:sod-eta}
\end{figure}

In these tests, a small time step determined by high spatial resolution leads to a large cell's Knudsen number ${\rm Kn}_c = \tau / \Delta t$. For the original UGKWP method, particles are sampled according to $e^{-\Delta t/ \tau}$ for the free transport particle with $t_f = \Delta t$ even in the uniform equilibrium regions.
For the AUGKWP method, the analytical wave formulation will be used in the equilibrium region.
Particle mass fraction given by the original UGKWP method and the AUGKWP method, the exponential function of cell's Knudsen number $e^{-\Delta t/\tau}$, and the gradient-length local Knudsen number ${\rm Kn}_{Gll}$ for Sod tube at ${\rm Kn}= 10^{-4}$ are plotted in Fig.~\ref{fig:sod-eta}. It shows the particles appear in the original UGKWP method even though the flow regime is continuum, while for the AUGKWP method, with the consideration of local Knudsen number, particles are sampled in the non-equilibrium region only, such as the
shock front region with ${\rm Kn}_{Gll} > 0.01$.
Therefore, a high computational efficiency is achieved in AUGKWP.
Tab.~\ref{tab:sod-time} is the comparison of the simulation times between the original UGKWP and AUGKWP methods at different Knudsen numbers. It shows AUGKWP method fully recovers the hydrodynamical solver without using particles in the equilibrium flow regime regardless the cell resolution used.
\begin{table}[H]
	\centering
	\caption{Comparison of simulation time between the original UGKWP and AUGKWP method.}
	\begin{tabular}{ccc}
	\hline
	Knudsen number & Original UGKWP (s) & AUGKWP (s)  \\ 	  \hline
	10             & 77.04              & 70.75   \\
	1	           & 78.76		        & 75.05   \\
	$10^{-1}$	   & 82.06		        & 76.37   \\
	$10^{-2}$      & 98.36		        & 81.12   \\
	$10^{-3}$      & 124.28		        & 42.89   \\
	$10^{-4}$      & 164.33		        & 3.45    \\
	$10^{-5}$      & 181.19		        & 0.41    \\ 	  \hline
	\end{tabular}
	\label{tab:sod-time}
\end{table}

\subsection{Shock Structure}
To validate the capability of AUGKWP method for describing strong non-equilibrium state, shock structure at upstream Mach number ${\rm Ma} = 4$ and 10 is investigated. The computational domain $[-25, 25]$ has a length of $50$ times of the particle mean free path and is divided by $100$ cells uniformly.
The left and right boundaries are treated as far field condition. The CFL number takes $0.5$.
The reference Knudsen number is ${\rm Kn}_{ref} = 0.001$. In this study, the upstream temperature is $T_1 = 50$ K. The rest parameters could be obtained from the non-dimensional initial condition.

In kinetic theory, the particle collision time depends on the particle velocity. In order to cope with this physical reality, the relaxation time of the high-speed particles is amended  in the UGKWP and AUGKWP methods \cite{xu2021modeling}
\begin{equation*}\label{eq:tauStar}
	\tau^*= \begin{cases}\tau, & \text { if }|\vec{u}-\vec{U}| \leq b \sqrt{RT}, \\
		    \frac{1}{ 1 + a^*|\vec{u}-\vec{U}| / \sqrt{RT} } \tau, & \text { if }|\vec{u}-\vec{U}|>b \sqrt{RT},\end{cases}
\end{equation*}
with two parameters $a= 0.1$ and $b=5$.

\begin{figure}[H]
	\centering
	\subfloat[]{\includegraphics[width=0.4\textwidth]{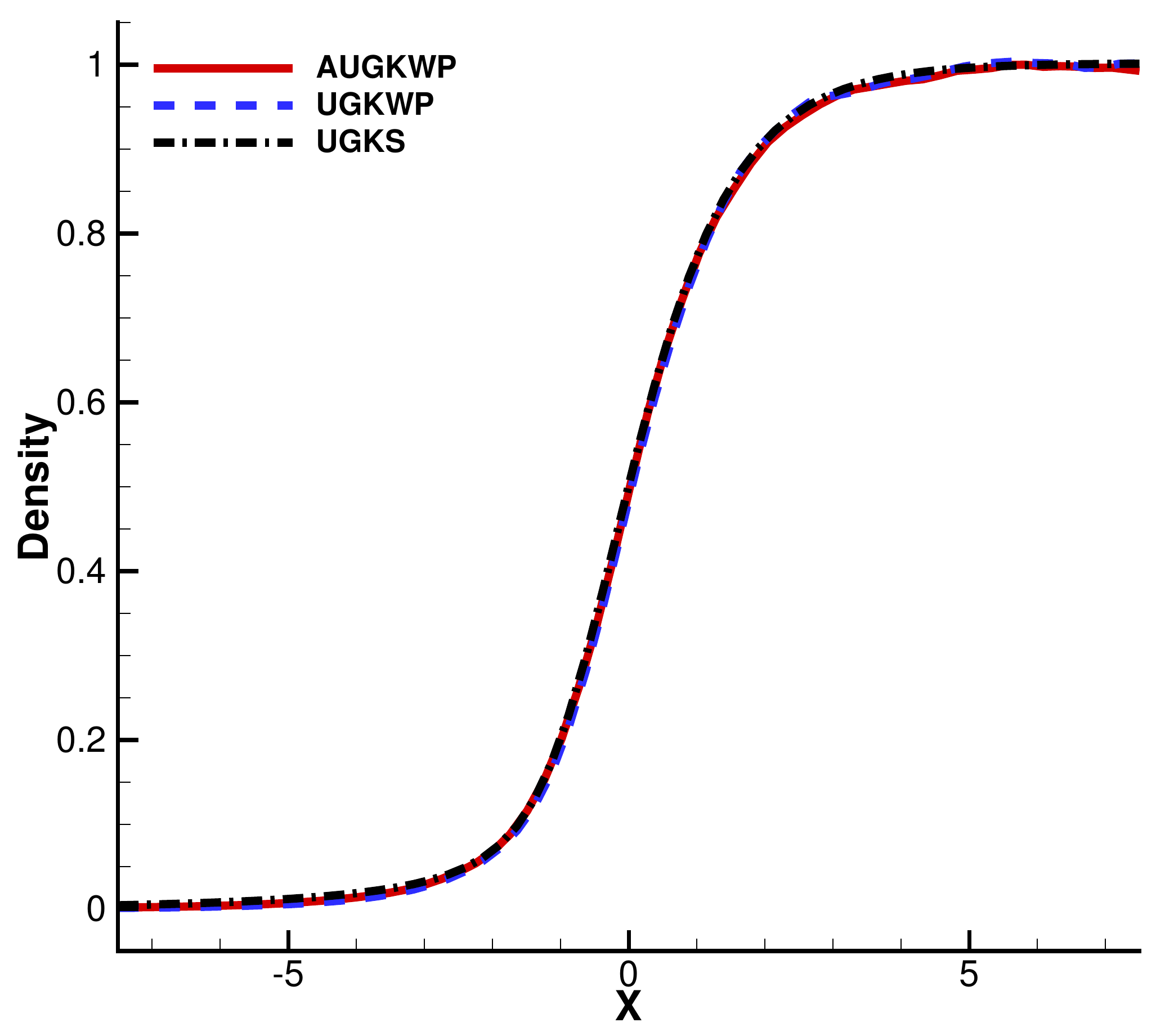}}
	\subfloat[]{\includegraphics[width=0.4\textwidth]{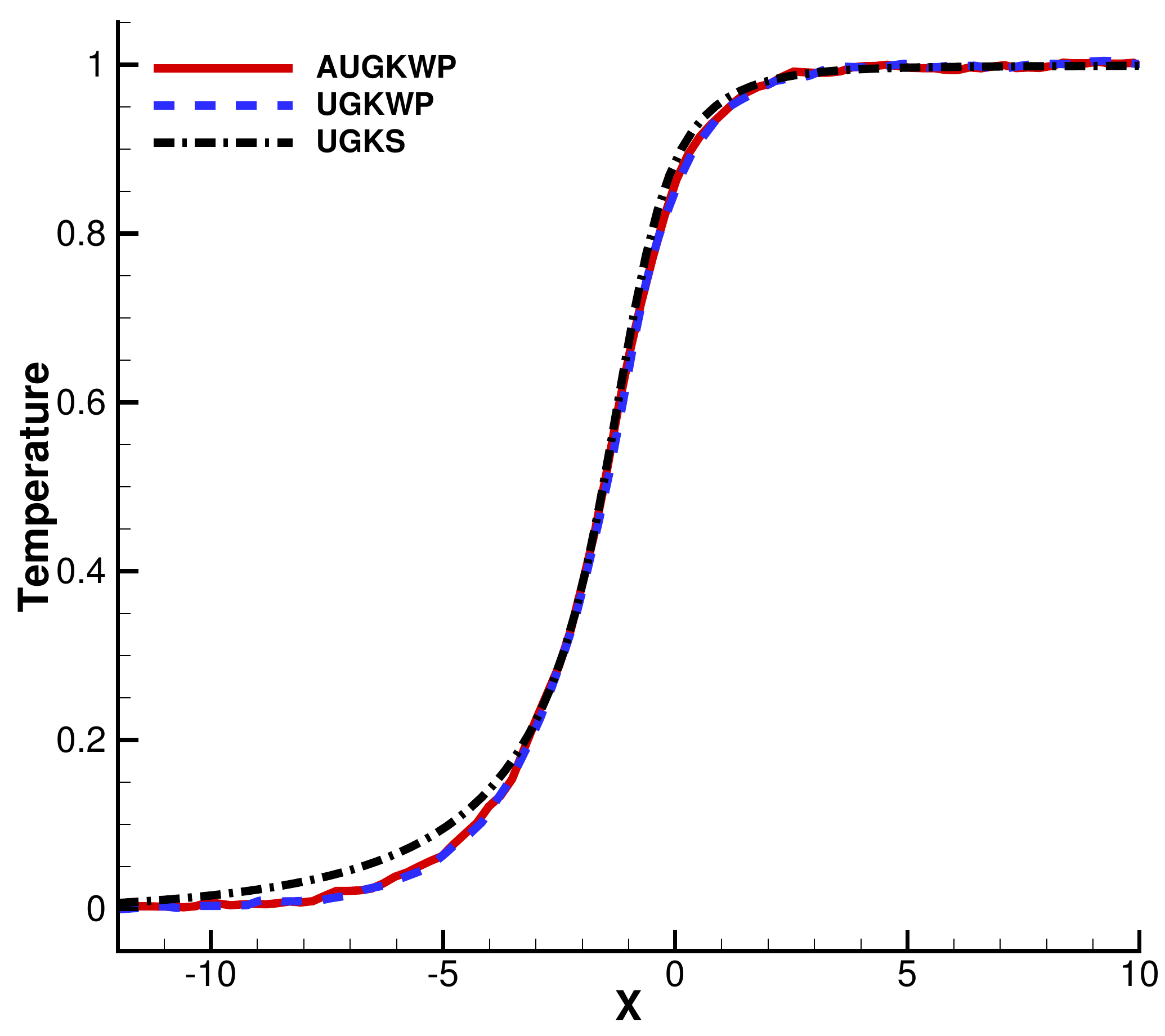}}
	\caption{Shock structure at ${\rm Ma} = 4$. (a) Density and (b) temperatures compared with the original UGKWP method and the UGKS. The modification of particle collision time in Eq.~\eqref{eq:tauStar} is not used in UGKS.  }
	\label{fig:shock-Ma4}
\end{figure}

\begin{figure}[H]
	\centering
	\subfloat[]{\includegraphics[width=0.4\textwidth]{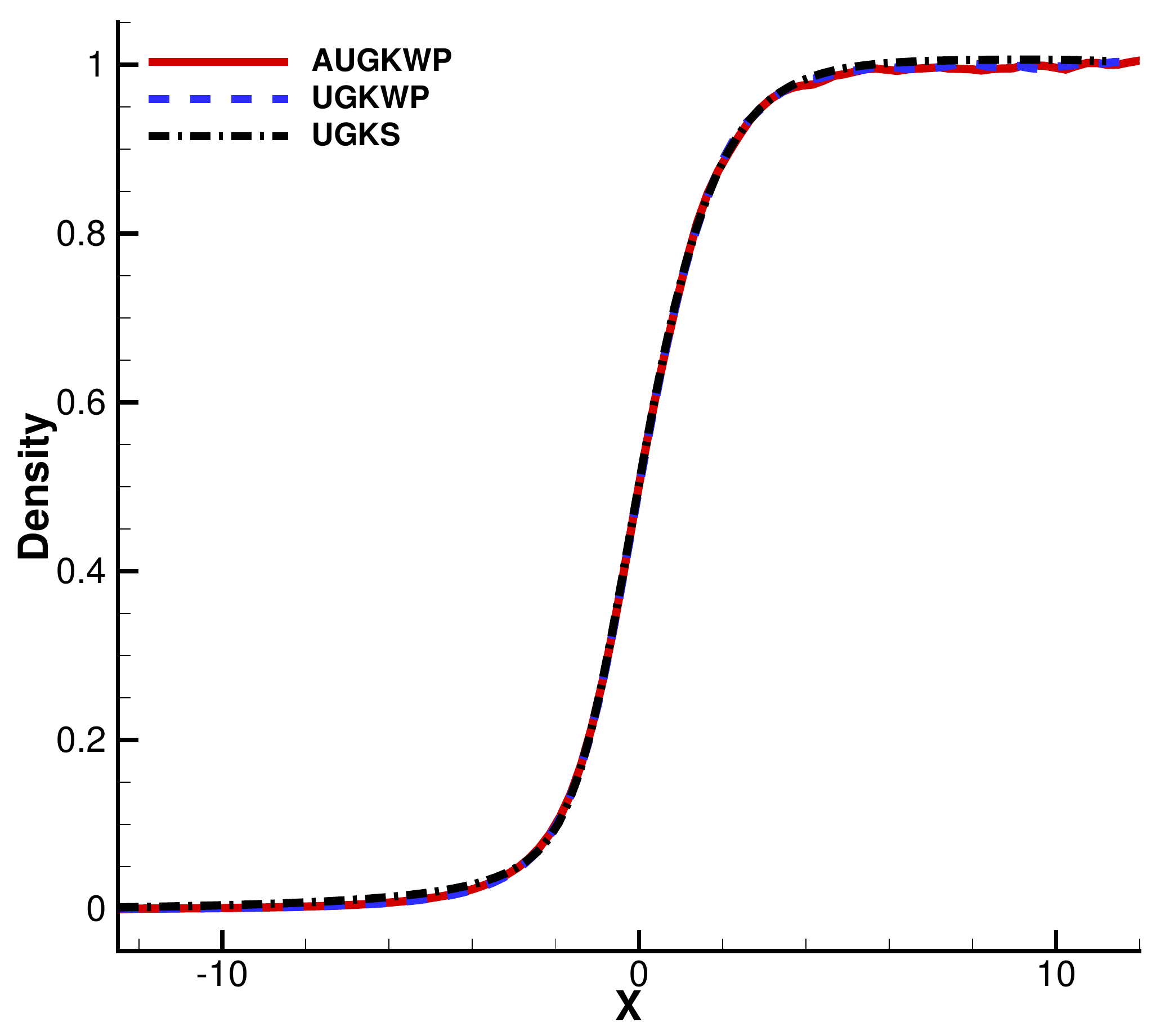}}
	\subfloat[]{\includegraphics[width=0.4\textwidth]{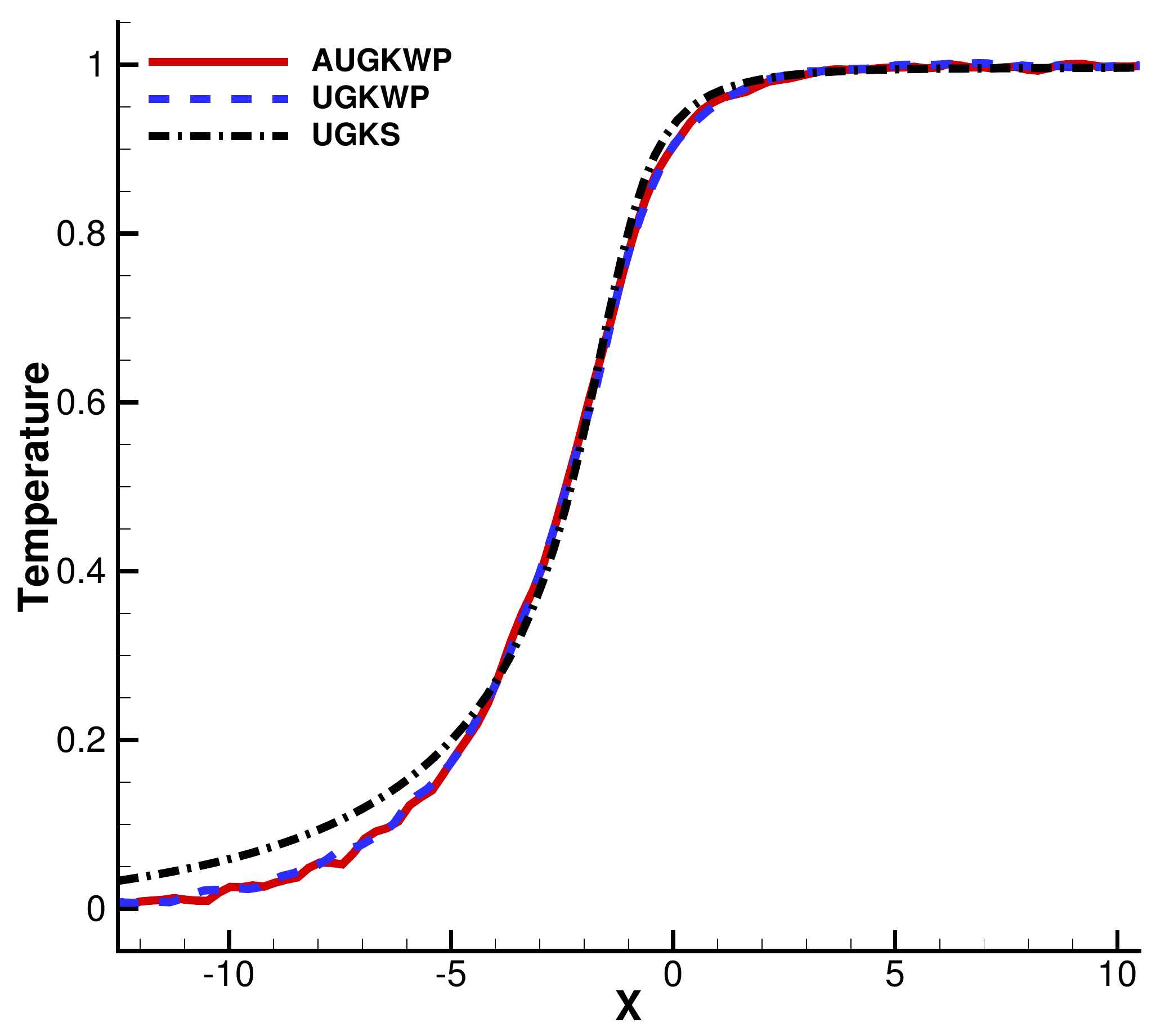}}
	\caption{Shock structure at ${\rm Ma} = 10$. (a) Density and (b) temperatures compared with the original UGKWP method and the UGKS. The modification of particle collision time in Eq.~\eqref{eq:tauStar} is not used in UGKS.}
	\label{fig:shock-Ma10}
\end{figure}

For the UGKWP and AUGKWP methods, $400$ simulation particles are used in each cell. To reduce the statistical noise, the time-averaging is taken from 2500th step over 12500 steps. The normalized density and temperature from the original UGKWP method, the AUGKWP method, and the UGKS are plotted in Fig.~\ref{fig:shock-Ma4}--\ref{fig:shock-Ma10}. The results from the current method have good agreement with that from the original UGKWP and the deterministic method.

\subsection{Flow around a circular cylinder}

\begin{figure}[H]
	\centering
	\subfloat[]{\includegraphics[width=0.45\textwidth]
	{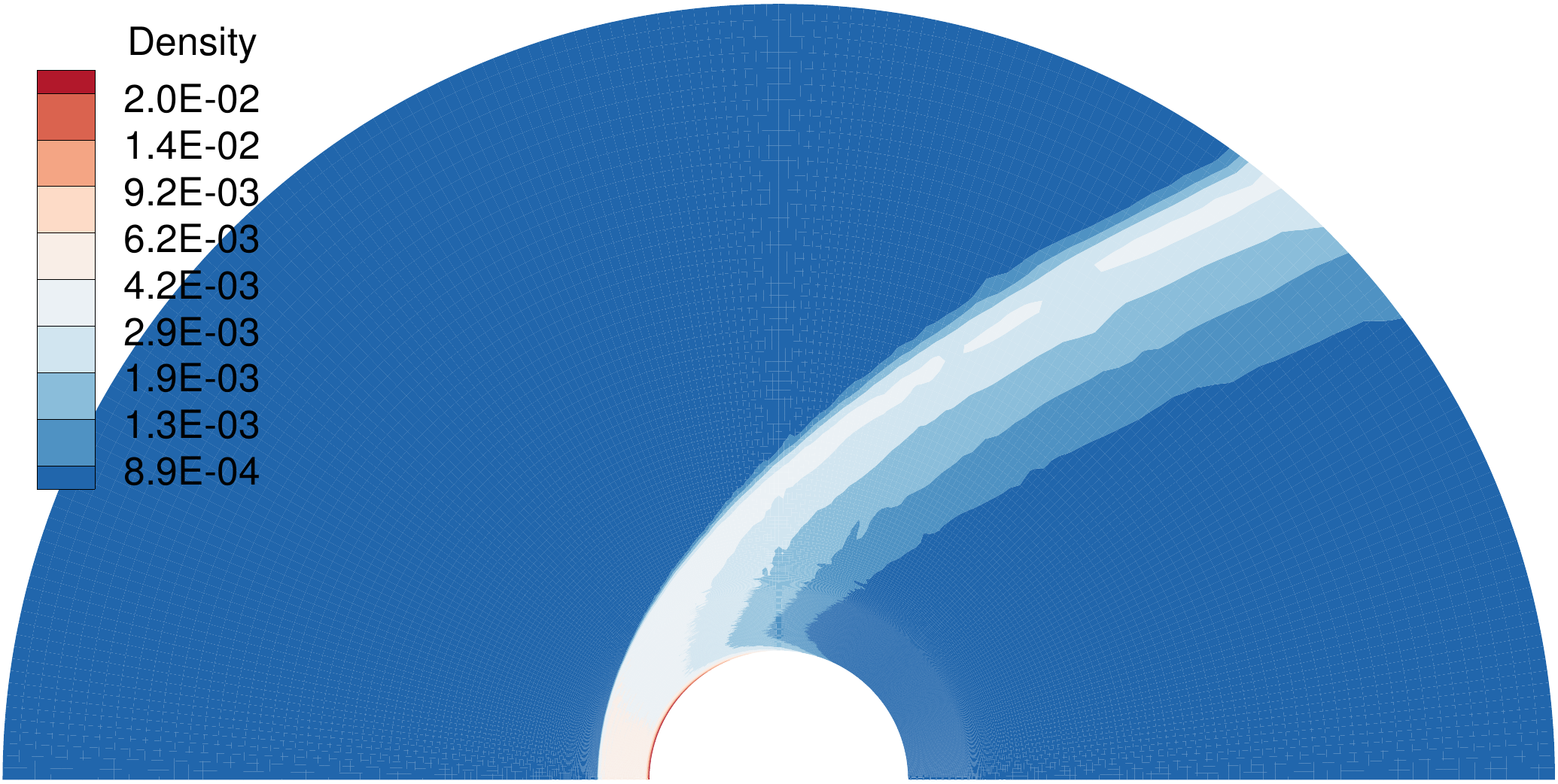}}
	\subfloat[]{\includegraphics[width=0.45\textwidth]
	{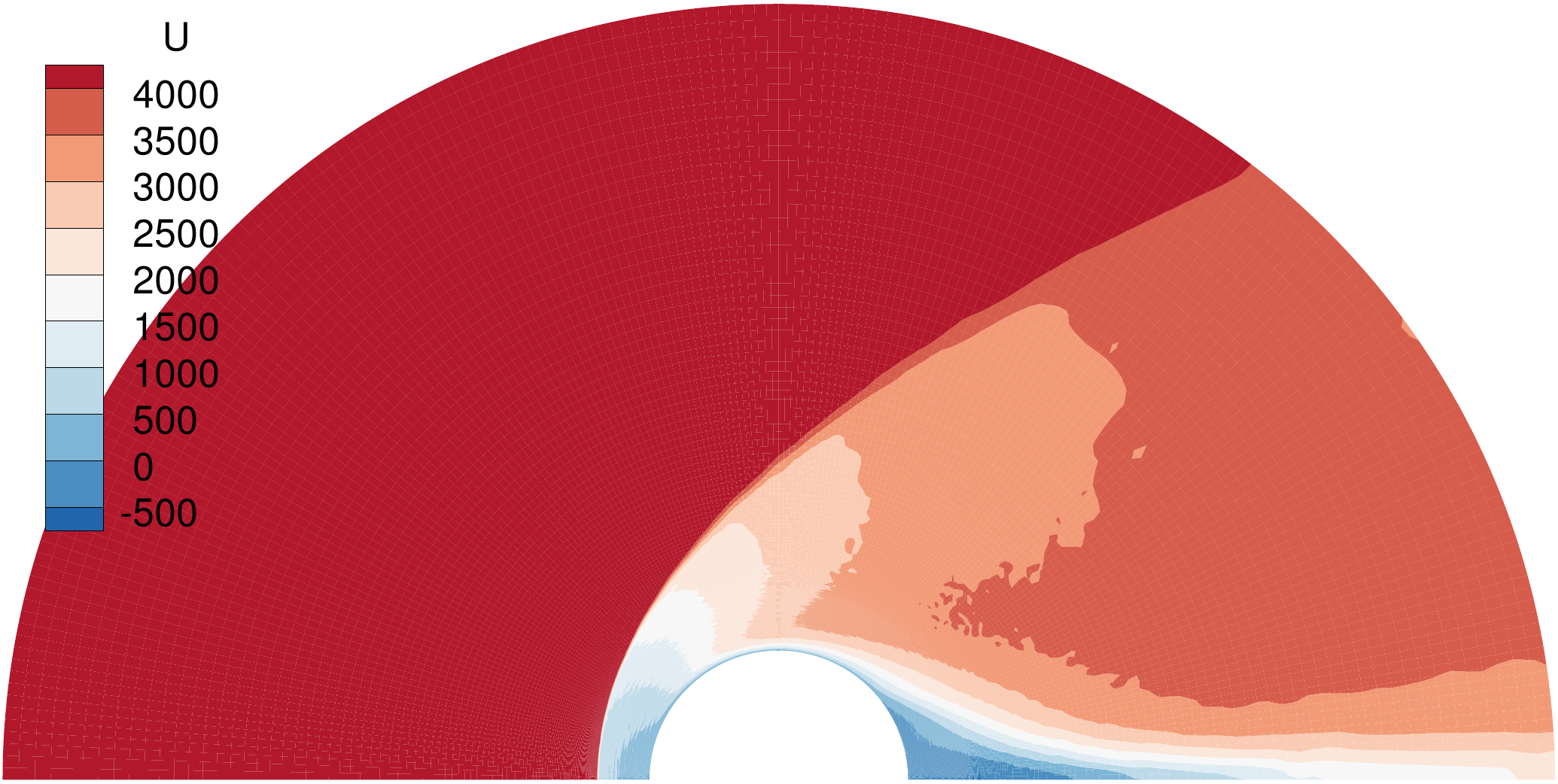}} \\
	\subfloat[]{\includegraphics[width=0.45\textwidth]
	{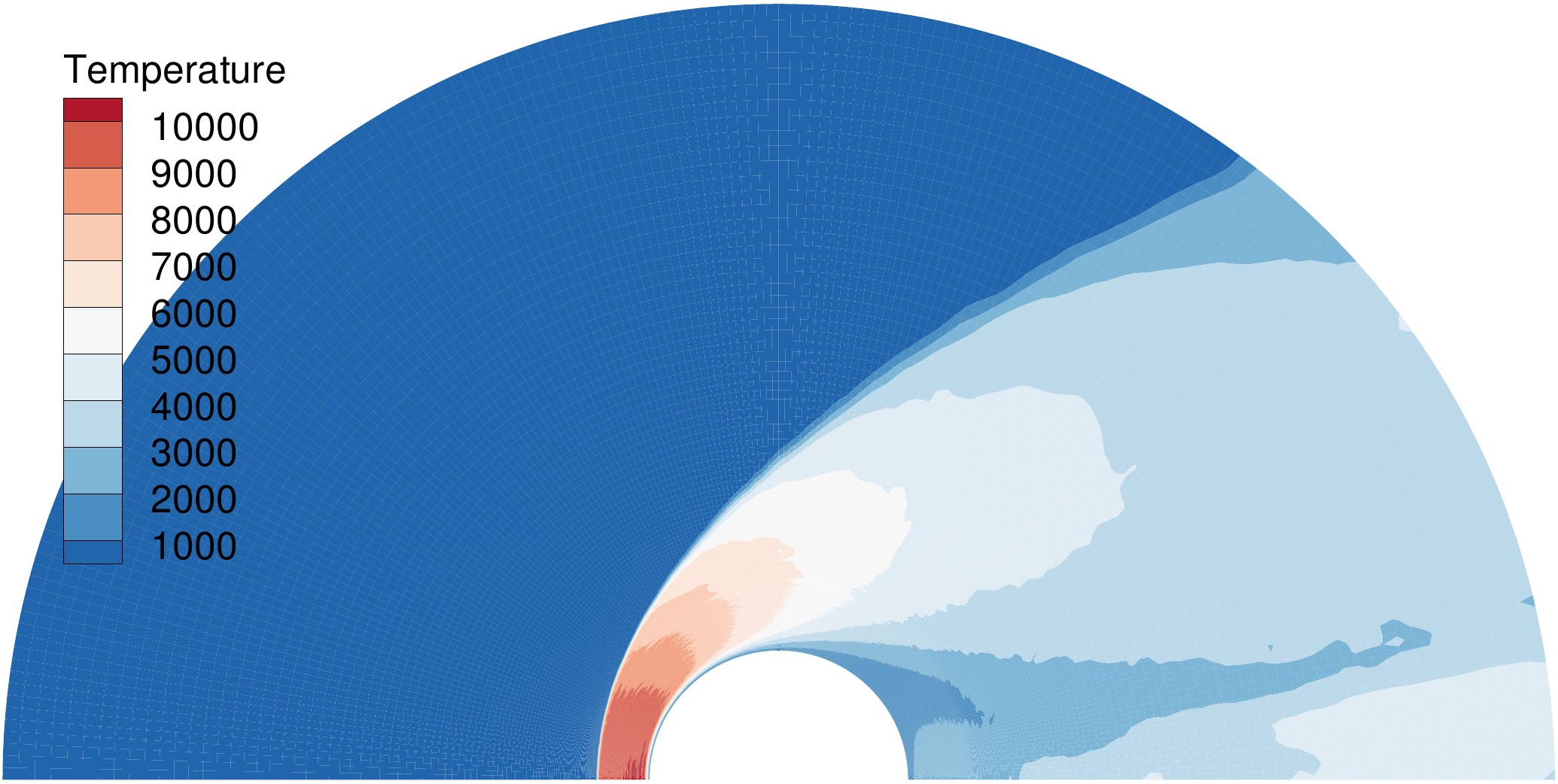}}
	\caption{Hypersonic flow at ${\rm Ma} = 15$ and ${\rm Kn} = 0.001$ around a semi-circular
	cylinder.
	(a) Density, (b) $x$ direction velocity, and (c) temperature contours.}
	\label{fig:cylinder}
\end{figure}

\begin{figure}[H]
	\centering
	\subfloat[]{\includegraphics[width=0.33\textwidth]{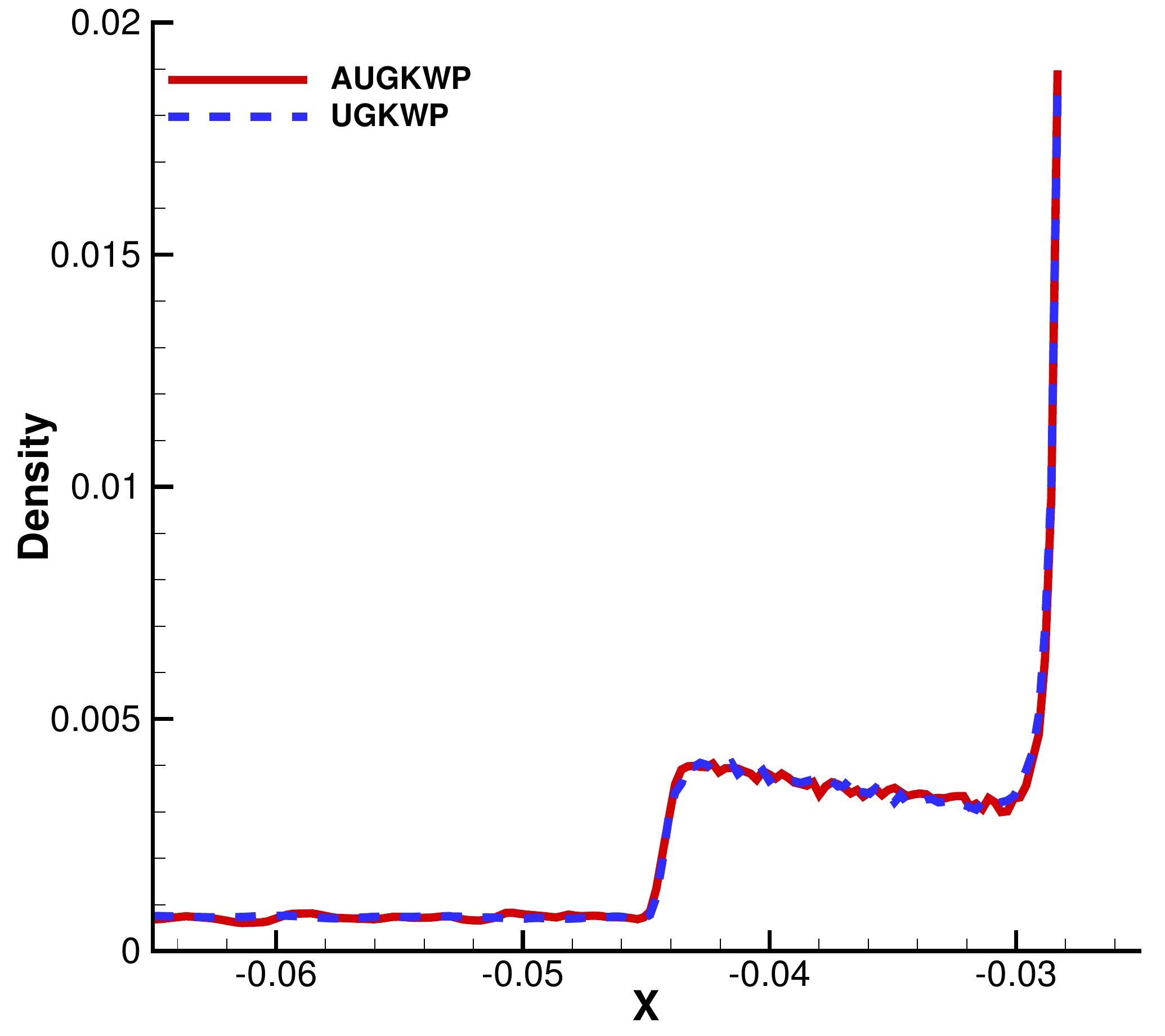}}
	\subfloat[]{\includegraphics[width=0.33\textwidth]{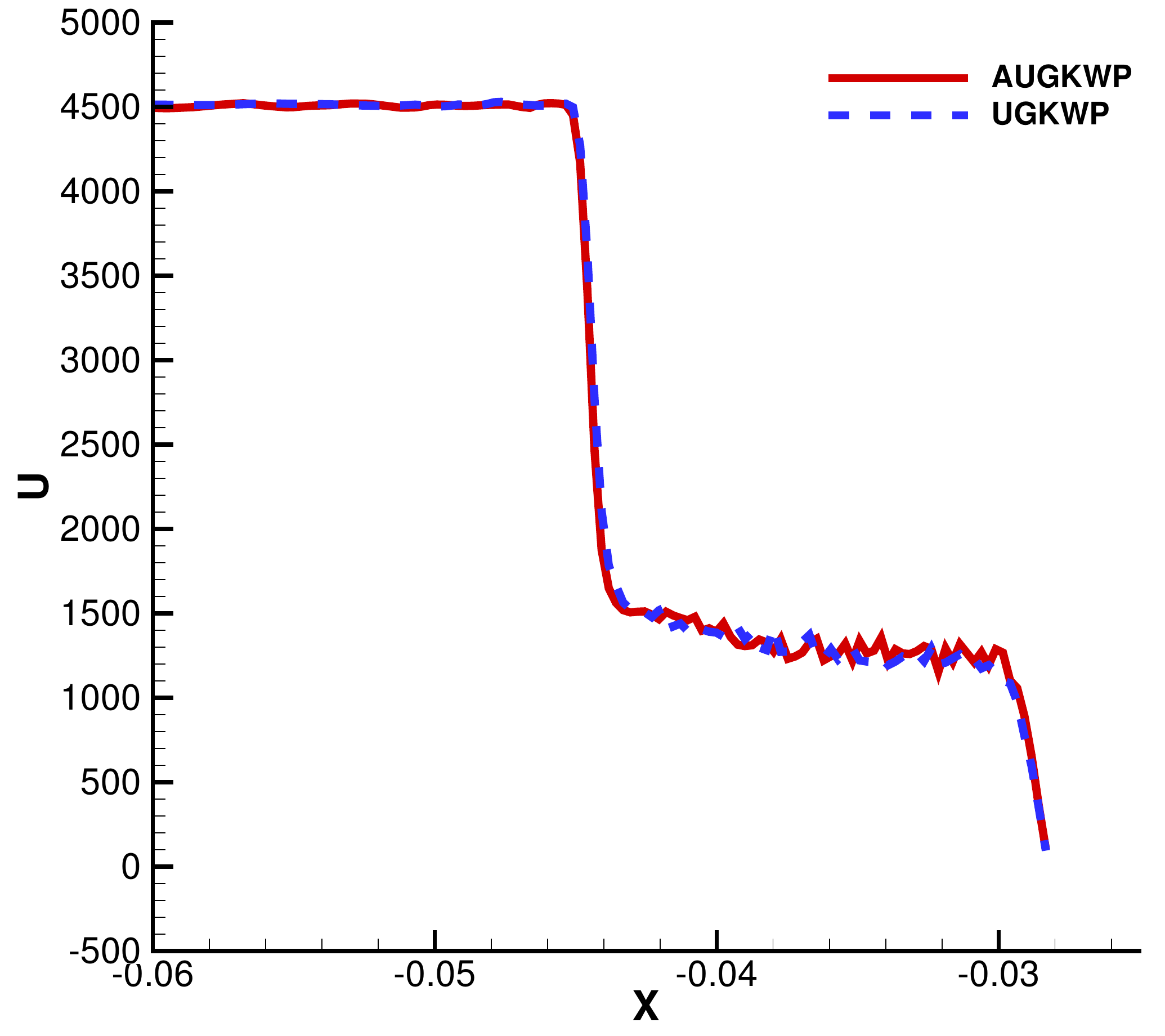}}
	\subfloat[]{\includegraphics[width=0.33\textwidth]{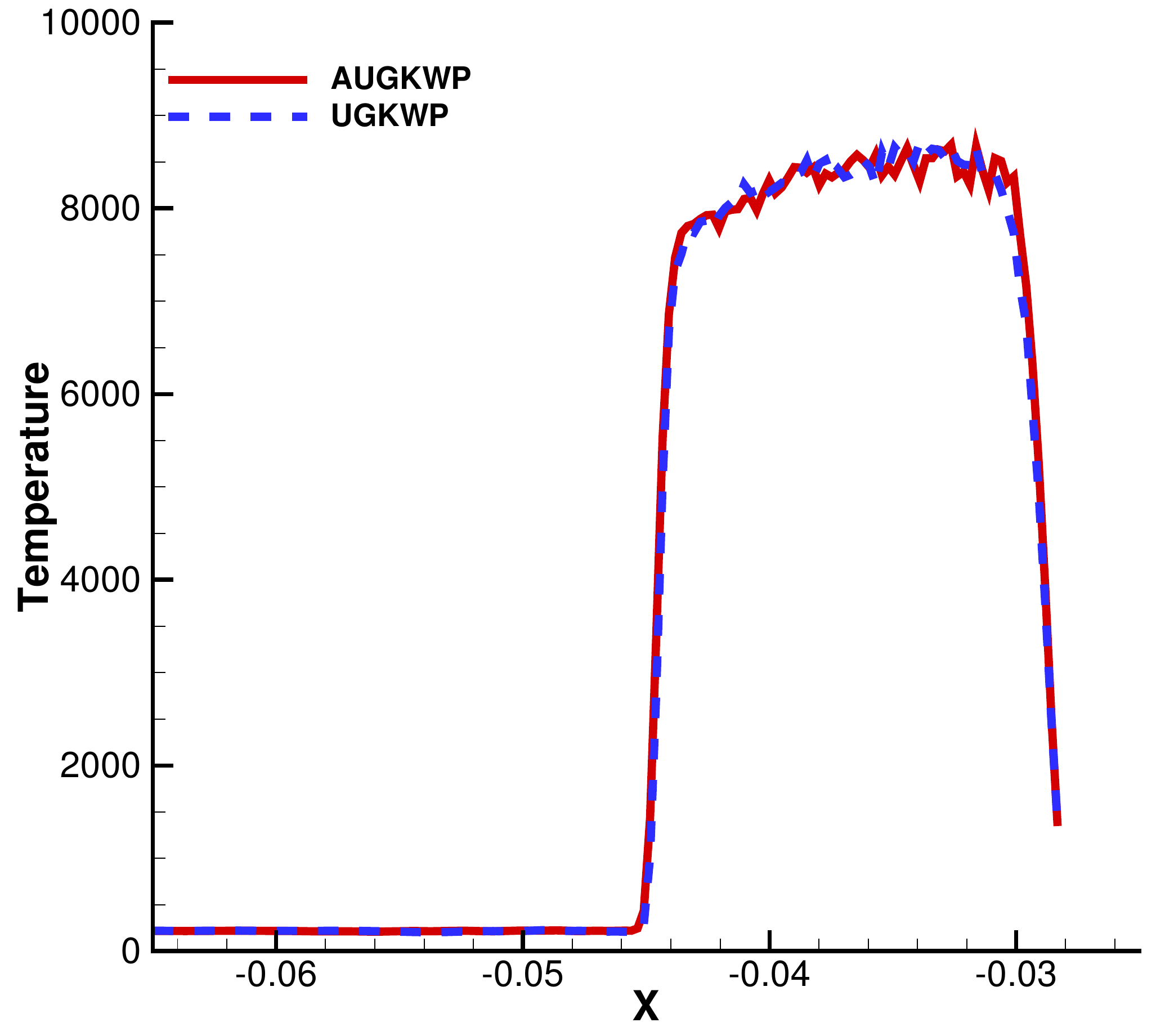}}
	\caption{Hypersonic flow at ${\rm Ma} = 15$  and ${\rm Kn} = 0.001$ around a semi-circular
	cylinder. (a) Density, (b) $x$ direction velocity, and (c) temperature distributions along the $45^{ \rm \circ }$ extraction line at ${\rm Ma} = 15$ and ${\rm Kn}=0.001$.}
	\label{fig:cylinder-T}
\end{figure}

High-speed flow passing over a semi-circular cylinder at a Mach number $15$ and ${\rm Kn} = 0.001$ is simulated. The diameter of the cylinder has $D = 0.08$ m. The Knudsen number is defined with respect to the diameter. The computational domain is discretized by $280 \times 200 \times 1$ quadrilateral cells. The initial reference number of particles $N_r$ is set as 2000. The initial temperature of free stream gives $T_\infty = 217.5$ K, and the isothermal wall temperature is fixed at $T_w = 1000$ K. The CFL number is $0.5$. The reference Knudsen number is ${\rm Kn}_{ref} = 0.01$. Fig.~\ref{fig:cylinder} plots the contours of flow field computed by the AUGKWP method, where an initial flow field provided by $10000$ steps of GKS calculation \cite{xu2001} is adopted, and $60000$ steps of the AUGKWP and UGKWP methods have been carried out to achieve a steady state. Fig.~\ref{fig:cylinder-T} shows the comparison between the AUGKWP and original UGKWP method for the density, velocity in $x$ direction, and temperature extracted along the $45^{\circ}$ line in the upstream. The results from AUGKWP method are reasonable. The particle mass fraction in Fig.~\ref{fig:cylinder} implies the efficiency of the new wave-particle decomposition method. The simulation time for the AUGKWP and UGKWP methods on Tianhe-2 with 3 nodes (72 cores, Intel Xeon E5-2692 v2, 2.2 GHz) are 15623 s and 30487 s, respectively.

\begin{figure}[H]
	\centering
	\subfloat[]{\includegraphics[width=0.45\textwidth]
	{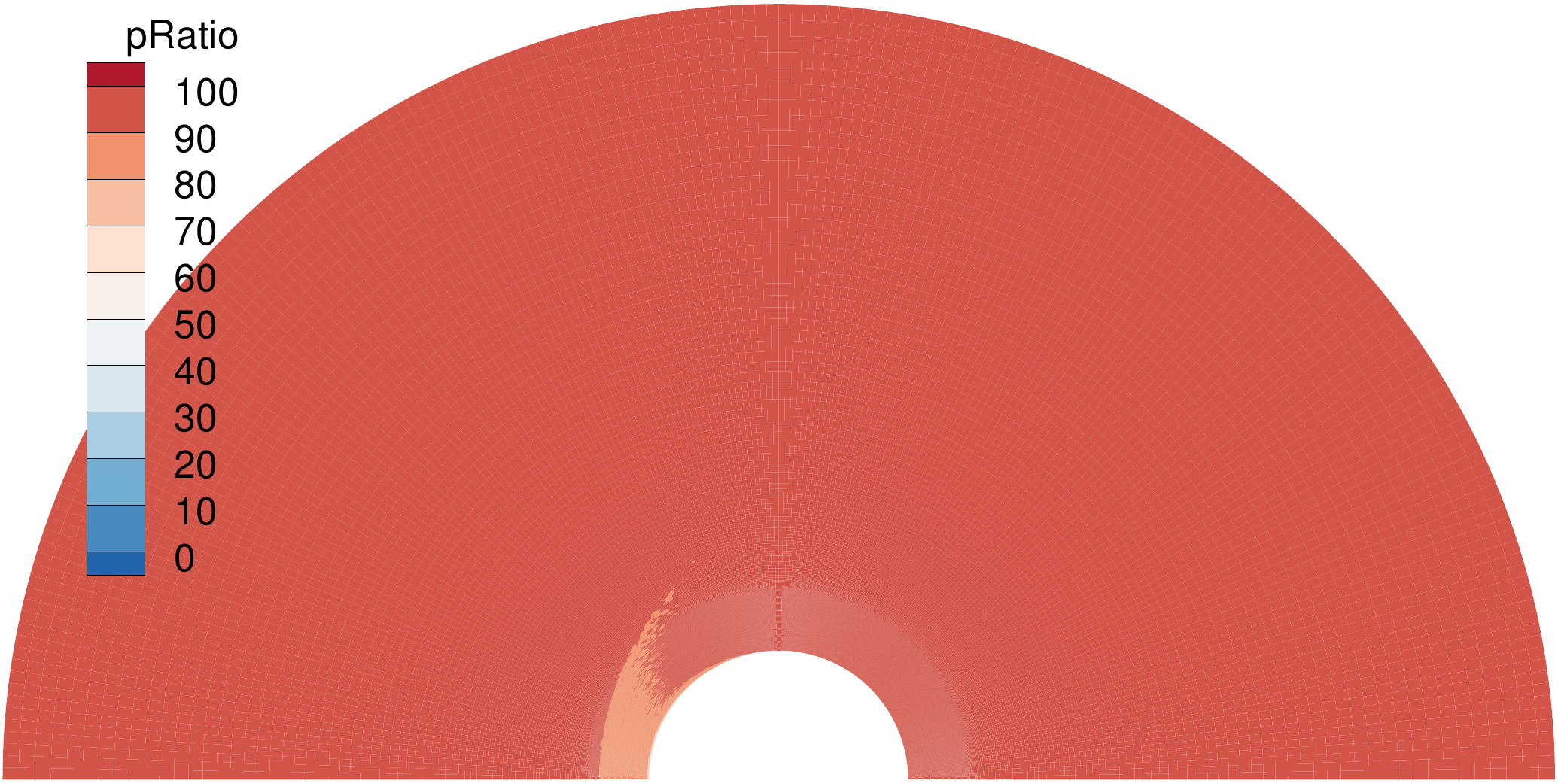}}
	\subfloat[]{\includegraphics[width=0.45\textwidth]
	{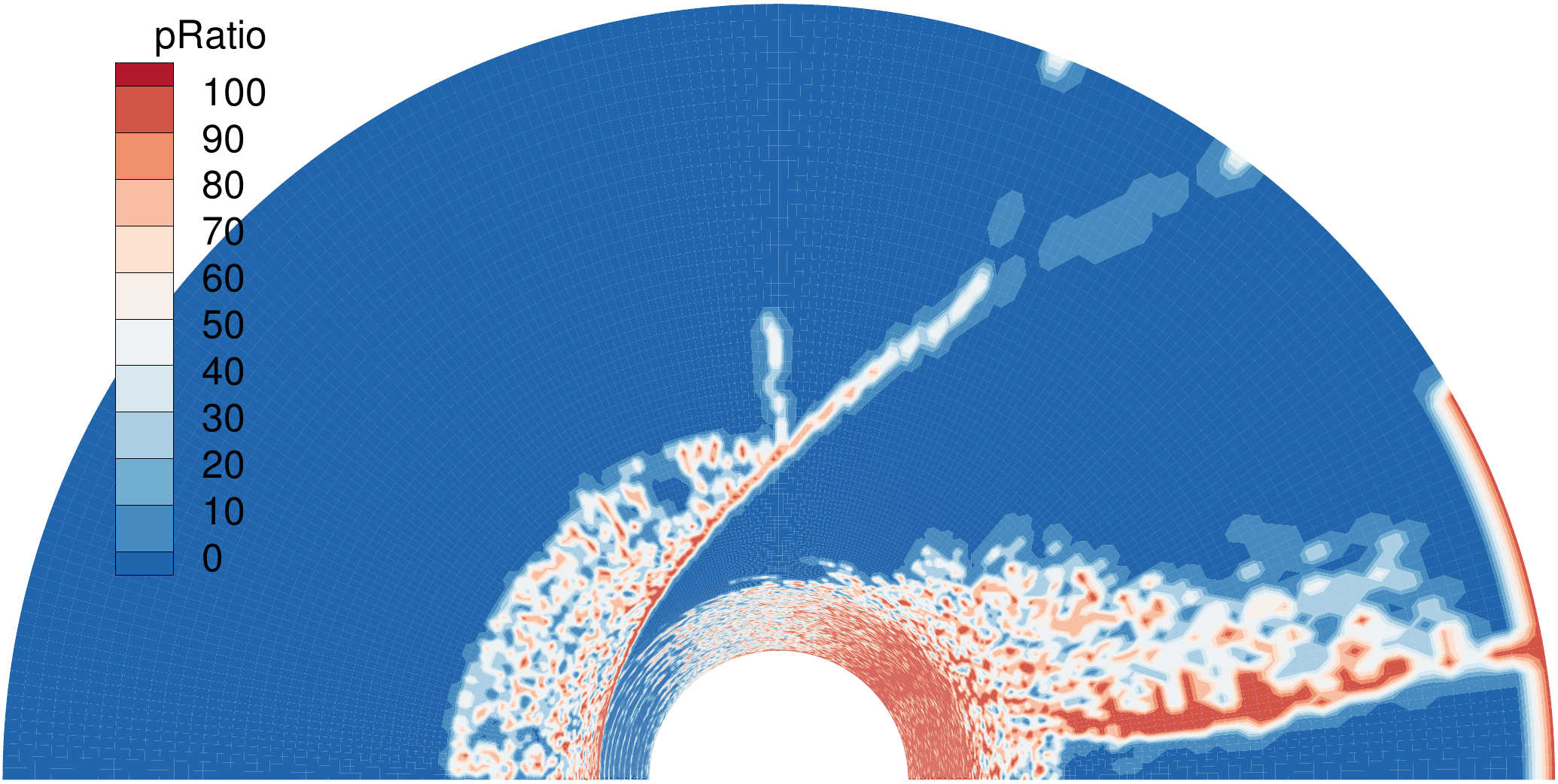}}
	\caption{Hypersonic flow at ${\rm Ma} = 15$ and ${\rm Kn} = 0.001$ around a semi-circular
	cylinder. Particle mass fraction distribution given by (a) the original UGKWP method and (b) the AUGKWP method.}
	\label{fig:cylinder-pRatio}
\end{figure}

\subsection{Flow around an Apollo reentry}
\begin{figure}[H]
	\centering
	\subfloat[]{\includegraphics[width=0.4\textwidth]{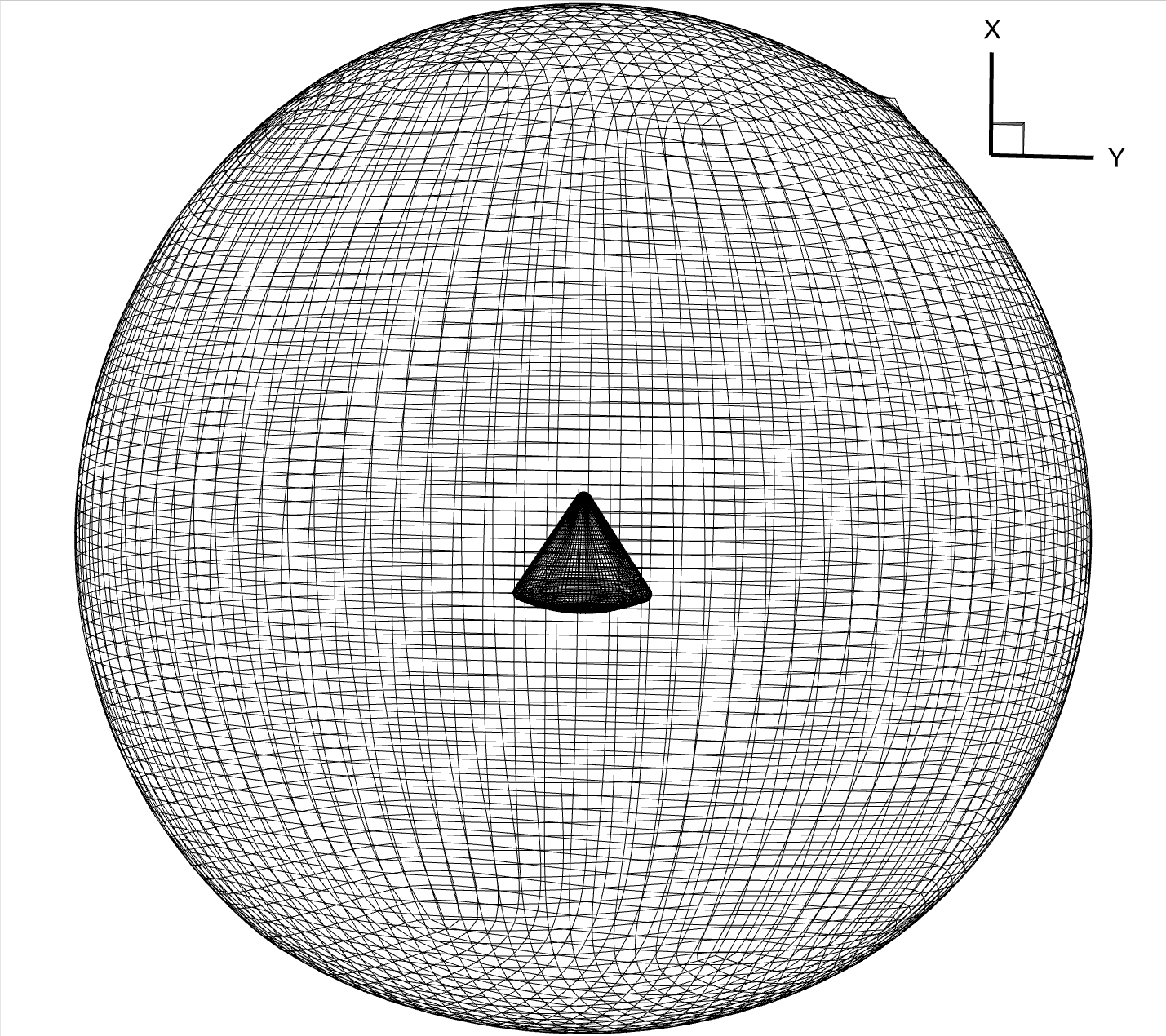}}
	\subfloat[]{\includegraphics[width=0.4\textwidth]{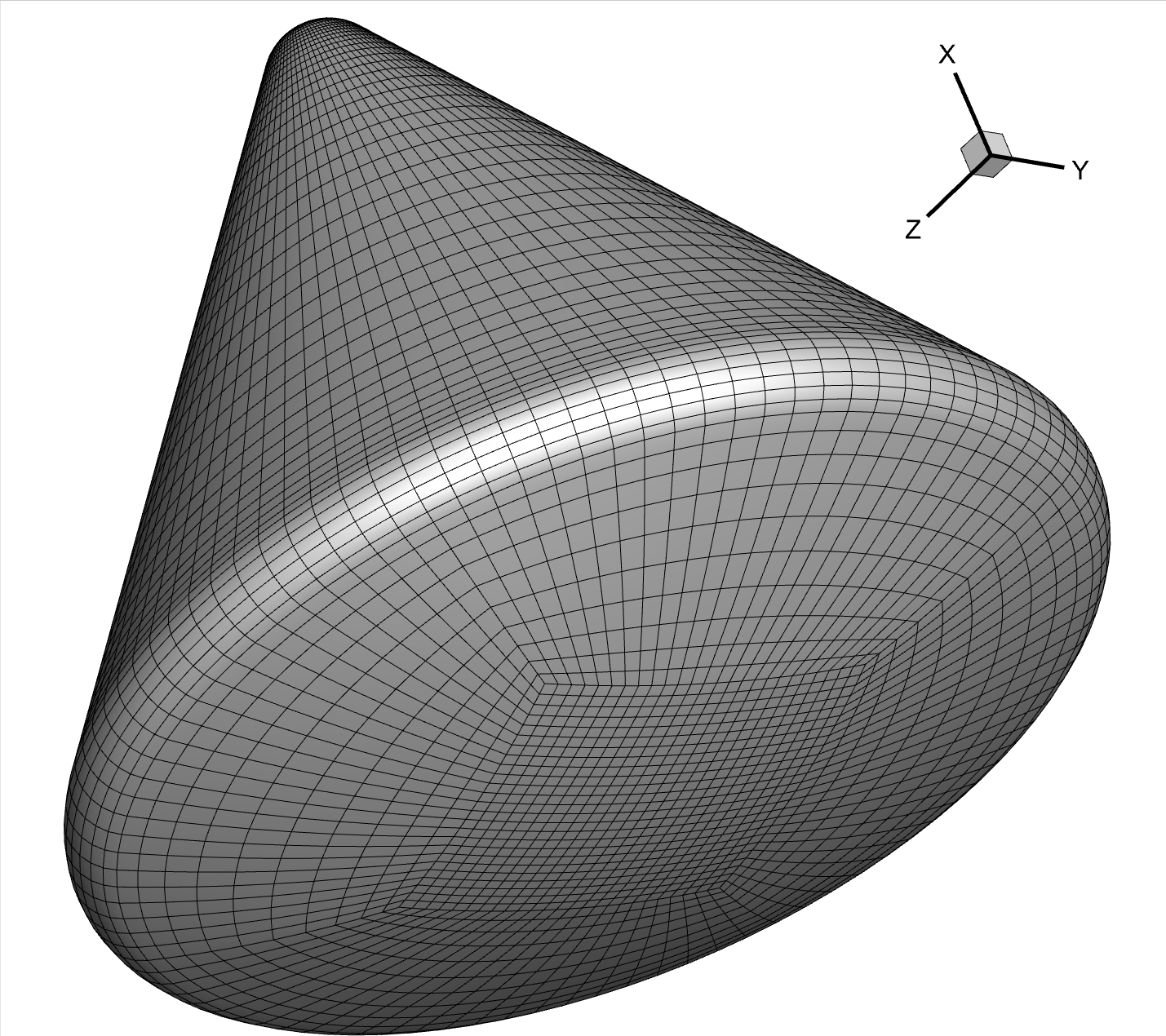}}
	\caption{Mesh of the Apollo reentry. (a) Global view and (b) local enlargement.}
	\label{fig:apollo-mesh}
\end{figure}

Hypersonic flow at ${\rm Ma} = 5$ and ${\rm Kn} = 10^{-3}$ passing over an Apollo reentry in the transition flow regime is simulated for nitrogen gas. This case shows the efficiency and capability of the AUGKWP method for simulating the large-scale three-dimensional hypersonic flow. The reference length for the definition of the Knudsen number is $L_{ref} = 3.912$ m. As shown in Fig.~\ref{fig:apollo-mesh}, the mesh consits of 372500 cells. The reference Knudsen number is set as ${\rm Kn}_{ref} = 0.01$ and the initial reference number of particles $N_r$ is 400 per cell. The flow at free stream has an initial temperature $T_\infty = 142.2$ K, and the reentry surface is treated as isothermal wall with a constant temperature $T_w = 500$ K. The angle of attack is $30^{\circ}$. The initial flow field is prepared by GKS calculation with 10000 steps.
Then, 25000 steps have been carried out to achieve a steady state solution.

\begin{figure}[H]
	\centering
	\subfloat[]{\includegraphics[width=0.48\textwidth]
	{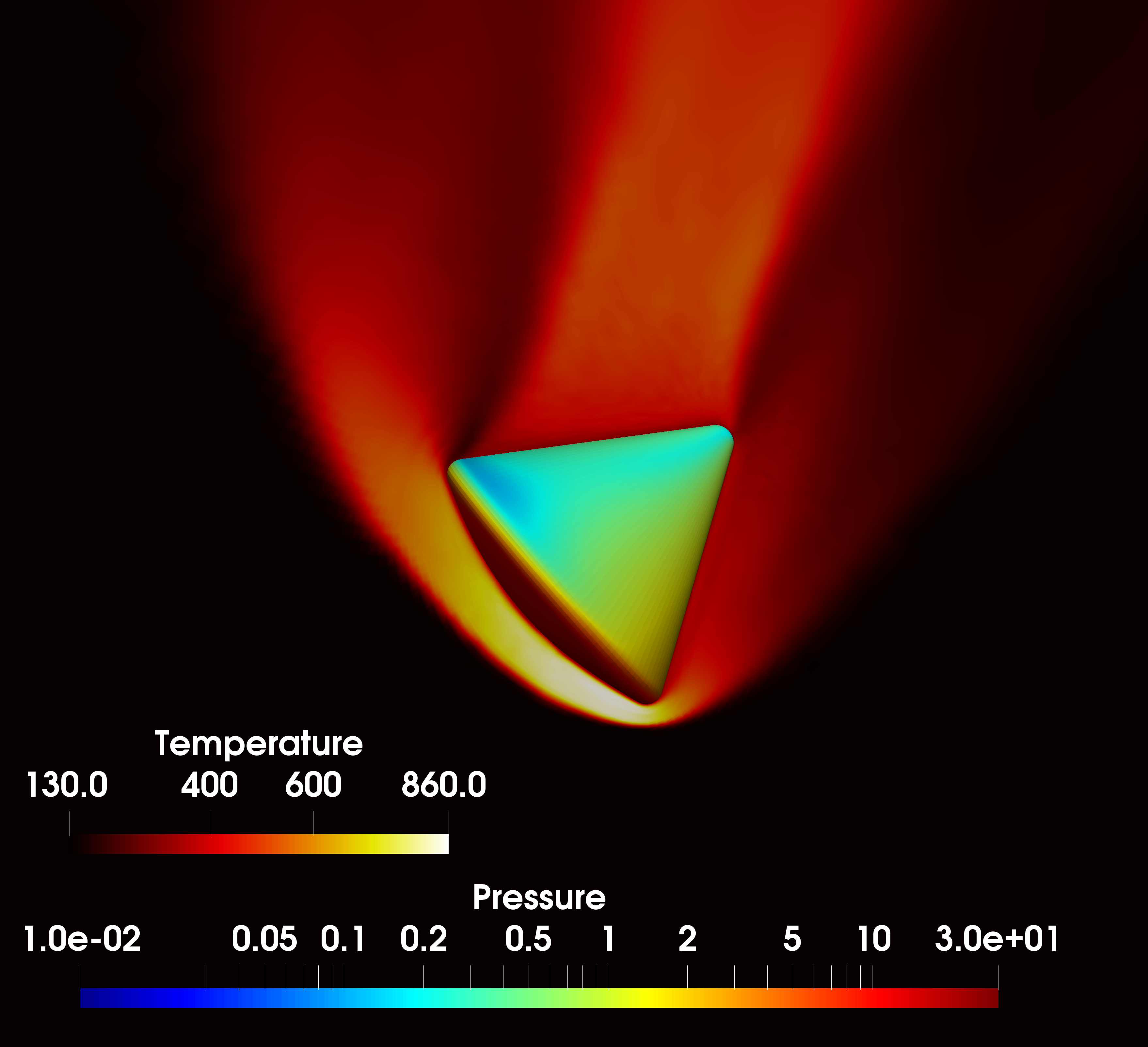}}
	\hspace{1mm}
	\subfloat[]{\includegraphics[width=0.48\textwidth]
	{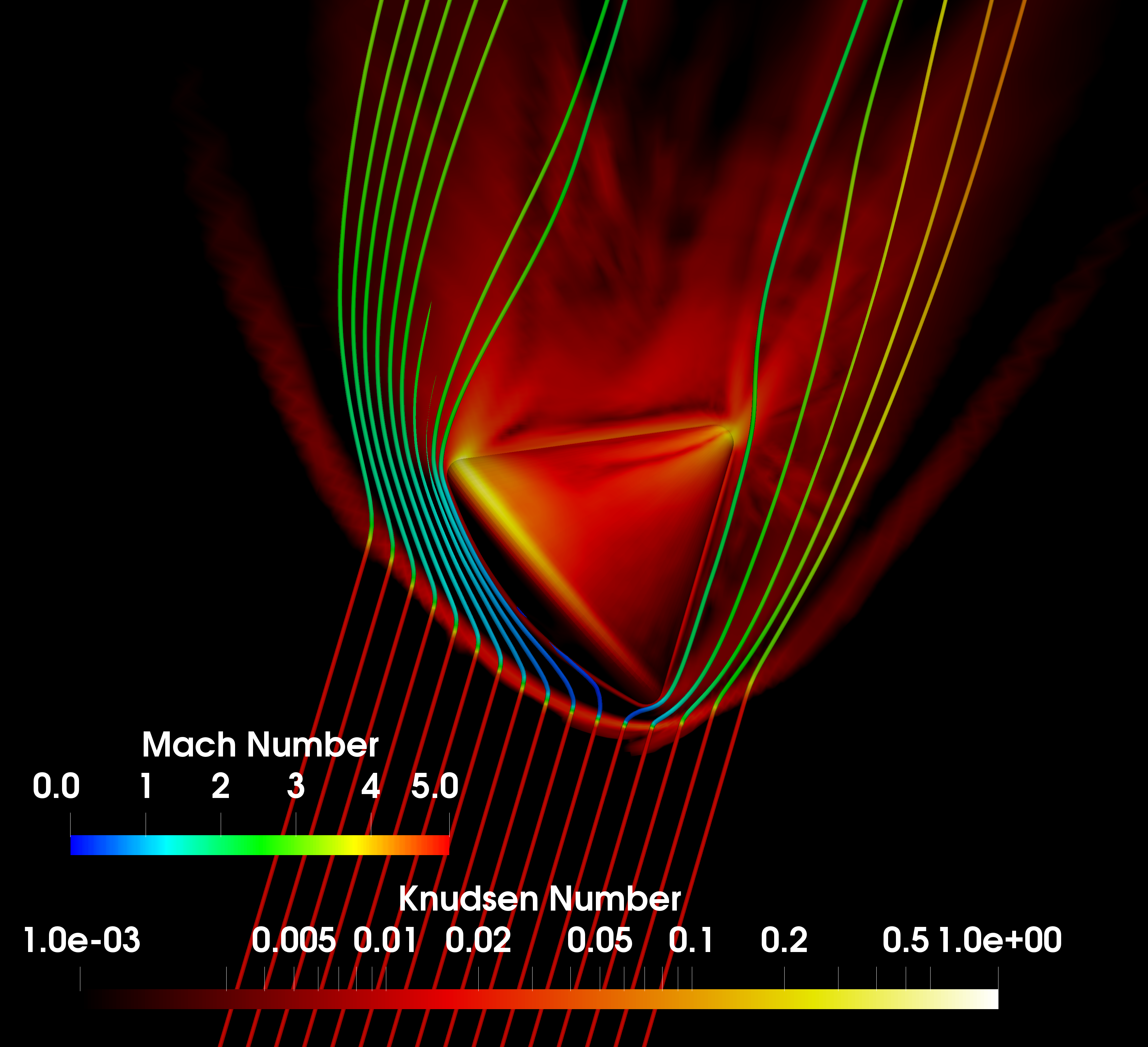}}
	\caption{Hypersonic flow at ${\rm Ma} = 5$ and ${\rm Kn} = 0.001$ around an Apollo reentry. (a) Temperature and pressure distributions, and (b) local Knudsen Number ${\rm Kn}_{Gll}$ distribution and Mach number along the streamline.}
	\label{fig:apollo}
\end{figure}

\begin{figure}[H]
	\centering
	\subfloat[]{\includegraphics[width=0.33\textwidth]{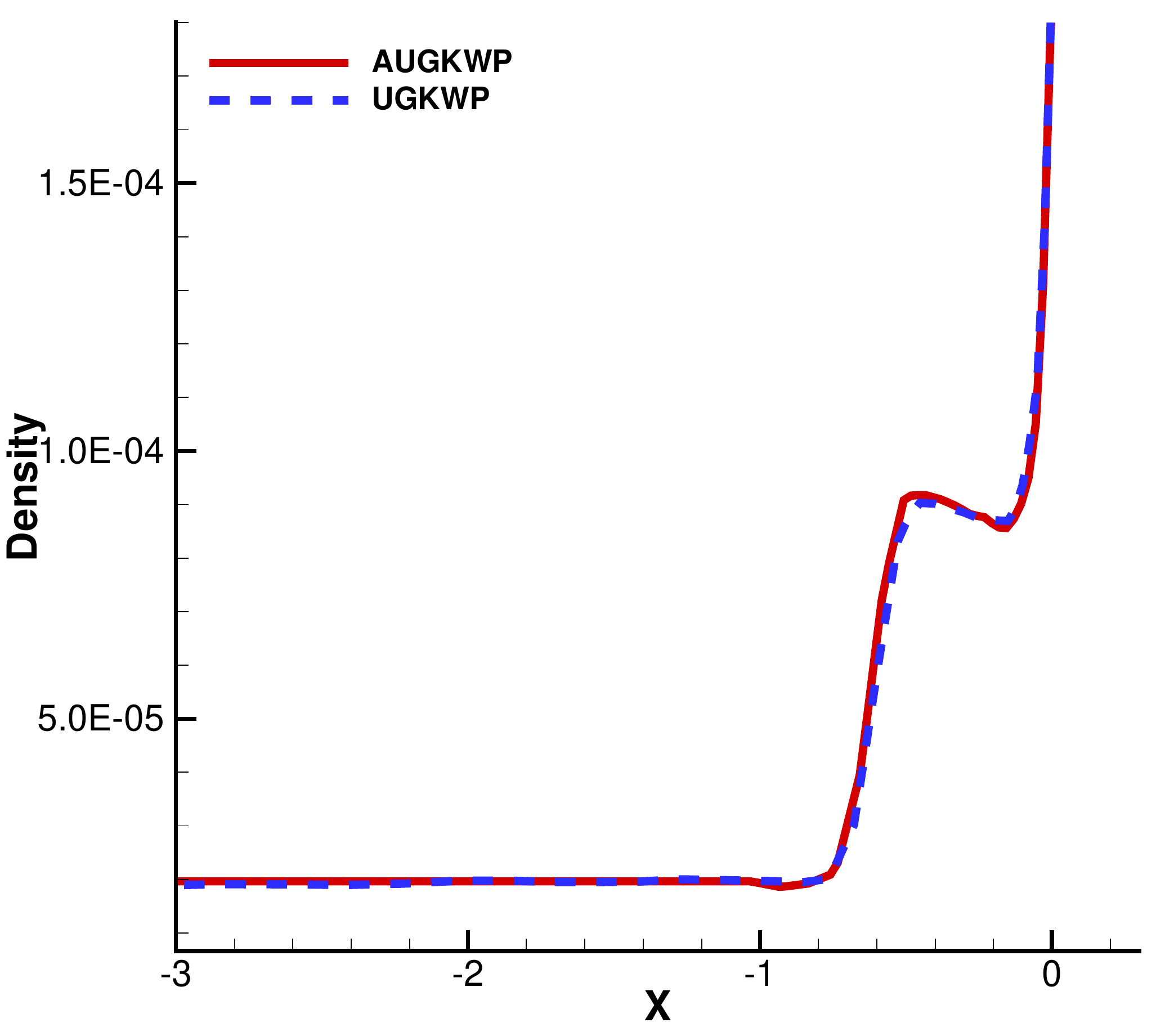}}
	\subfloat[]{\includegraphics[width=0.33\textwidth]{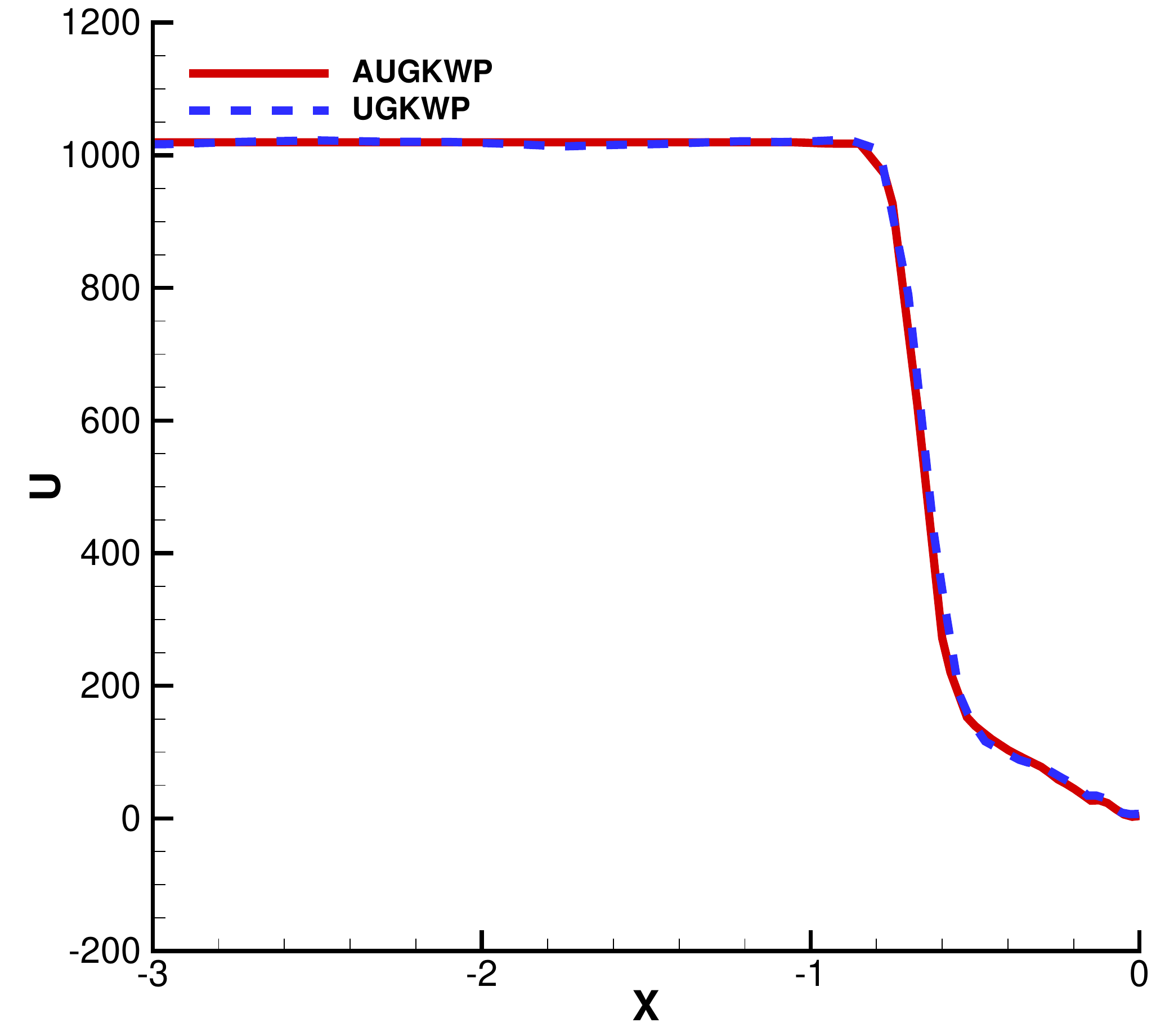}}
	\subfloat[]{\includegraphics[width=0.33\textwidth]{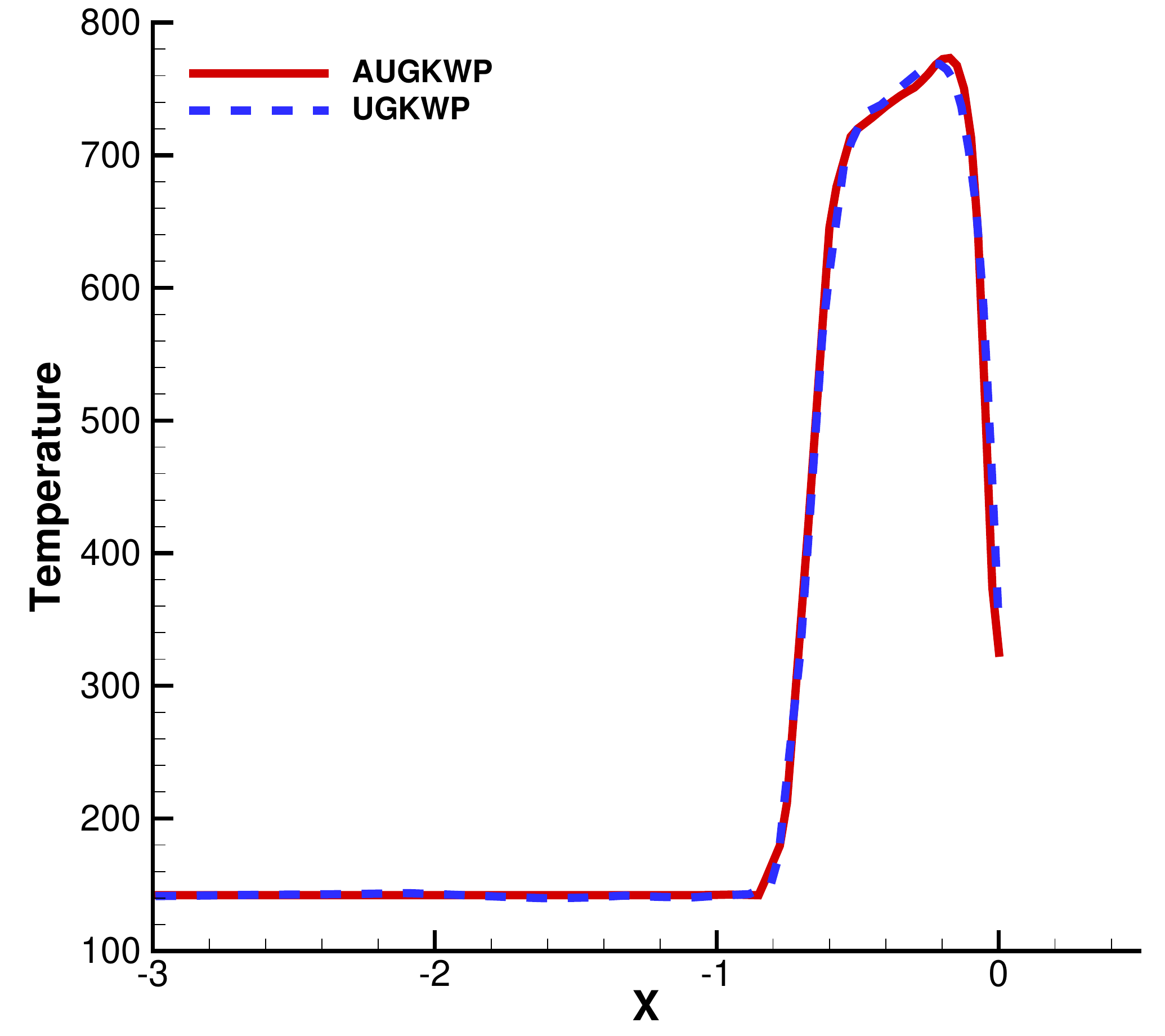}}
	\caption{(a) Density, (b) $x$ direction velocity, and (c) temperature distributions along the windward central axis at ${\rm Ma} = 5$ and ${\rm Kn}=0.001$.}
	\label{fig:apollo-line}
\end{figure}

\begin{figure}[H]
	\centering
	\subfloat[]{\includegraphics[width=0.48\textwidth]
	{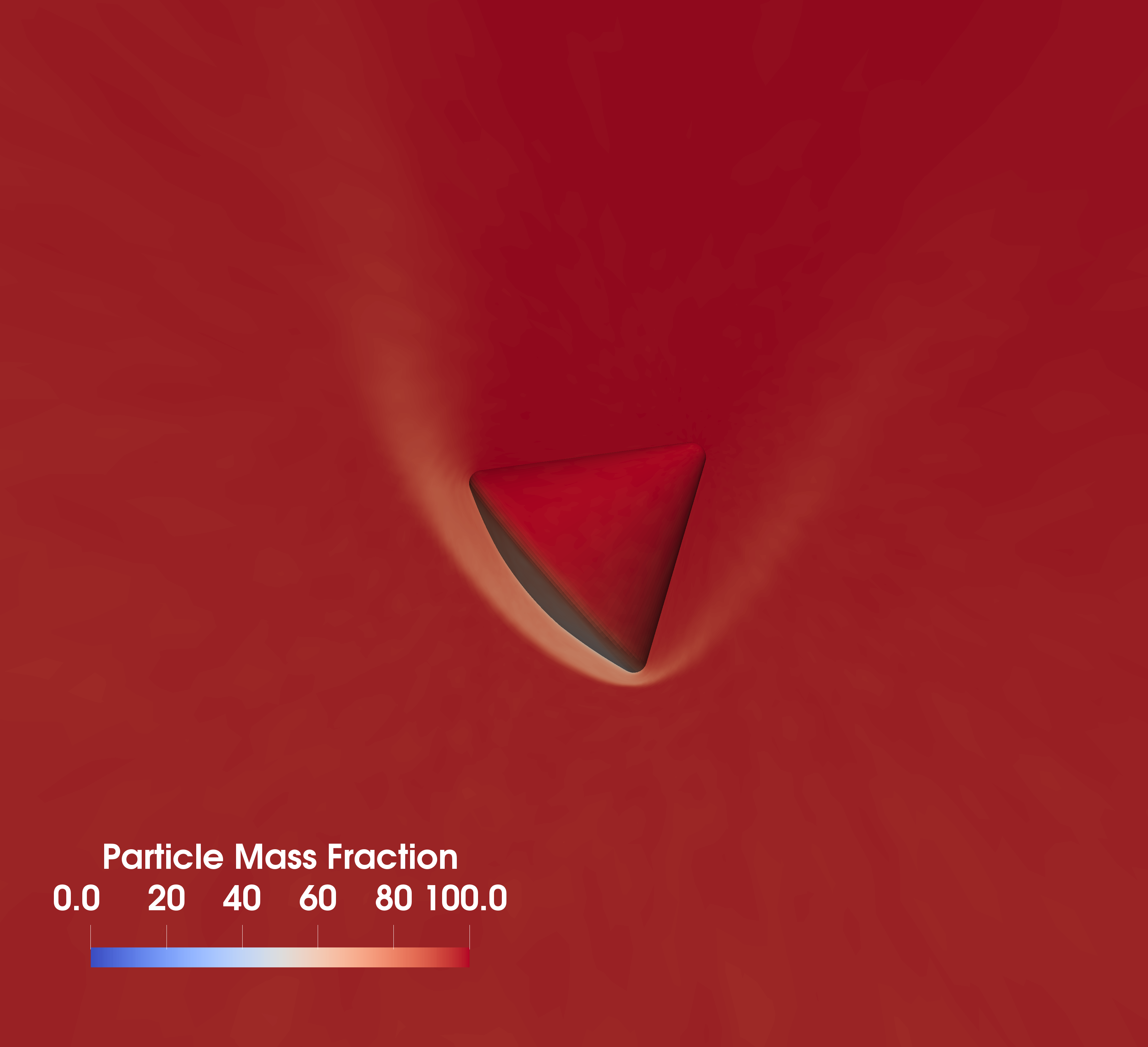}}
	\hspace{1mm}
	\subfloat[]{\includegraphics[width=0.48\textwidth]
	{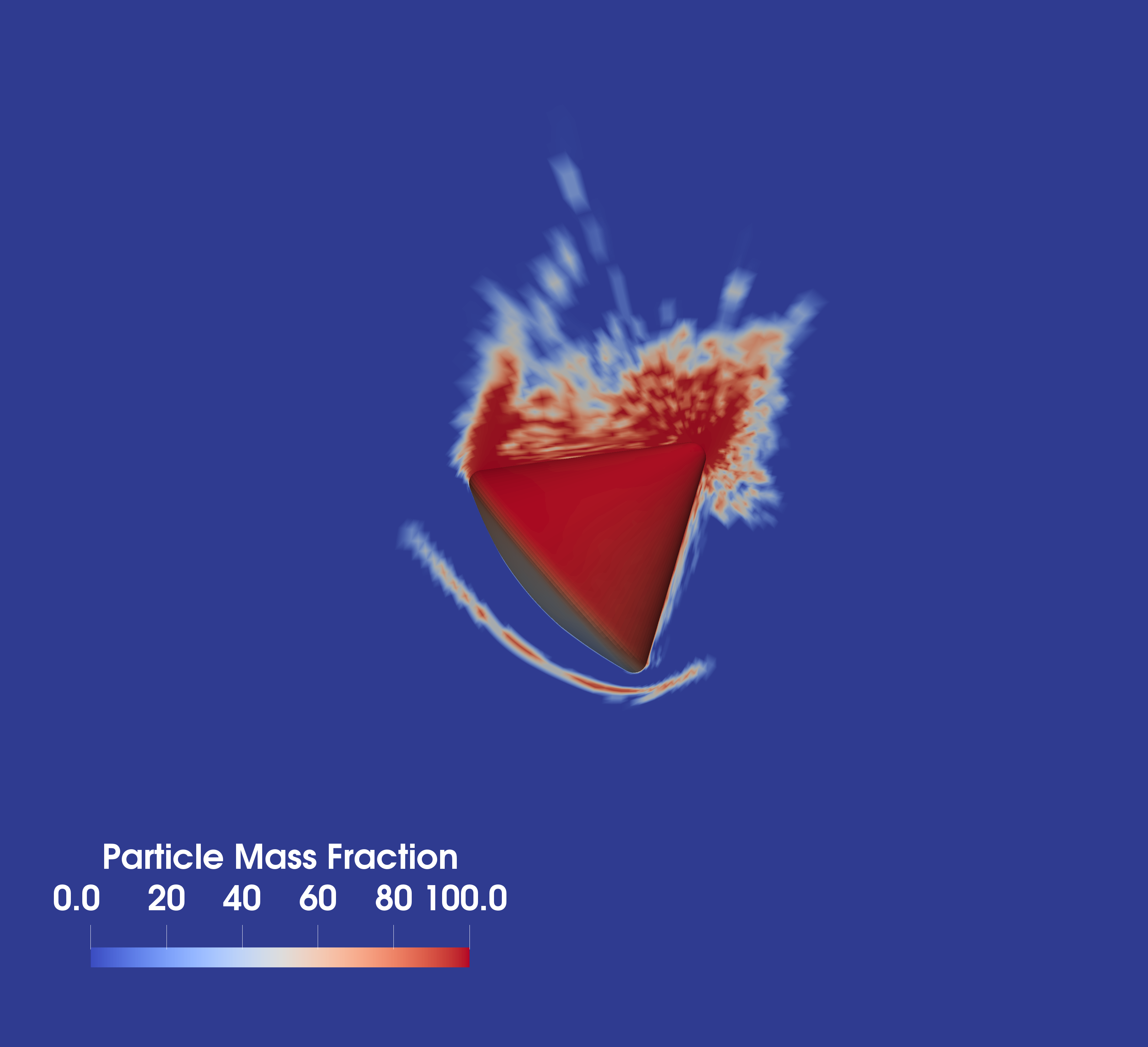}}
	\caption{Hypersonic flow at ${\rm Ma} = 5$ and ${\rm Kn} = 0.001$ around an Apollo reentry. Particle mass fraction distribution given by (a) the original UGKWP method and (b) the AUGKWP method.}
	\label{fig:apollo-pRatio}
\end{figure}

Figure~\ref{fig:apollo} shows the distributions of temperature, pressure, local Knudsen number and Mach number computed by AUGKWP method. The quantitive comparison between the AUGKWP and original UGKWP method for density, velocity in $x$ direction, and temperature extracted from central axis along the windward are plotted in Fig.~\ref{fig:apollo-line}. Good agreements have been obtained.
The efficiency of AUGKWP has been much improved due to its well controlled mass fraction of particles, as shown in Fig.~\ref{fig:apollo-pRatio}. Particles in the non-equilibrium are presented. The computational time for the AUGKWP and UGKWP methods are 24545 s and 42154 s on Tianhe-2 with 5 nodes (120 cores, Intel Xeon E5-2692 v2, 2.2 GHz). The AUGKWP becomes an indispensable tool for large-scale three-dimensional hypersonic rarefied flow simulation.

\subsection{Nozzle plume flow into a background vacuum}

In this case, the AUGKWP method is applied to the $CO_2$ expansions into a background vacuum. The unsteady and multiscale process of this plume flow is hard to compute with acceptive accuracy by conventional DSMC-CFD hybrid methods with time-dependent buffer zone, especially in the initial flow expansion stage. The UGKS, as a multiscale method, is hard to simulate such a flow with the requirement of covering a wide range discretized particle velocity space to capture the high Mach number jet in flow acceleration process through the nozzle.
The AUGKWP method treats the multiscale and multispeed expansion flow systematically through the dynamically adaptive wave-particle decomposition in each numerical cell, and makes the simulation acceptable in the memory requirement.

\begin{figure}[H]
	\centering
	\subfloat[]{\includegraphics[width=0.45\textwidth]{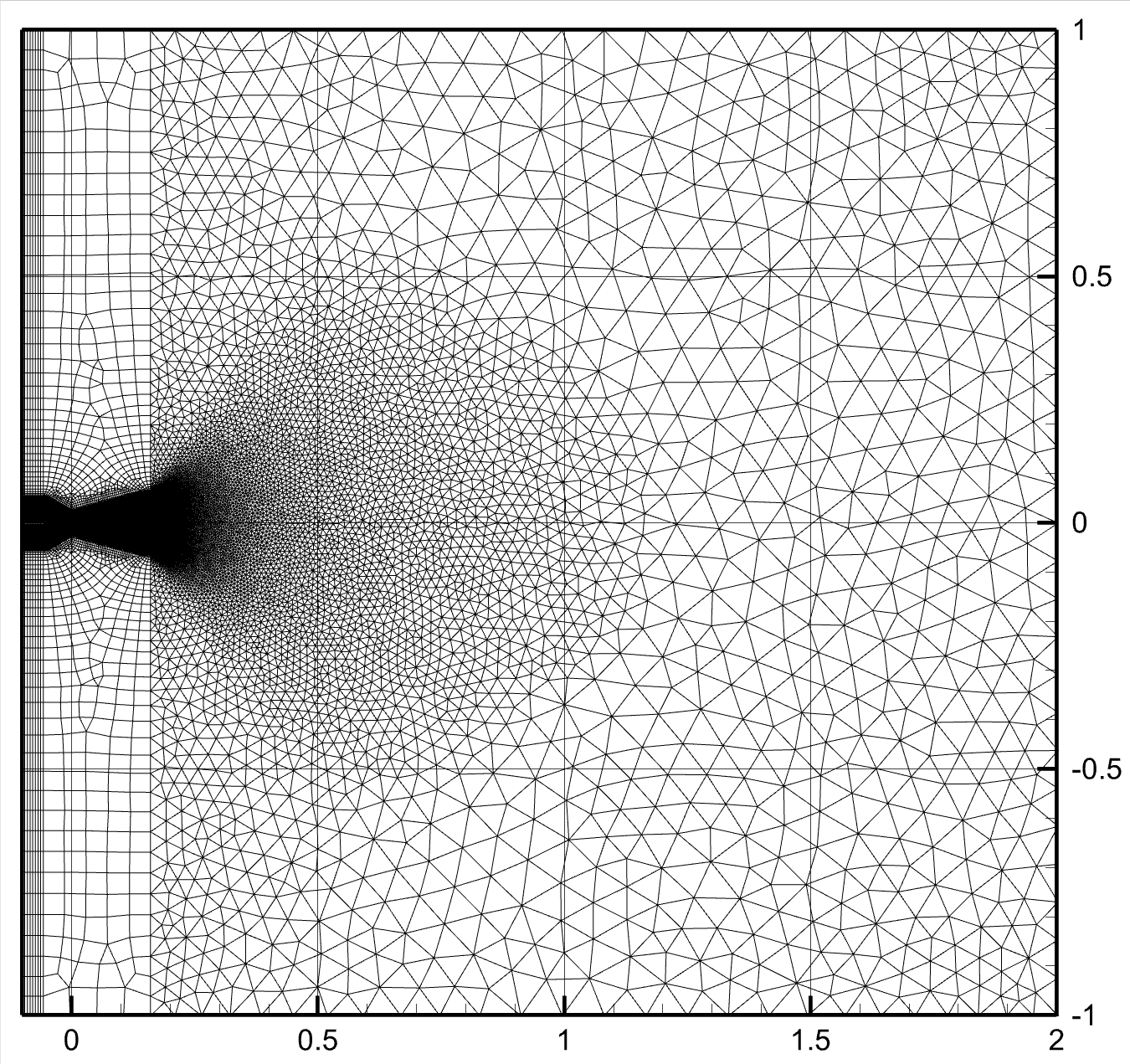}}
	\hspace{1mm}
	\subfloat[]{\includegraphics[width=0.45\textwidth]{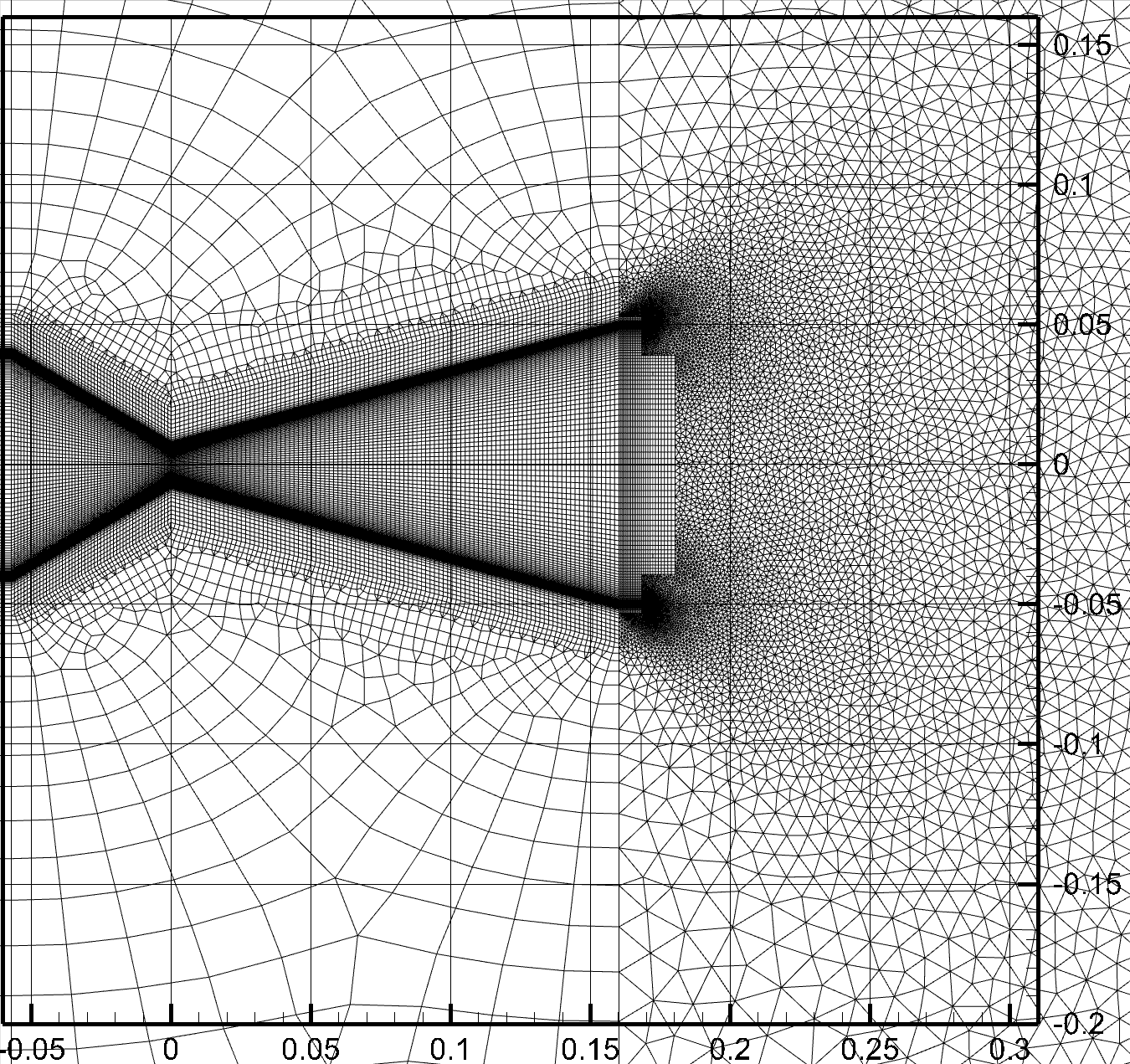}} \\
	\caption{Mesh of the nozzle. (a) Grid used in computation and (b) the geometry of nozzle.}
	\label{fig:nozzle-mesh}
\end{figure}

The geometry of the nozzle and the two-dimensional mesh with 47186 cells are used in the simulation, as shown in Fig.~\ref{fig:nozzle-mesh}. The inlet boundary condition is set with a temperature $710$ K and pressure 36.5 torr. The gas $CO_2$ is employed with the molecular mass $m=7.31 \times 10^{-26}$ kg, the ratio of specific heats $\gamma = 1.4$, and $\omega = 0.67$. The nozzle wall is treated as isothermal one with $T_w = 300$ K. The background environment is set with a temperature $T_B = 300 K$ and low pressure $P_B = 0.01$ Pa. The CFL number is $0.5$. The reference number of particles is $N_r=400$. For the AUGKWP method, the reference Knudsen number is ${\rm Kn}_{ref} = 0.01$.

\begin{figure}[H]
	\centering
	\subfloat[]{\includegraphics[width=0.48\textwidth]{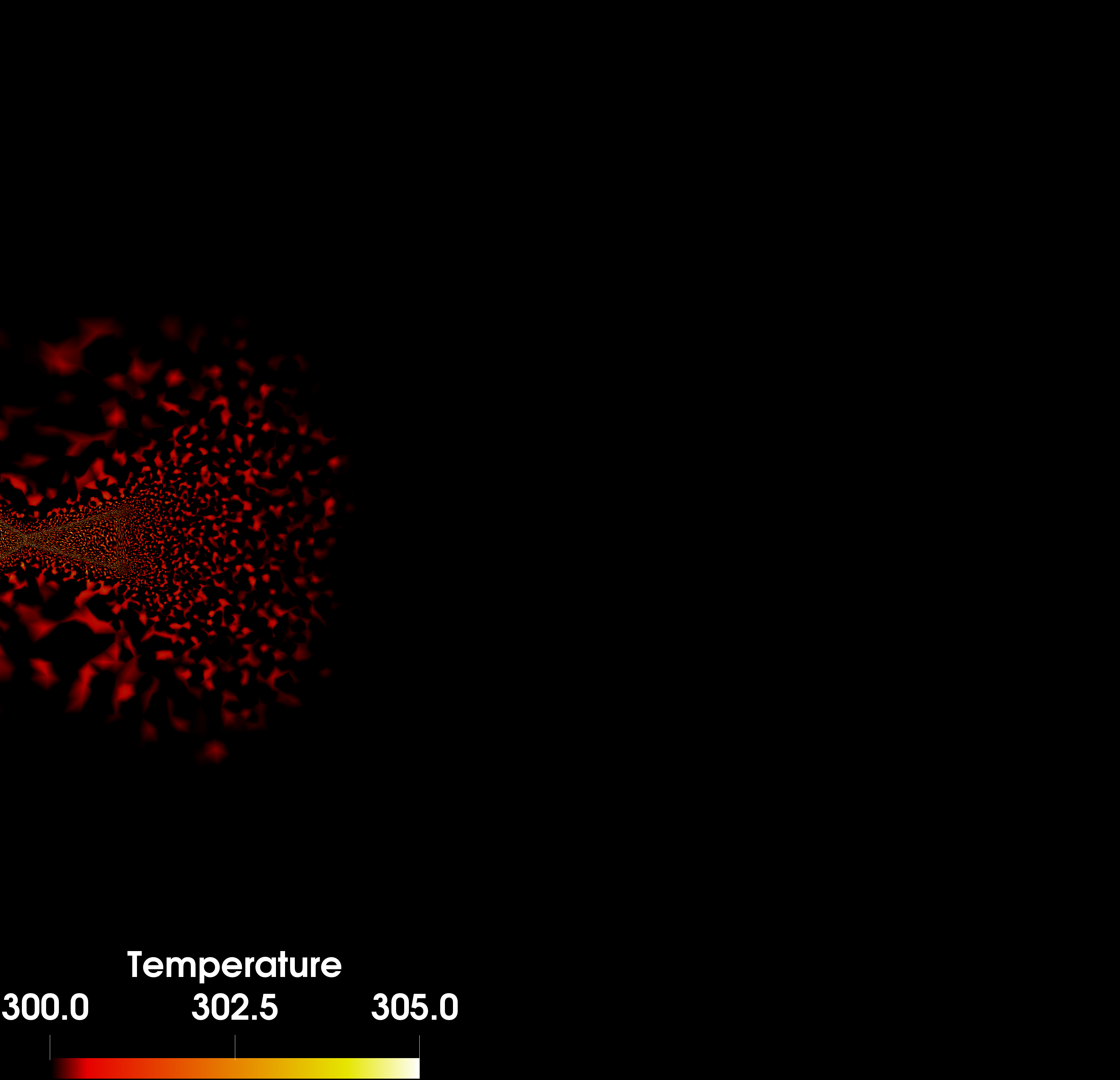}}
	\hspace{1mm}
	\subfloat[]{\includegraphics[width=0.48\textwidth]{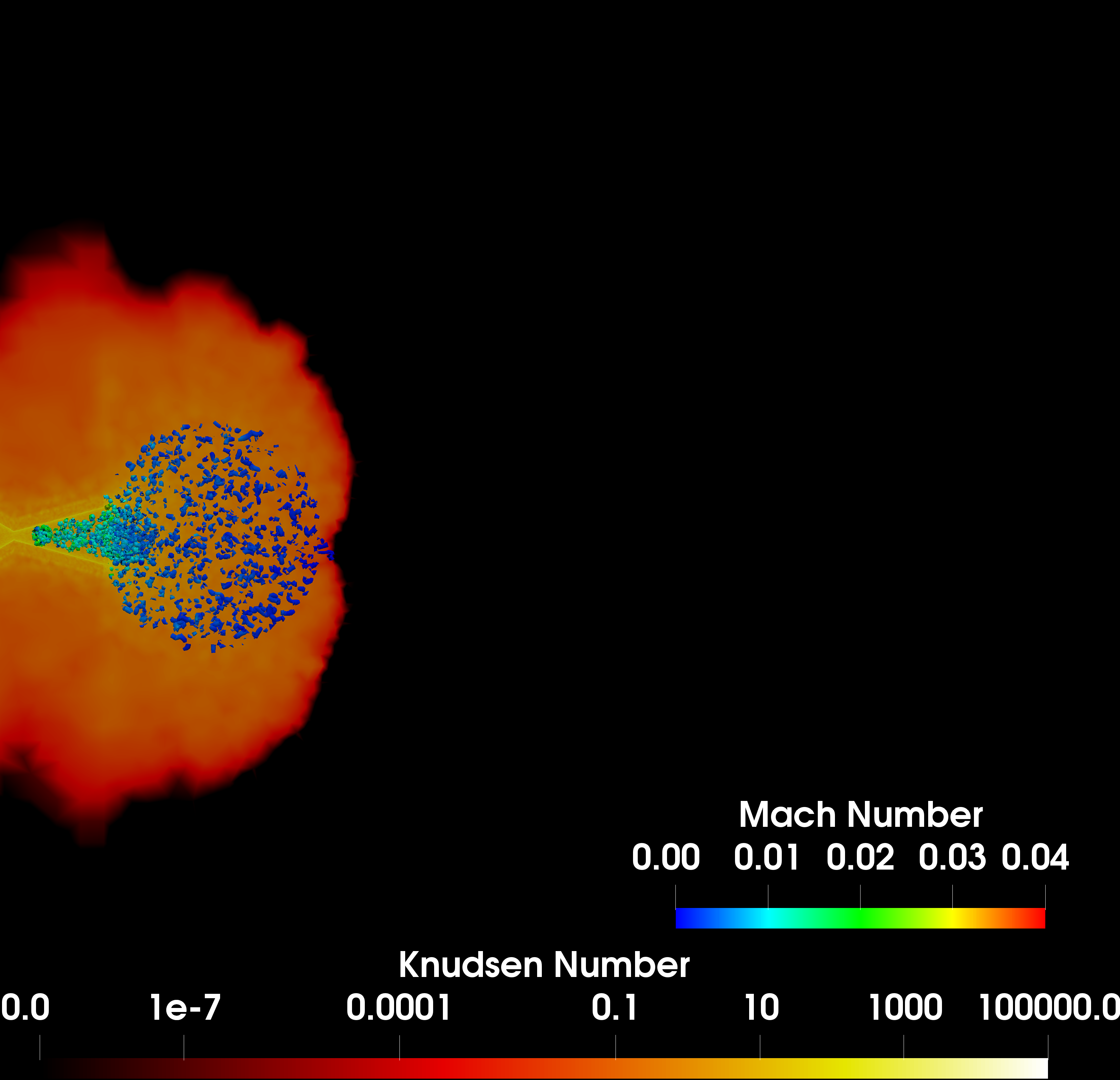}} \\
	\subfloat[]{\includegraphics[width=0.48\textwidth]{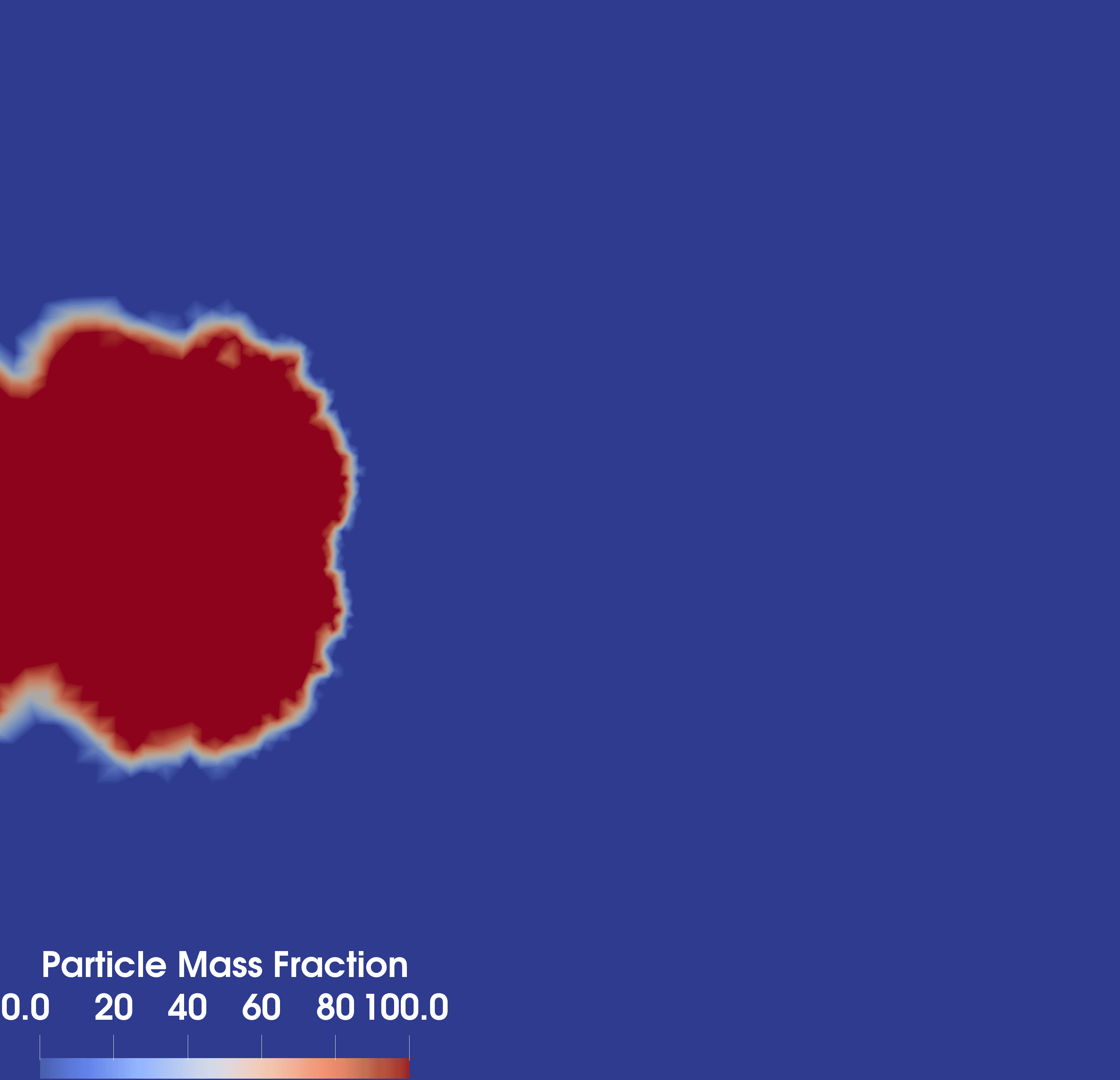}}
	\hspace{1mm}
	\subfloat[]{\includegraphics[width=0.48\textwidth]{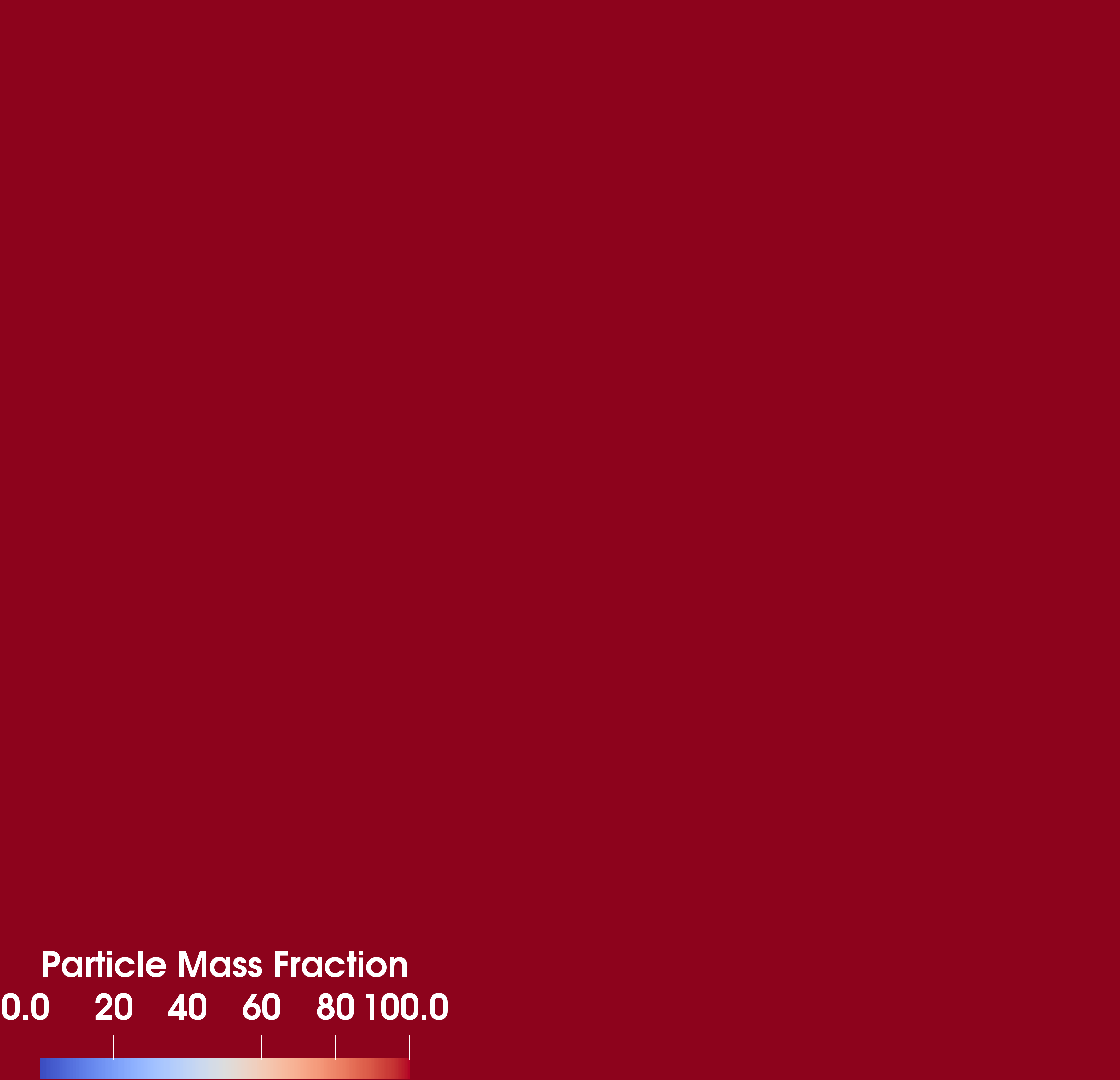}}
	\caption{Nozzle plume flow to a background vacuum at initial stage. Distributions of (a) Temperature, (b) local Knudsen number and Mach number along the streamline, (c) particle mass fraction of AUGKWP, and (d) particle mass fraction of UGKWP.}
	\label{fig:nozzle-init}
\end{figure}

\begin{figure}[H]
	\centering
	\subfloat[]{\includegraphics[width=0.48\textwidth]{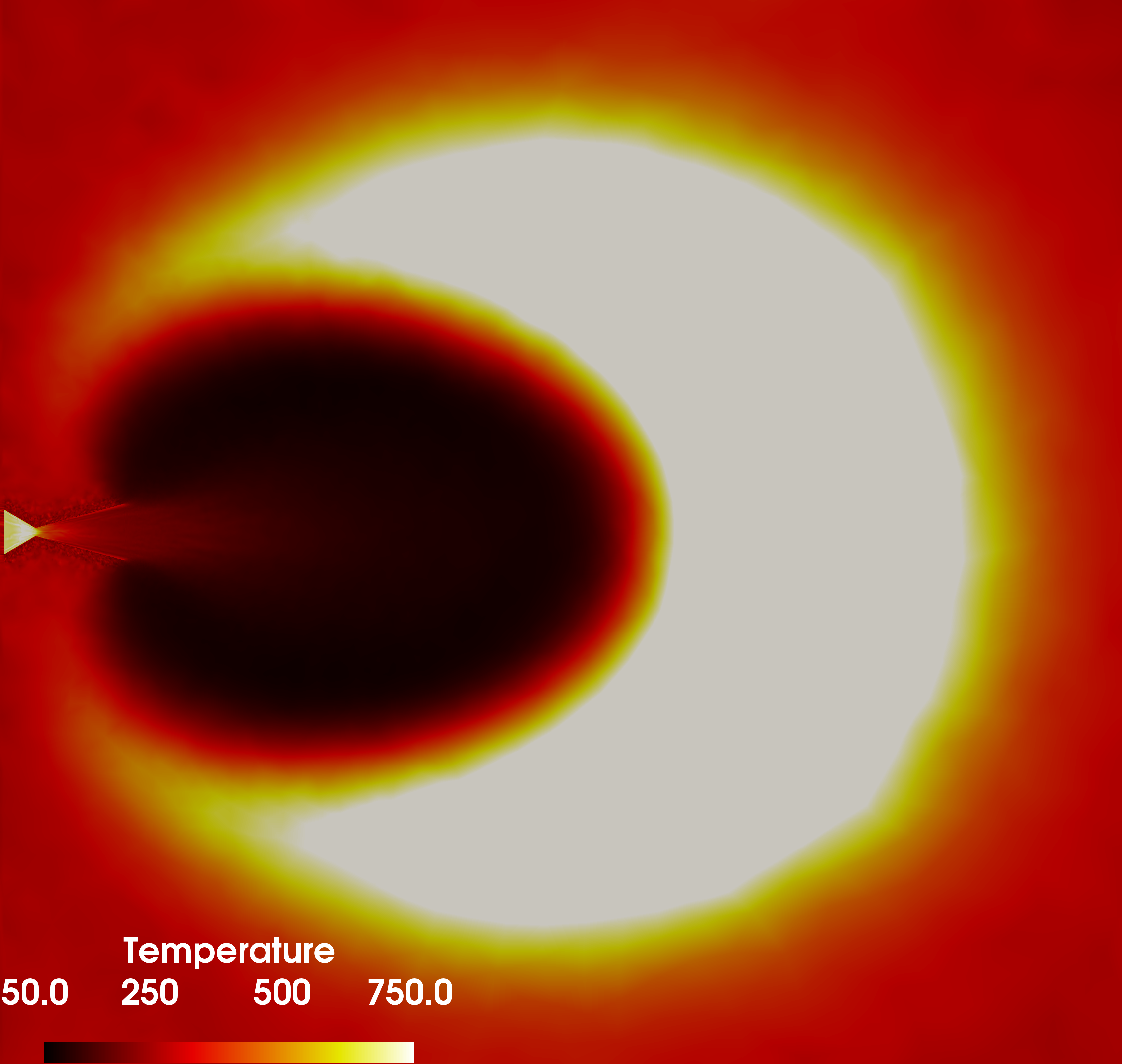}}
	\hspace{1mm}
	\subfloat[]{\includegraphics[width=0.48\textwidth]{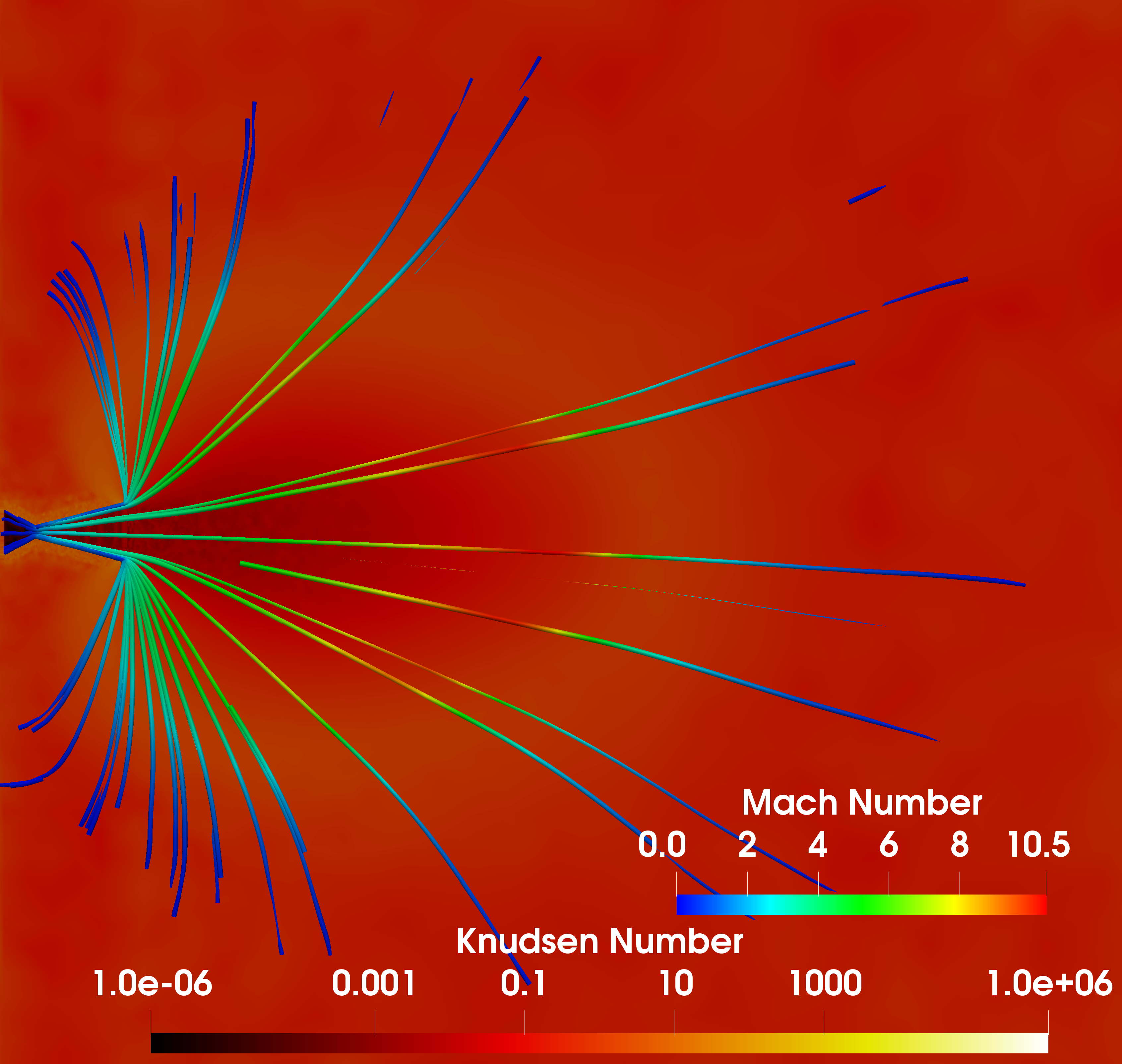}} \\
	\subfloat[]{\includegraphics[width=0.48\textwidth]{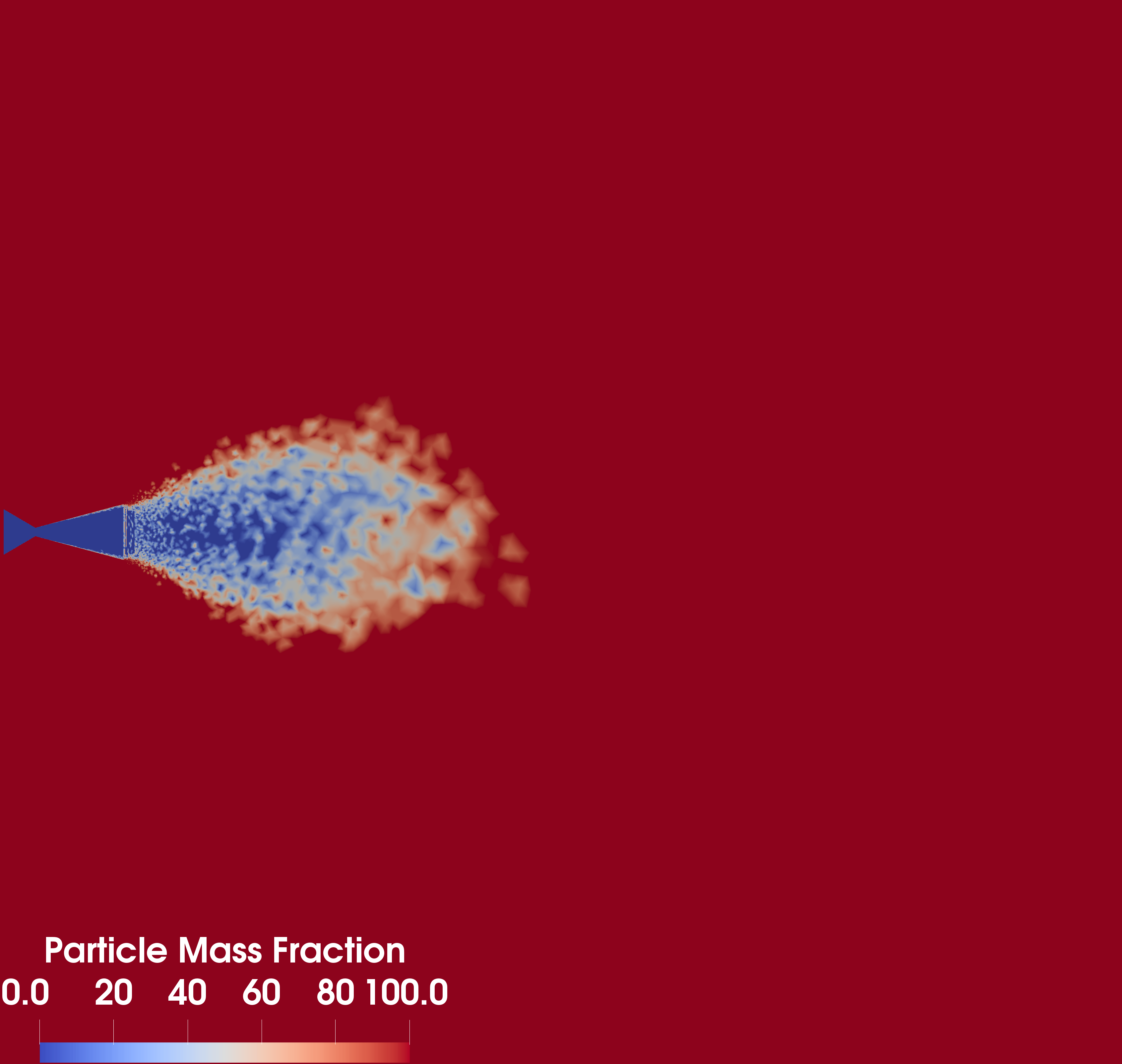}}
	\hspace{1mm}
	\subfloat[]{\includegraphics[width=0.48\textwidth]{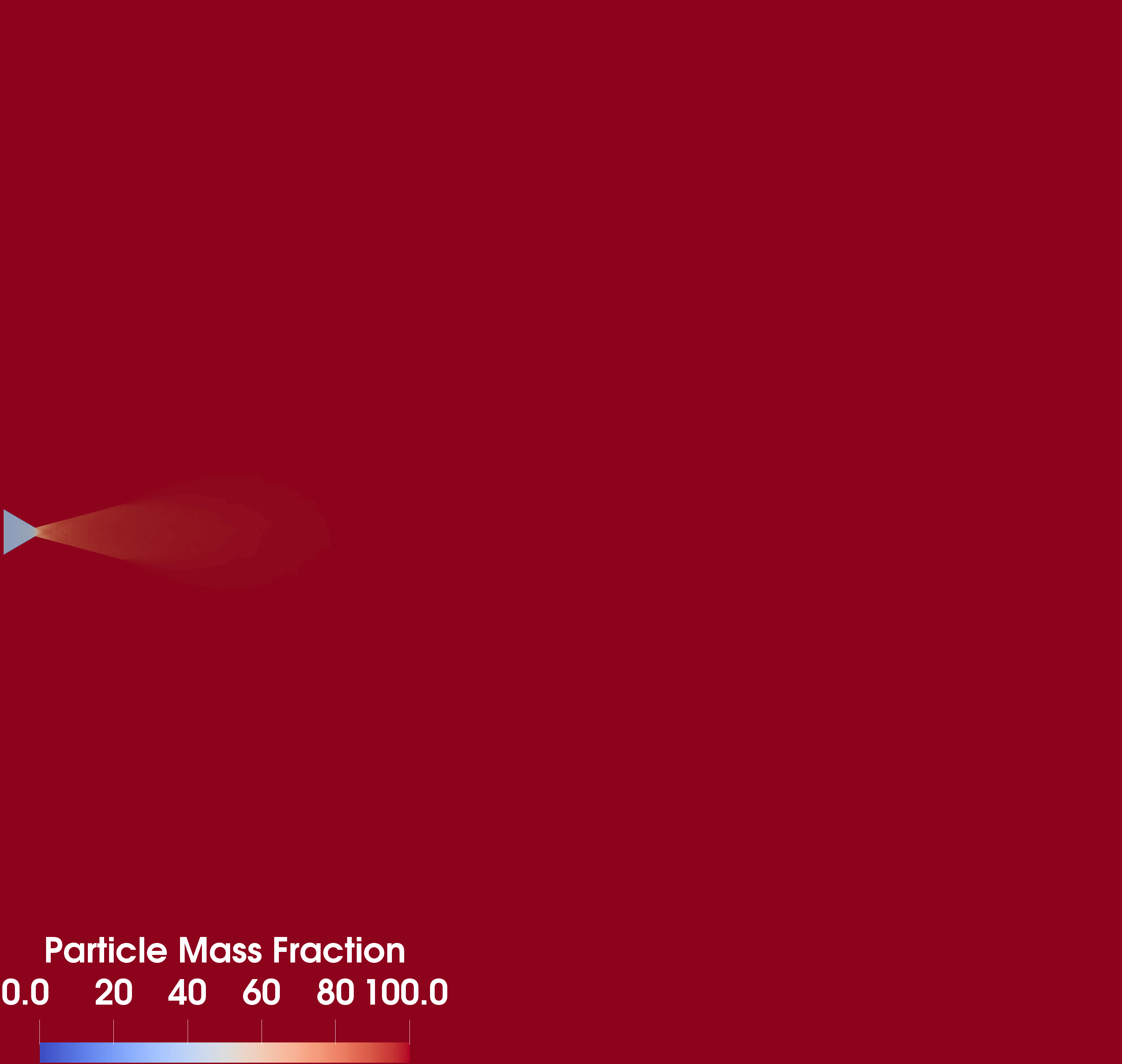}}
	\caption{Nozzle plume flow to a background vacuum at developing stage. Distributions of (a) Temperature, (b) local Knudsen number and Mach number along the streamline, (c)  particle mass fraction of AUGKWP, and (d) particle mass fraction of UGKWP.}
	\label{fig:nozzle-develop}
\end{figure}

\begin{figure}[H]
	\centering
	\subfloat[]{\includegraphics[width=0.48\textwidth]{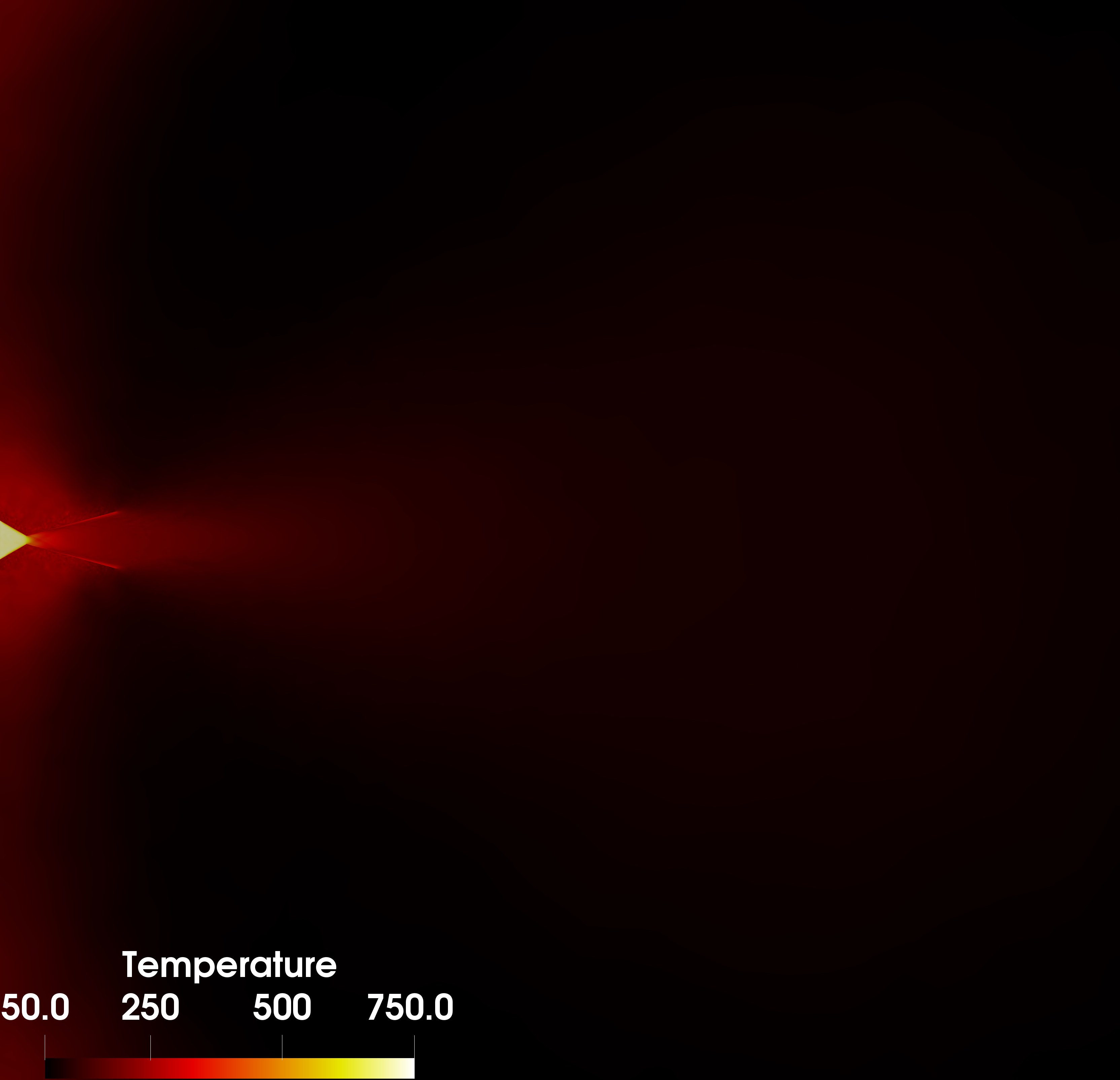}}
	\hspace{1mm}
	\subfloat[]{\includegraphics[width=0.48\textwidth]{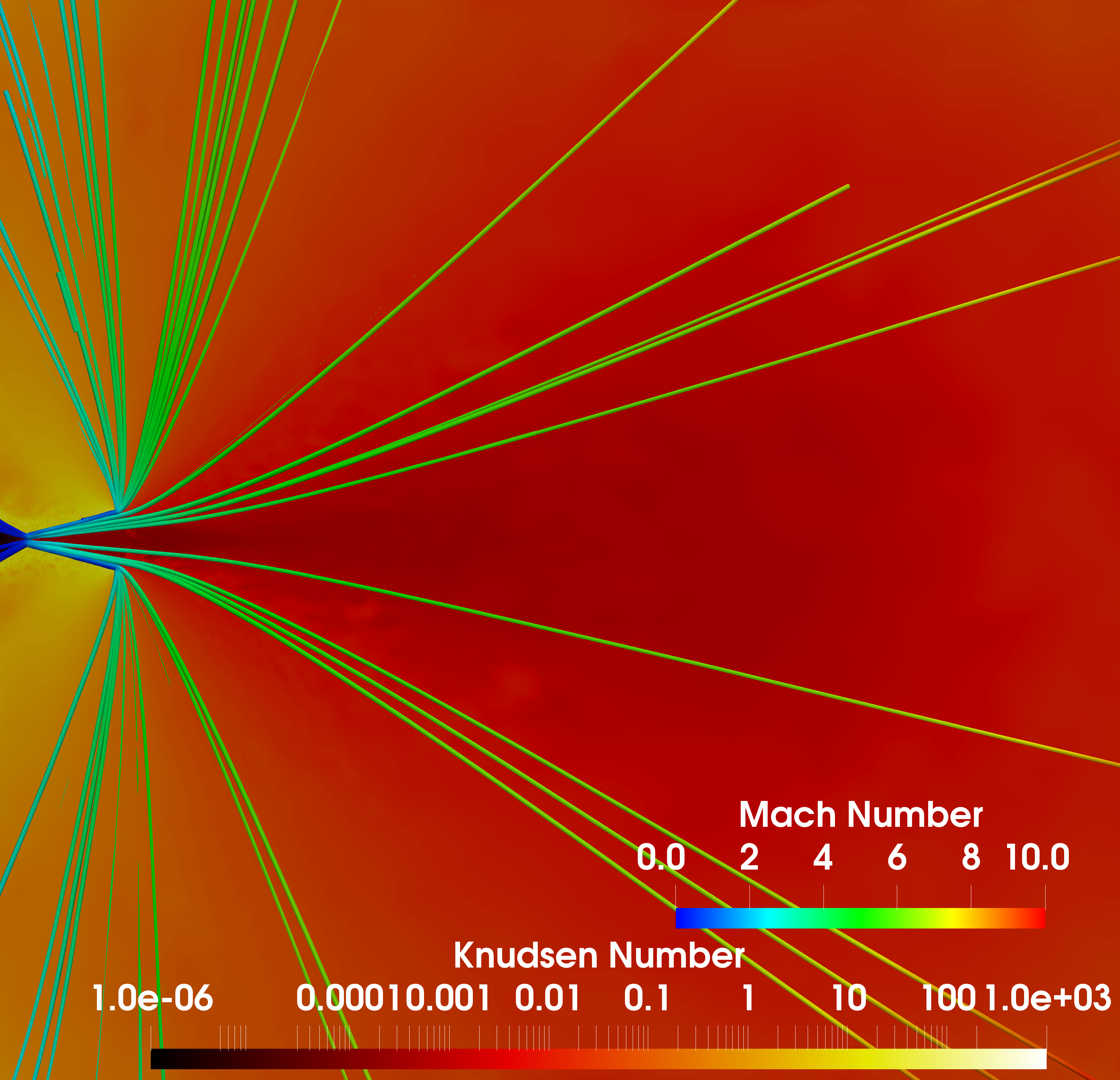}} \\
	\subfloat[]{\includegraphics[width=0.48\textwidth]{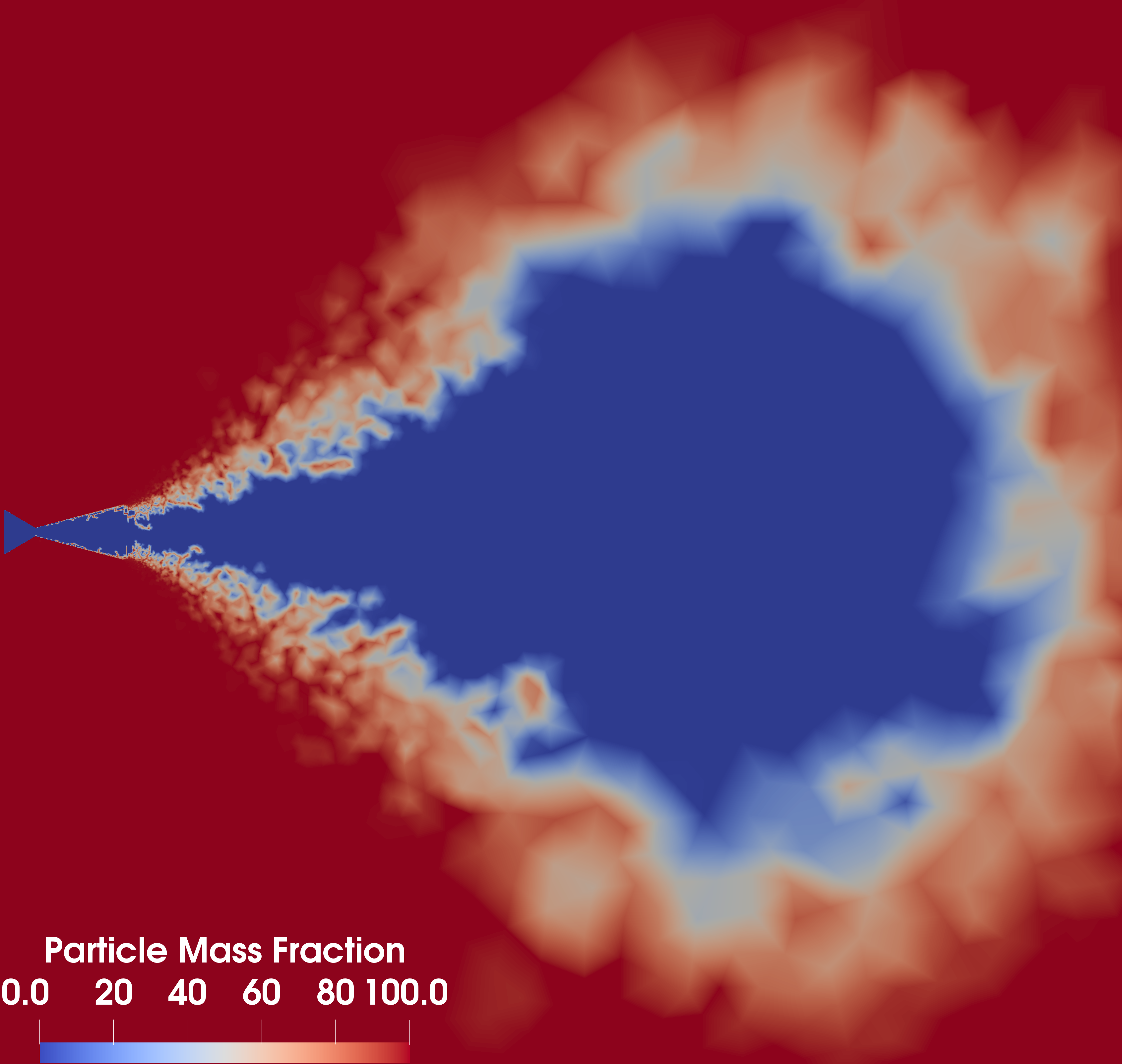}}
	\hspace{1mm}
	\subfloat[]{\includegraphics[width=0.48\textwidth]{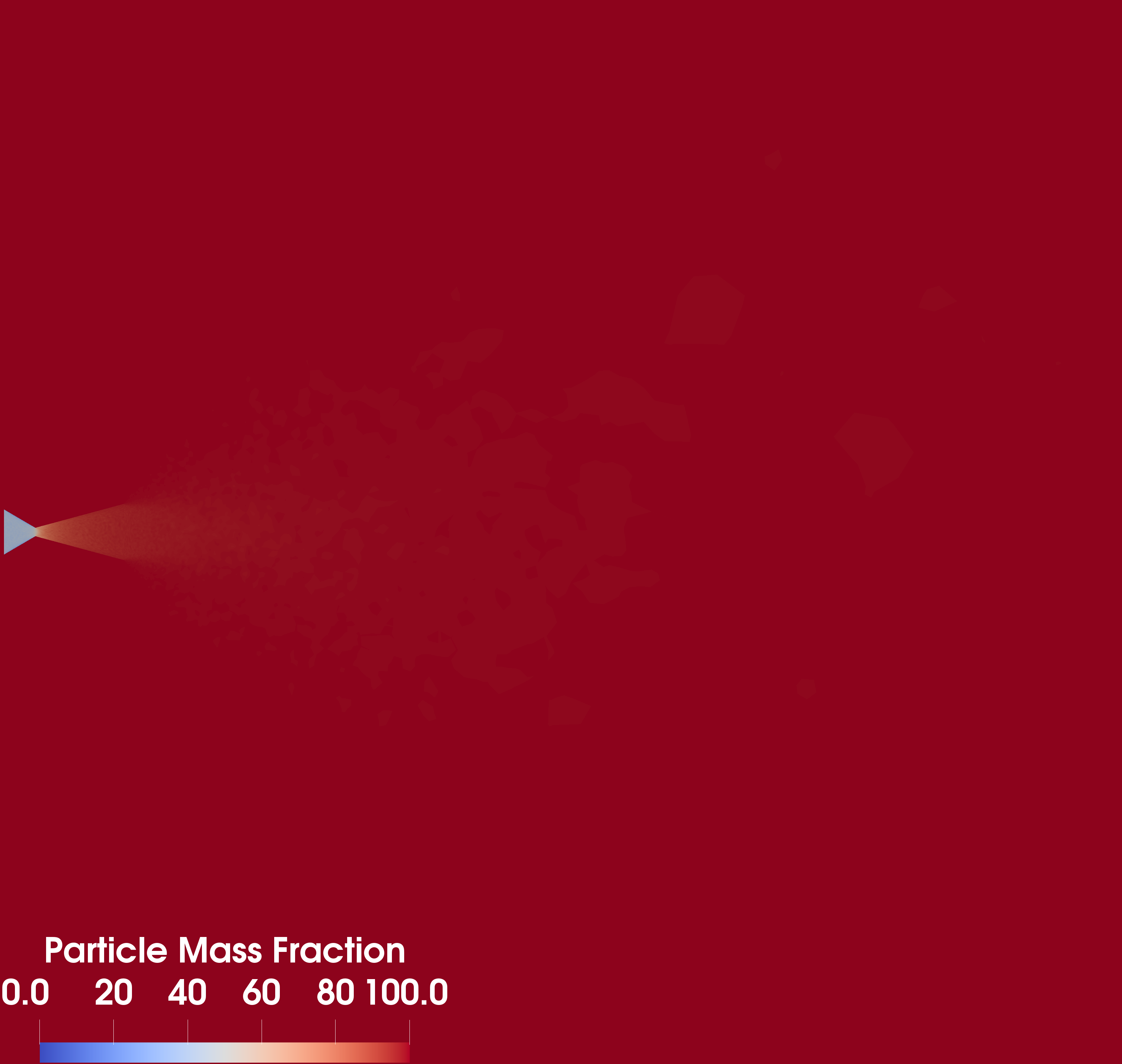}}
	\caption{Nozzle plume flow to a background vacuum at steady stage. Distributions of (a) Temperature, (b) local Knudsen number and Mach number along the streamline, (c) particle mass fraction of AUGKWP, and (d) particle mass fraction of UGKWP.}
	\label{fig:nozzle-steady}
\end{figure}

The simulation covers the whole gas expansion process into a background vacuum through three stages, such as the initial, developing, and steady stages. Fig.~\ref{fig:nozzle-init}--\ref{fig:nozzle-steady} show the distributions of temperature, gradient-length dependent local Knudsen number, and Mach number along the streamline, and particle mass fraction of the AUGKWP and UGKWP methods in each stage.

The flow field in the initial stage is shown in Fig.~\ref{fig:nozzle-init}.
Particles with large mean free path transport first to the background vacuum.
The expansion gas forms a non-equilibrium central region. The AUGKWP method employs particles in this highly expansion region only, where the
analytical wave is used in other region.
However, the UGKWP adapts particles everywhere, even in the uniformly undisturbed background equilibrium region.

In the developing stage (see Fig.~\ref{fig:nozzle-develop}), a continuum flow regime appears near the nozzle exit, a transition regime forms around the high temperature expansion region, and a free molecular flow remains in the front of the plume.
The gradient-length local Knudsen number shows a variation with 12 orders of magnitude in the whole computational domain.
The simulation of this unsteady multiscale transport requires a method with the capability of capturing continuum and rarefied flow simultaneously
at any moment by following the plume flow.
The unified treatment in AUGKWP with a wave-particle decomposition in each control volume allows an instant and adaptive description in each cell in this expansion process.

\begin{figure}[H]
	\centering
	\subfloat[]{\includegraphics[width=0.33\textwidth]{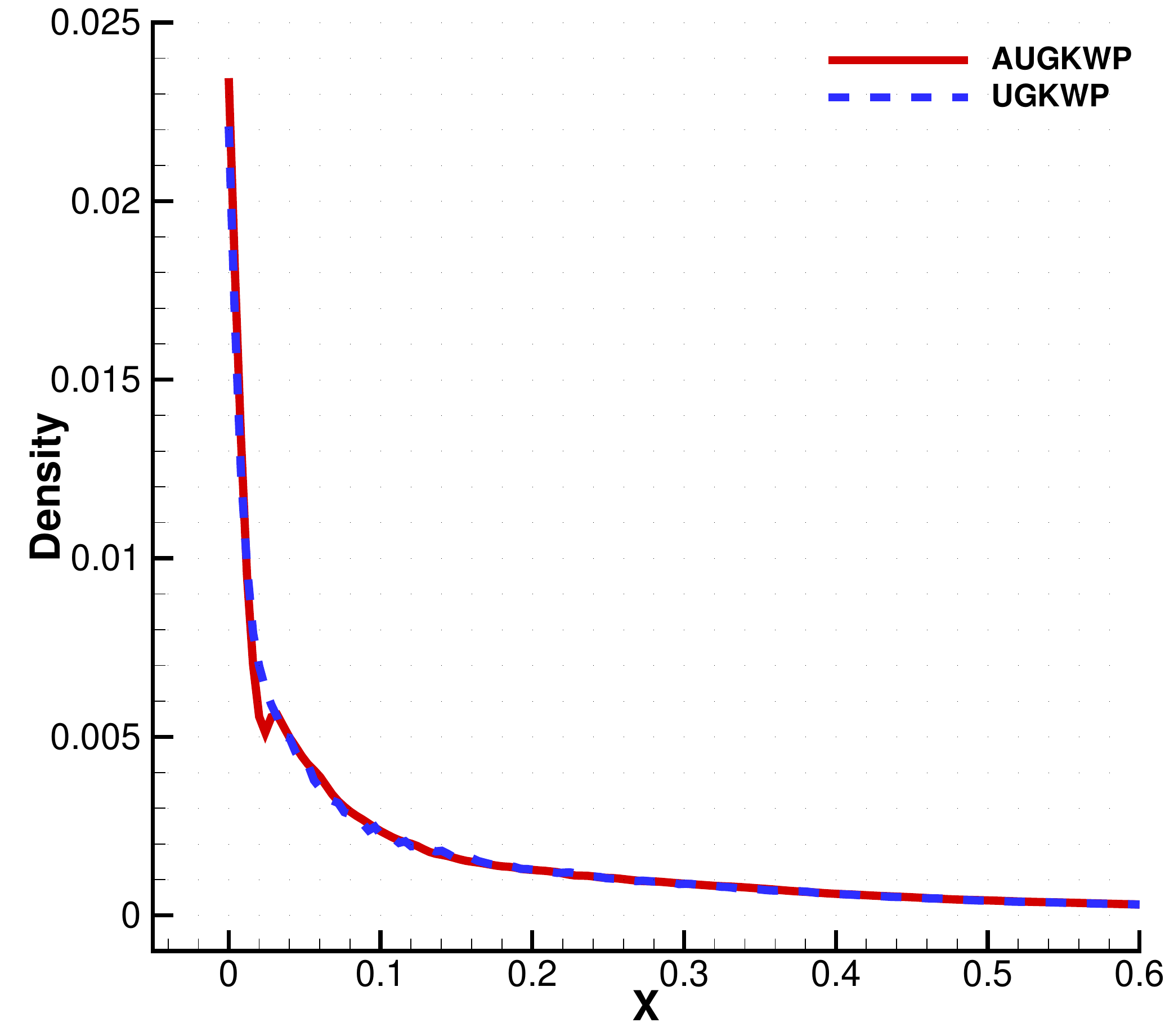}}
	\subfloat[]{\includegraphics[width=0.33\textwidth]{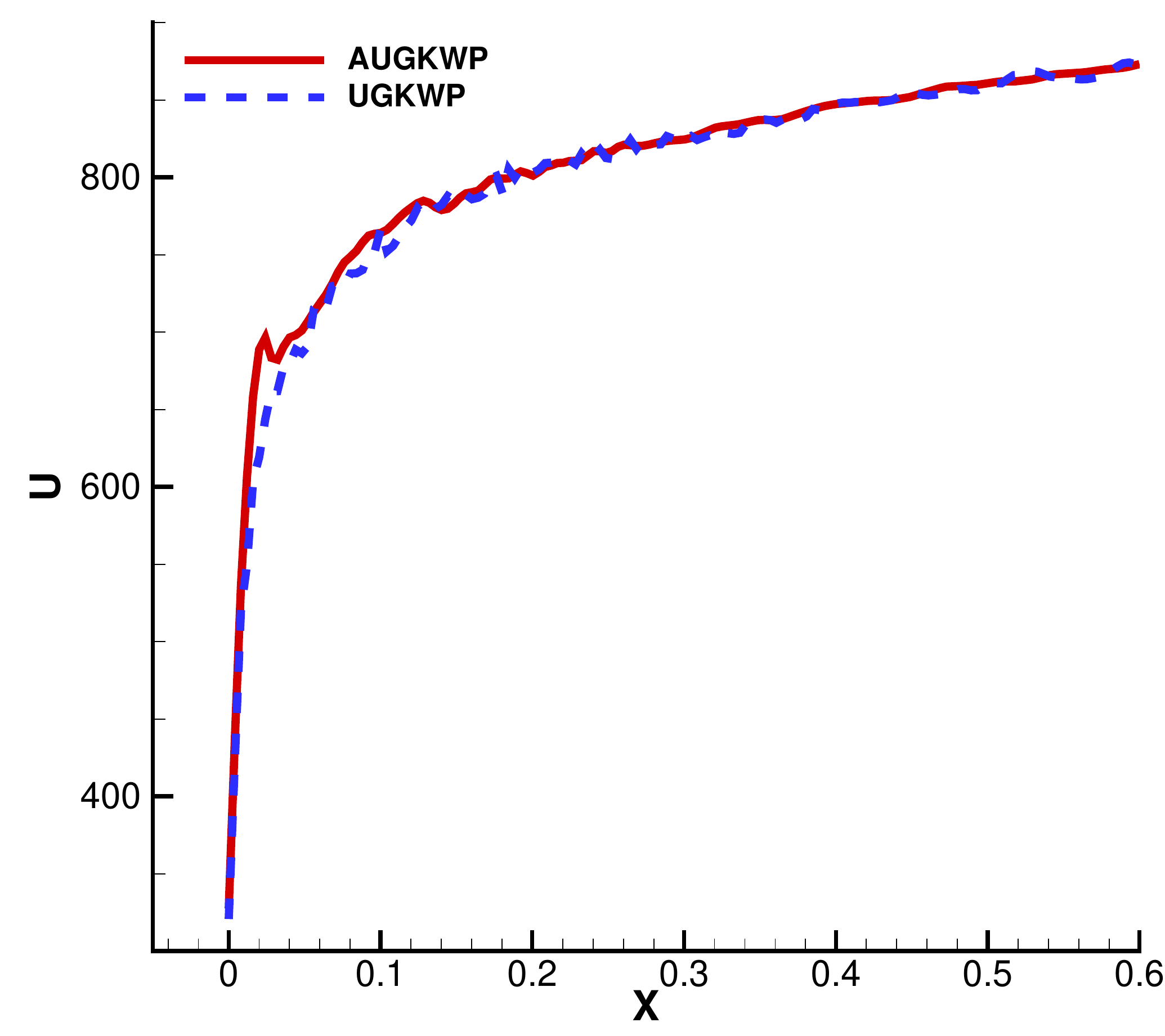}}
	\subfloat[]{\includegraphics[width=0.33\textwidth]{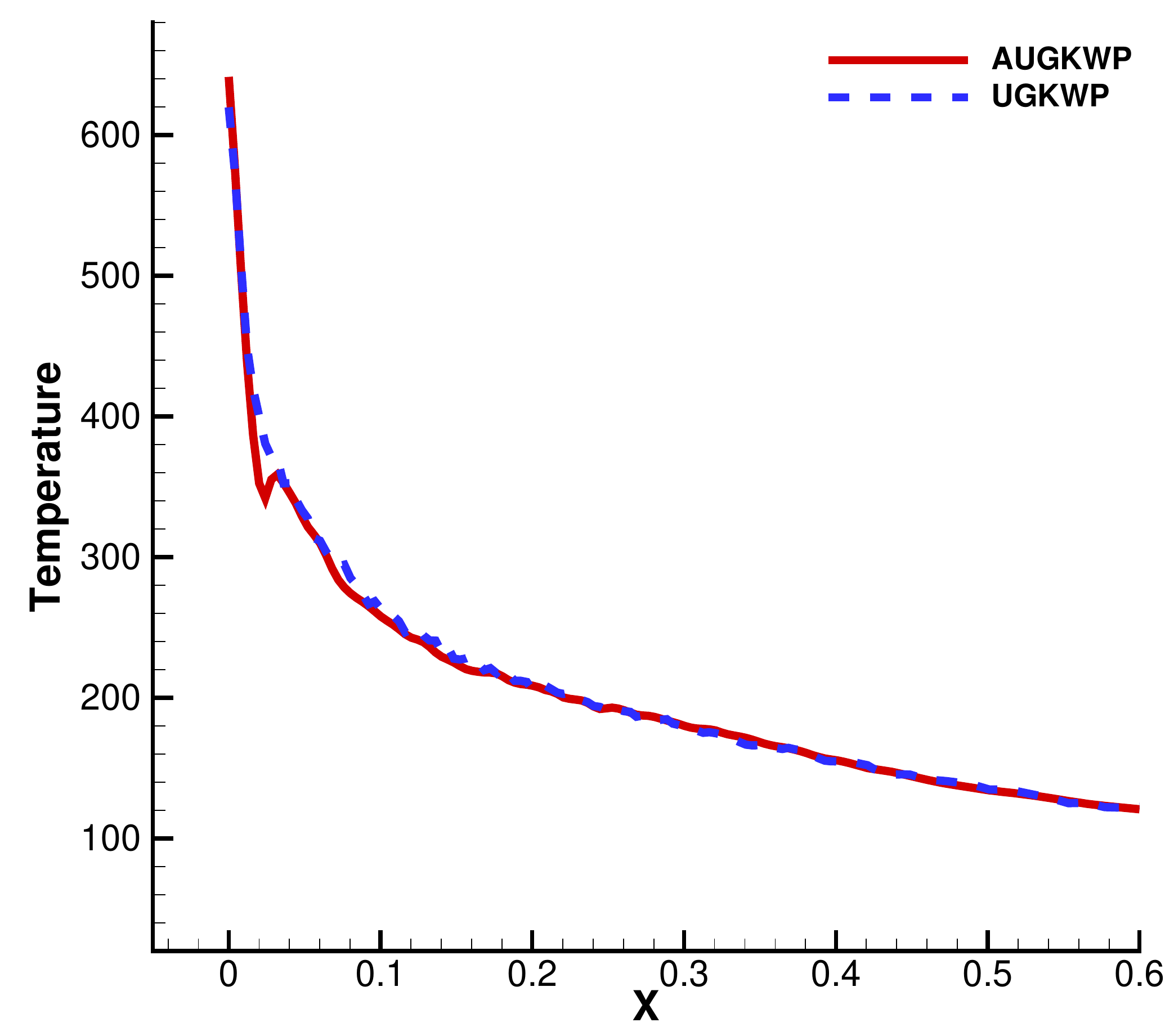}}
	\caption{Nozzle plume flow to a background vacuum at steady stage. (a) Density, (b) $x$ direction velocity, and (c) temperature from
UGKWP and AUGKWP methods along the centerline.}
	\label{fig:nozzle-line}
\end{figure}

Figure~\ref{fig:nozzle-steady} presents the plume flow approaching a steady state, where the gas is fully expanded with a steady flow pattern.
The AUGKWP method provides a clear separation of different flow regimes, i.e., a continuum flow in the expansion region, a free molecular free in the background flow, and a transition flow between them (see Fig.~\ref{fig:nozzle-steady}(c)).
In the original UGKWP, the numerical time step determined by the smallest cell size leads the particles representation in the almost whole computational domain. The quantitative comparison of density, velocity in $x$ direction, and the temperature along the centerline at the steady stage is plotted in Fig.~\ref{fig:nozzle-line}. It shows agreements between the results given by AUGKWP and UGKWP methods.

Overall, the AUGKWP method gives a non-equilibrium state guided wave-particle decomposition.
The computational times in the current studies on Tianhe-2 with 20 nodes (480 cores)
are 12.6 hours and 36.1 hours for AUGKWP and UGKWP methods, respectively.

\section{Conclusion}\label{sec:conclusion}

The UGKWP is a multiscale method for flow simulation in all regimes.
The UGKWP method adopts a wave-particle decomposition to recover the multiscale transport uniformly in each control volume.
In this paper, an adaptive unified gas-kinetic wave-particle (AUGKWP) method is developed to further optimize the wave-particle decomposition
in the original UGKWP.
In order to concentrate particles in the non-equilibrium region only, instead of using the cell's Knudsen number ${\rm Kn}_c = \tau / \Delta t$ only in the original UGKWP method,
the AUGKWP method introduces a flow gradient-related local Knudsen number as well for the decomposition of wave and particle.
As a result, the AUGKWP avoids using particles in the highly dilute background equilibrium region and in the continuum flow simulation with extremely small numerical time step.
The AUGKWP method provides a more physically reliable wave-particle decomposition and guarantees the appearance of particle
in the non-equilibrium region, regardless of the mesh resolution.
Many test cases are used to validate the efficiency and accuracy of the AUGKWP method.
In comparison with the original UGKWP method, due to significant reduction of particles the AUGKWP can speed up computation, free memory requirements, and maintain the same accuracy for multiscale solution.
The AUGKWP will become a useful and indispensable tool in the simulation of high-speed rarefied and continuum flow in aerodynamic applications.

\section*{Author's contributions}

All authors contributed equally to this work.

\section*{Acknowledgments}
This work was supported by National Key R$\&$D Program of China (Grant Nos. 2022YFA1004500), National Natural Science Foundation of China (12172316),
and Hong Kong research grant council (16208021,16301222).

\section*{Data Availability}

The data that support the findings of this study are available from the corresponding author upon reasonable request.





\newpage
\bibliographystyle{elsarticle-num}
\bibliography{etaKn.bib}







\end{document}